\documentclass[aps,twocolumn,prc,superscriptaddress,showpacs,nofootinbib,floatfix,amssymb,amsfonts,amsmath]{revtex4-1}
\pdfoutput=1
\usepackage{graphicx}
\usepackage{dcolumn}
\usepackage{bm}
\usepackage{amsmath}    
\usepackage{amsfonts}   
\usepackage{amssymb}
\usepackage{graphicx}   

\begin{document}

\title{Extension of nuclear landscape to hyperheavy nuclei}

\author{S.\ E.\ Agbemava}
\affiliation{Department of Physics and Astronomy, Mississippi
State University, MS 39762}

\author{A.\ V.\ Afanasjev}
\affiliation{Department of Physics and Astronomy, Mississippi
State University, MS 39762}
\affiliation{Yukawa Institute of Theoretical Physics, Kyoto University, 
                Japan}

\author{A. Taninah}
\affiliation{Department of Physics and Astronomy, Mississippi
State University, MS 39762}

\author{A.\ Gyawali}
\affiliation{Department of Physics and Astronomy, Mississippi
State University, MS 39762}

\date{\today}

\begin{abstract}
   The properties of hyperheavy nuclei and the extension of nuclear landscape
to hyperheavy nuclei are extensively studied within covariant density functional
theory. Axial reflection symmetric and reflection asymmetric relativistic 
Hartree-Bogoliubov (RHB) calculations are carried out. The role of triaxiality is 
studied within triaxial RHB and triaxial relativistic mean field + BCS frameworks. 
With increasing proton number beyond $Z\sim 130$ the transition from ellipsoidal-like 
nuclear shapes to toroidal ones takes place. The description of latter shapes 
requires the basis which is typically significantly larger than the one employed for 
the description of ellipsoidal-like shapes. Many hyperheavy nuclei with toroidal 
shapes are expected to be unstable towards multifragmentation. However, three 
islands of stability of spherical hyperheavy nuclei have been predicted for the first
time in Ref.\ \cite{AAG.18}. Proton and neutron densities, charge radii, neutron skins 
and underlying shell structure of the nuclei  located in the centers of these islands 
have  been investigated  in detail. Large neutron shell gaps at $N=228, 308$ 
and 406 define approximate centers of these islands in neutron number. On the
contrary,  large proton gap appear only at $Z=154$ in the $(Z\sim 156, N\sim 310)$ 
island. As a result, this is the largest island of stability of spherical hyperheavy
nuclei found in the calculations. The calculations indicate the stability of the nuclei in 
these islands with respect of octupole and triaxial distortions. The shape evolution of 
toroidal shapes along the fission path and the stability of  such shapes with respect of 
fission have been studied. Fission barriers in neutron-rich superheavy nuclei  are studied 
in triaxial RHB framework; the impact of triaxiality on the heights of fission barriers is 
substantial in some parts of this region.  Based on the results obtained in the 
present work, the extension of nuclear landscape to hyperheavy  nuclei is provided.
\end{abstract}

\maketitle

\section{Introduction}
\label{introduction}
 
   One of the main focuses of modern low-energy  physics is the
limits of the existence of finite nuclei. New generation of facilities
such as FRIB, FAIR, RIKEN, and GANIL  will explore such limits in 
neutron-rich nuclei. SHE-factory and  similar facilities will attempt to 
extend the limits of our knowledge on superheavy nuclei. However, 
already now it is clear that there are significant restrictions on what could be 
achieved by these new facilities: many neutron-rich medium mass, 
heavy and superheavy nuclei will be beyond their experimental reach 
\cite{AA.16}. In such a situation, theoretical predictions became the 
only tool to investigate such limits. Indeed, a significant progress has 
been achieved in understanding of the limits of nuclear landscape for 
the $Z<120$ nuclei (see Refs.\ \cite{ Eet.12,AARR.13,AARR.14}) and 
more or less consistent picture has been obtained using the combination 
of  different theoretical tools. In addition, systematic theoretical 
uncertainties \cite{Eet.12,AARR.14,AARR.15} and statistical errors 
\cite{GDKTT.13,KENBGO.13,AAT.18} in the predictions of the properties 
of neutron-rich nuclei and the positions of two-proton and two neutron-drip 
lines have been evaluated.

  However, the nuclear landscape  is not restricted to the $Z<120$
nuclei. Although there were some attempts to investigate higher 
$Z$ nuclei \cite{DK-prl.98,DBDW.99,BNR.01,Denisov.05,GBG.15,IEAA.16}, 
these systematic studies were restricted to spherical symmetry.  Our recent 
study (Ref.\  \cite{AAG.18}) based on systematic axial Relativistic Hartree-Bogoliubov
(RHB) calculations and triaxial RHB as well as triaxial relativistic mean field + 
BCS (RMF+BCS) calculations for a reasonable large
set of selected nuclei has invalidated many conclusions of these 
studies\footnote{The effects of axial and triaxial deformations have also been 
studied for a few hyperheavy nuclei in Refs.\ \cite{BBDGK.01,Warda.07}
and Ref.\ \cite{StaW.09}, respectively. Somewhat larger set of the nuclei was 
studied with triaxiality included in Ref.\ \cite{BS.13} but according
to Ref.\ \cite{AAG.18} the deformation range employed in this work is 
not sufficient for $Z \geq 130$ nuclei.}.  In addition, it provided a new
vision on the properties of hyperheavy nuclei and on the extension 
of nuclear landscape to the $Z>120$ region. These results are briefly 
summarized below. The increase of proton number beyond $Z=120$ leads 
to the dominance of highly deformed and superdeformed oblate ground 
states.  However, these states with ellipsoidal-like shapes become unstable 
with respect of fission in  the $Z\sim 130$ region (see also Ref.\ \cite{BS.13} 
for the results obtained for fission barriers in non-relativistic theories). This 
triggers the transition to the states with toroidal shapes; the lowest in energy 
solutions in the $Z=140-180$ nuclei have such shapes in axial RHB calculations. 
It was illustrated that some of such states could be stable against fission. In 
addition,  some regions of stability of spherical hyperheavy nuclei have been 
predicted for the first time in Ref.\ \cite{AAG.18}.  Although these states are 
highly excited with respect of the lowest in energy states with toroidal shapes
(as obtained in axial RHB  calculations),  they will become the ground states 
if toroidal states are not stable with respect of multifragmentation (which 
according to present understanding (see Ref.\  \cite{Wong.73}) is quite 
likely scenario).

   Note that only in hyperheavy nuclei the states with toroidal shapes
could become the lowest in energy. The toroidal shapes in atomic nuclei 
have been investigated in a number  of the papers 
\cite{Warda.07,StaW.09,SW.14,IMMI.14,KSW.17,NBCHKRV.02}.
However, in absolute majority of the cases such shapes correspond to 
highly excited states either at spin zero \cite{StaW.09,KSW.17} or at extreme
values of angular momentum \cite{SW.14,IMMI.14,SWK.17}.  In the former 
case, such states are unstable against returning to the shape of sphere-like 
geometry (Ref.\ \cite{KSW.17}).  In the latter case, calculated angular momenta 
at which toroidal shapes appear substantially exceed the values of angular 
momentum presently  achievable at the state-of-art experimental facilities 
\cite{PhysRep-SBT}.

\begin{figure*}[htb]
\includegraphics[angle=0,width=8.5cm]{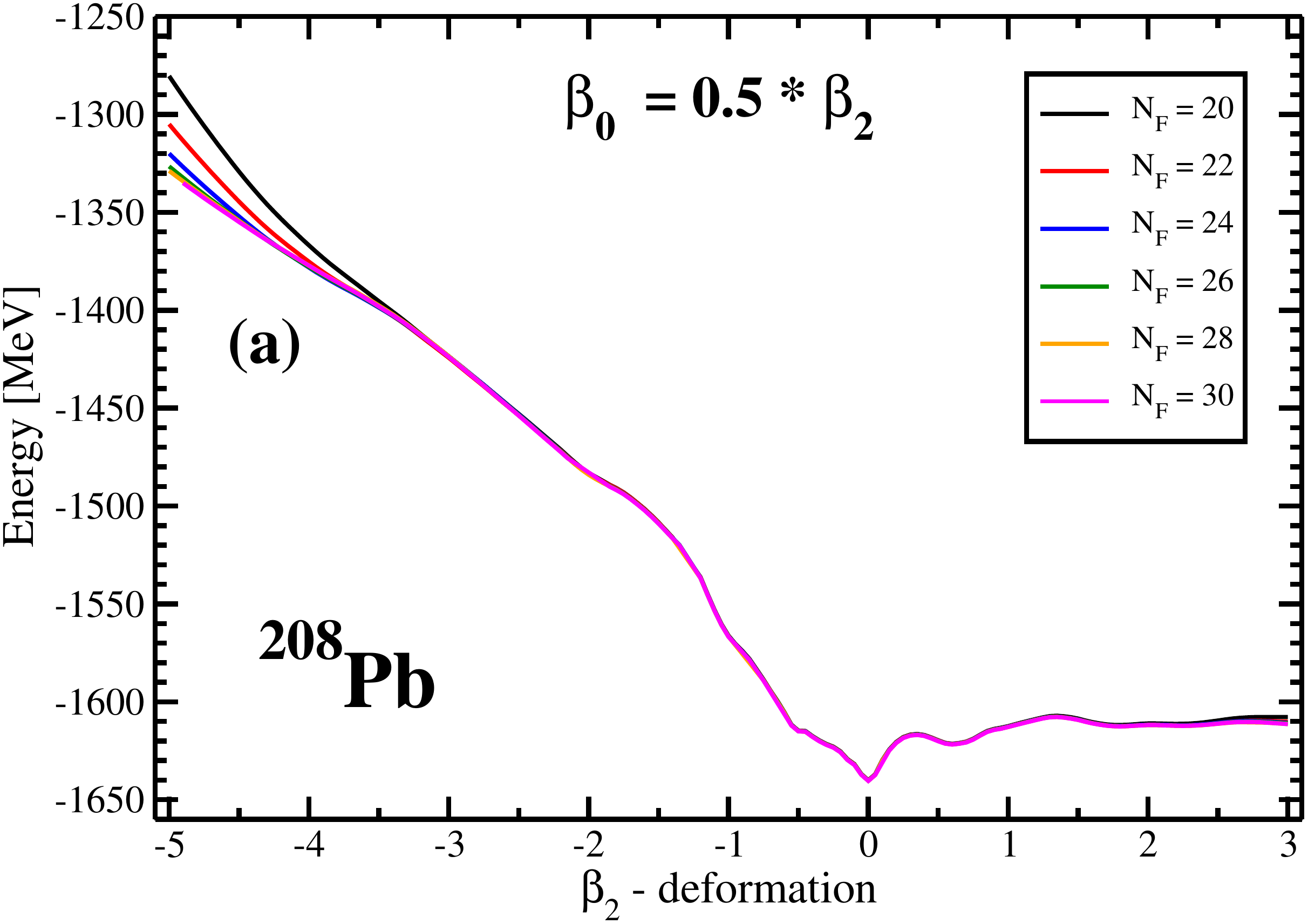}   
\includegraphics[angle=0,width=8.5cm]{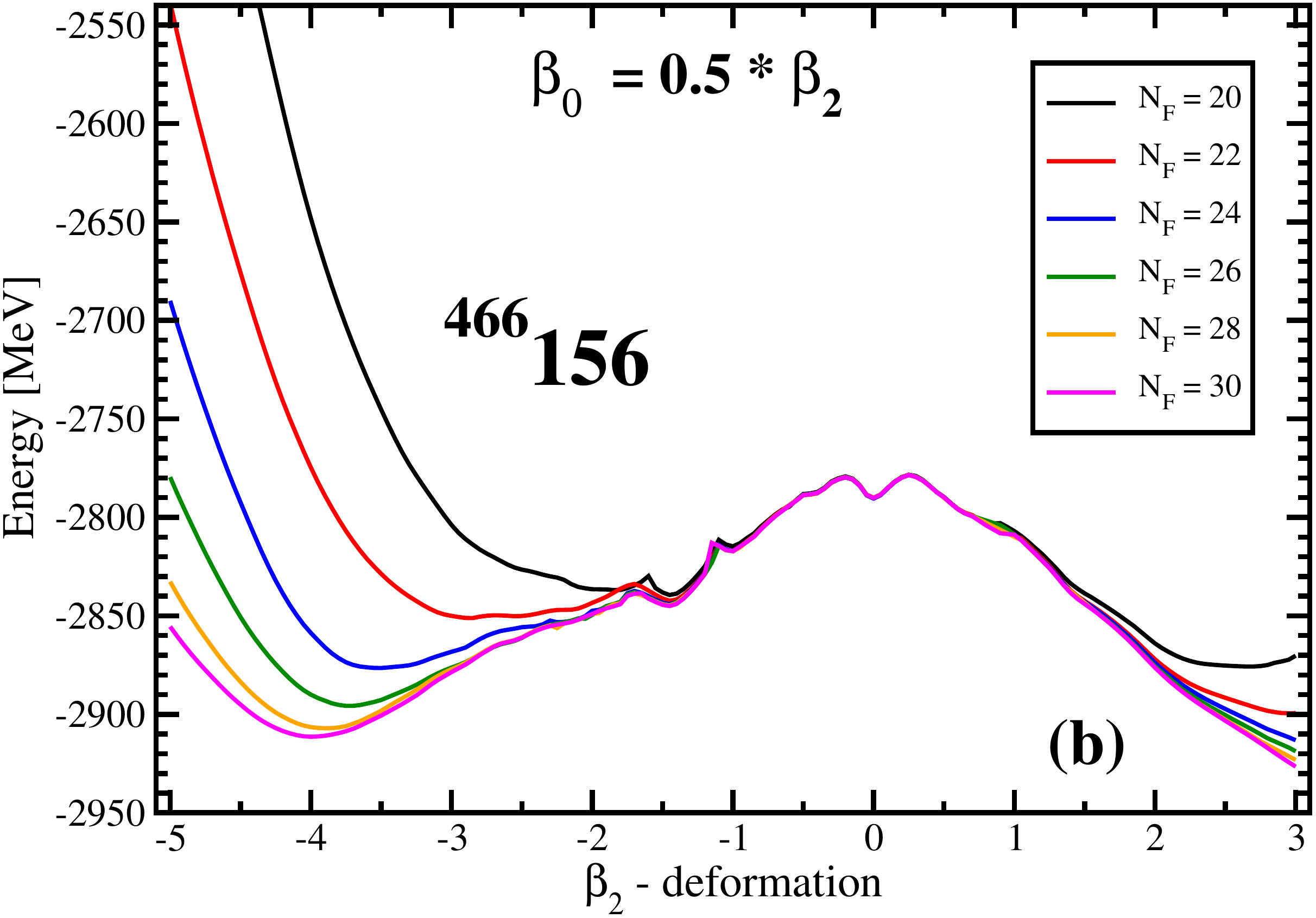}   
\includegraphics[angle=0,width=8.5cm]{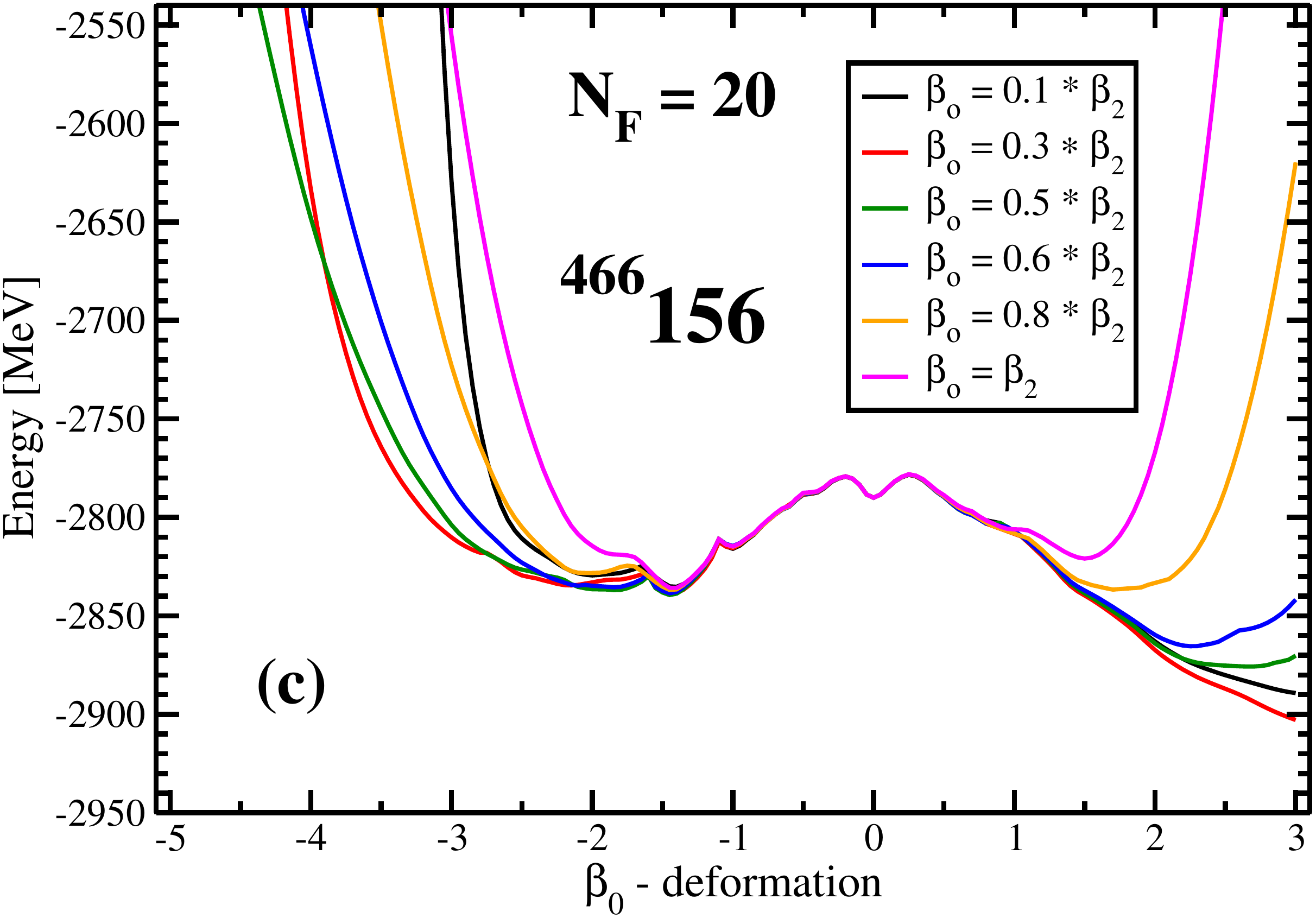}   
\includegraphics[angle=0,width=8.5cm]{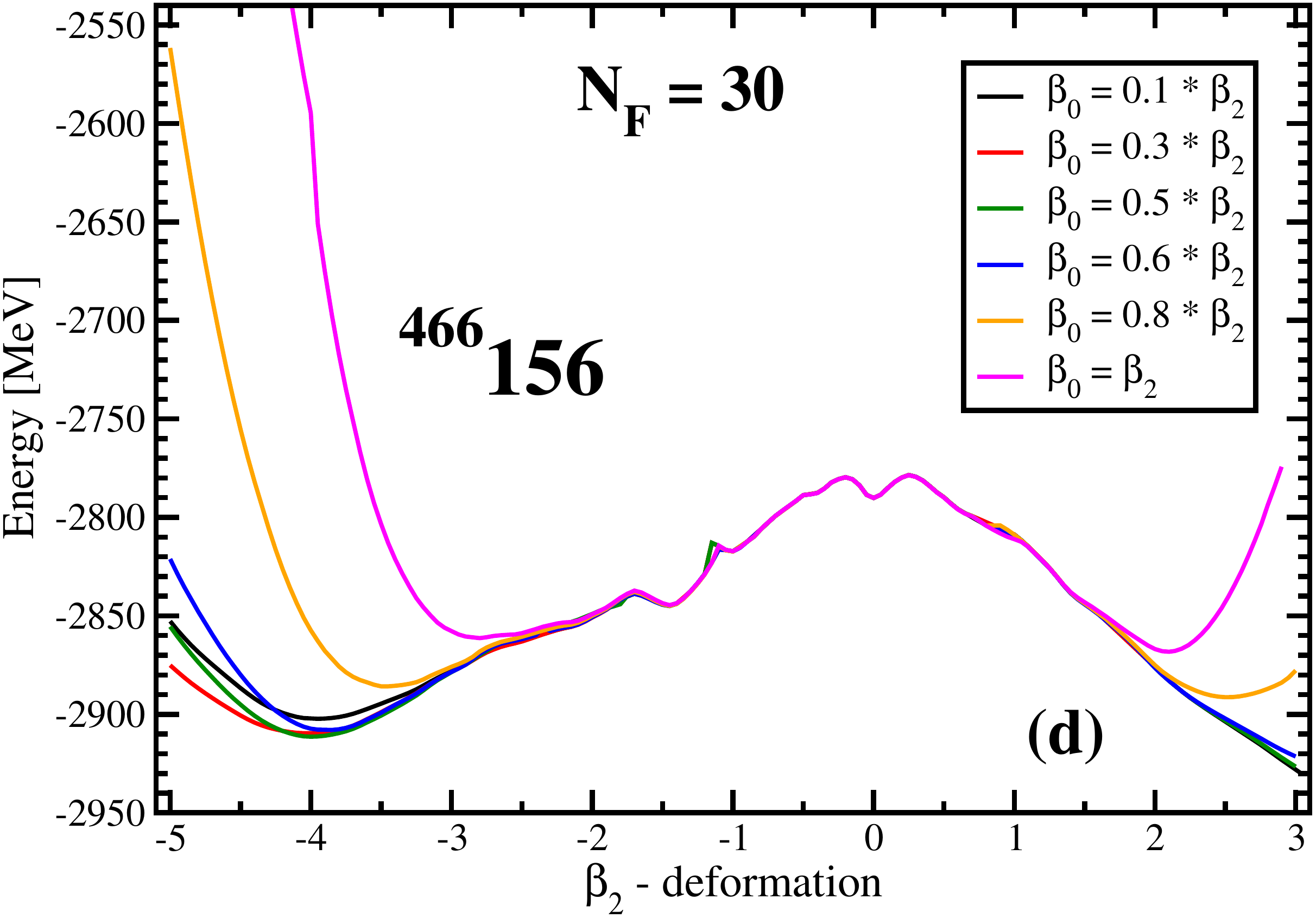}   
\caption{(Color online) The dependence of total binding energy
on the truncation of the basis and on the deformation of basis $\beta_0$ in the 
$^{208}$Pb and $^{466}$156 nuclei. Total binding energies are shown 
as a function of the $\beta_2$ values. Panels (a) and (b) show the 
dependence of  total binding energies on the number of fermionic
shells $N_F$ for the deformation of basis $\beta_0 = 0.5 \beta_2$. 
Panels (c) and (d) show the dependence of total binding energies
of hyperheavy $^{466}$156 nucleus on the deformation of basis 
$\beta_0$ for $N_F=20$ and $N_F=30$, respectively.}
\label{trunc_basis}
\end{figure*}

\begin{figure*}[htb]
\includegraphics[angle=0,width=5.9cm]{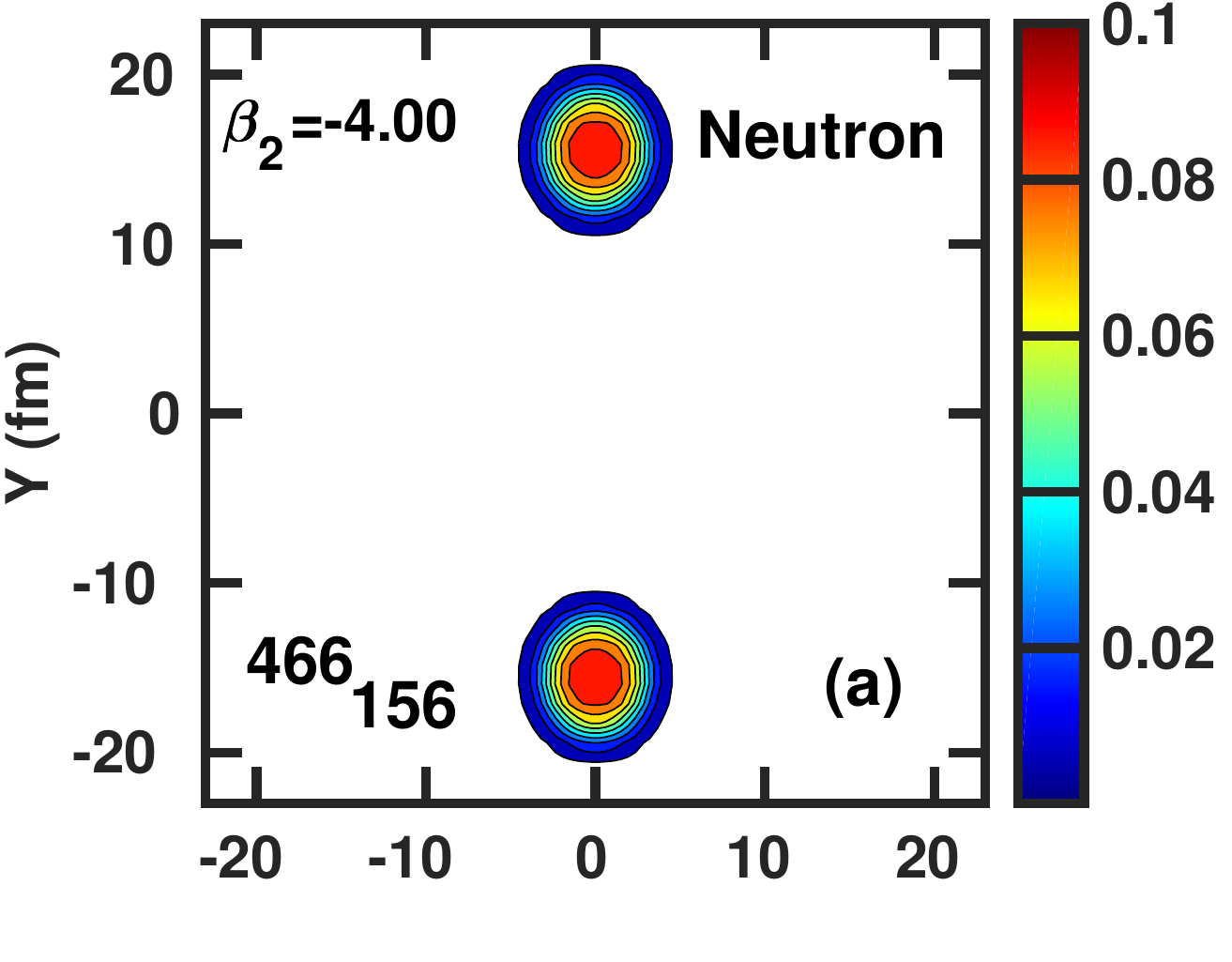}
\includegraphics[angle=0,width=5.9cm]{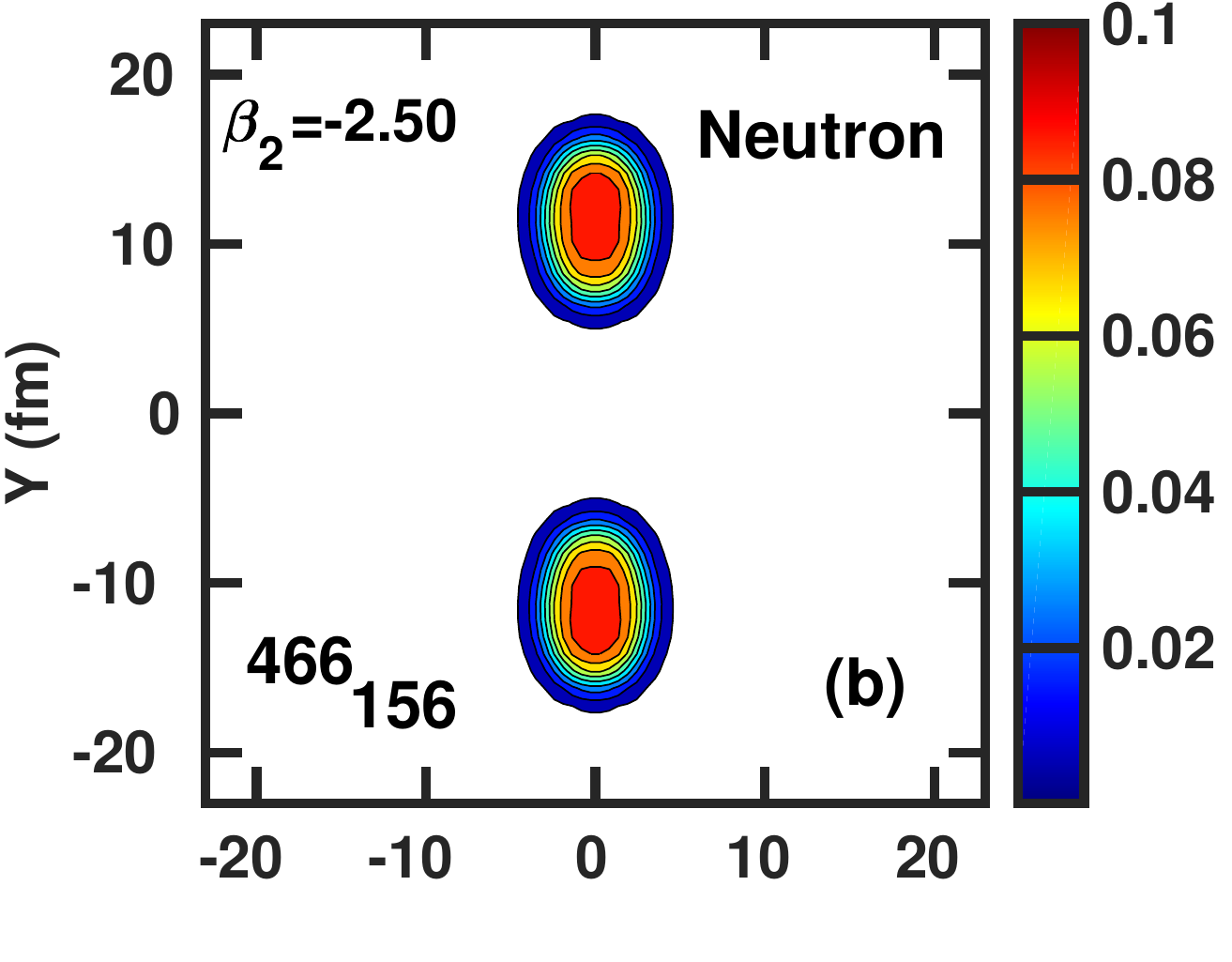}
\includegraphics[angle=0,width=5.9cm]{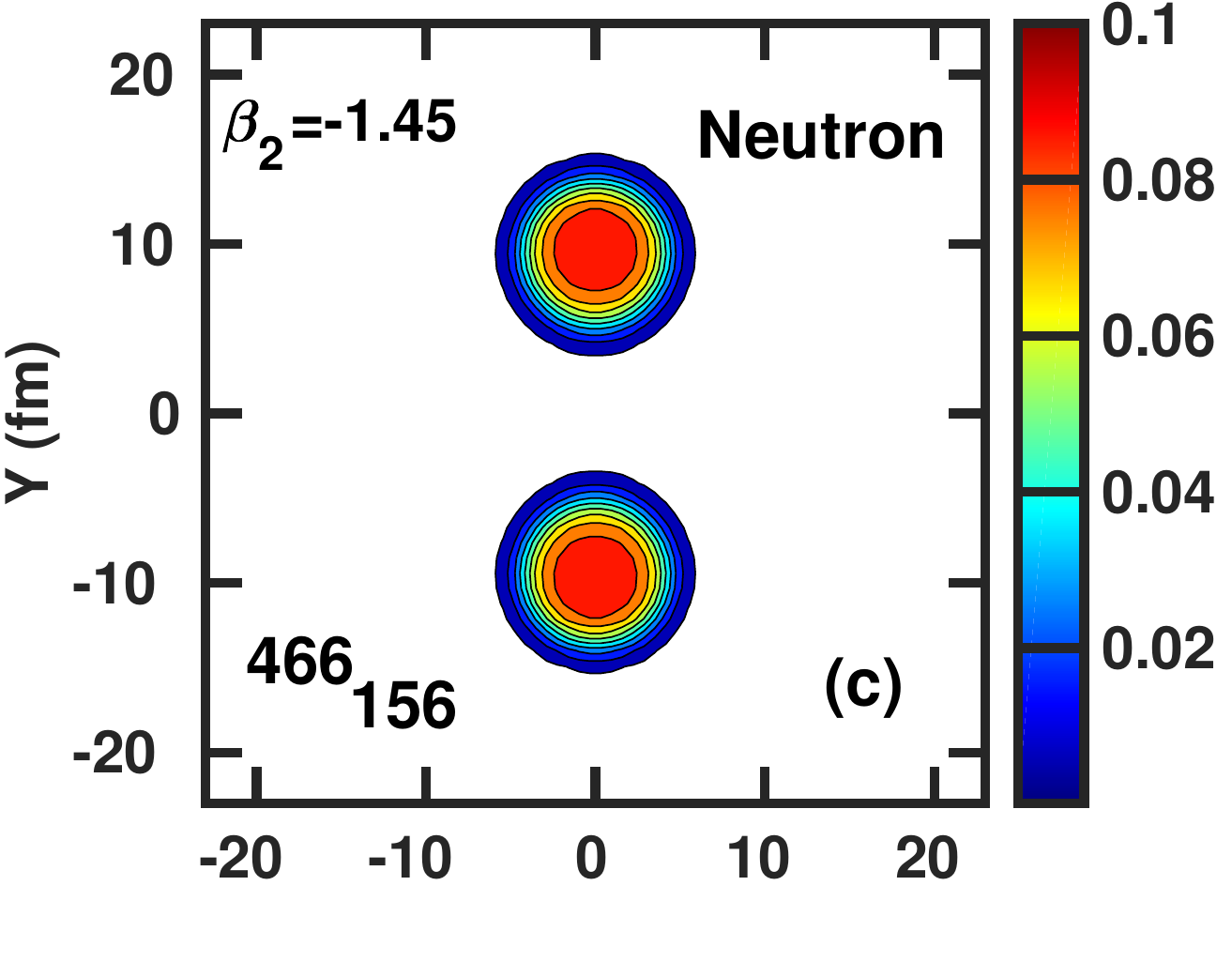}
\includegraphics[angle=0,width=5.9cm]{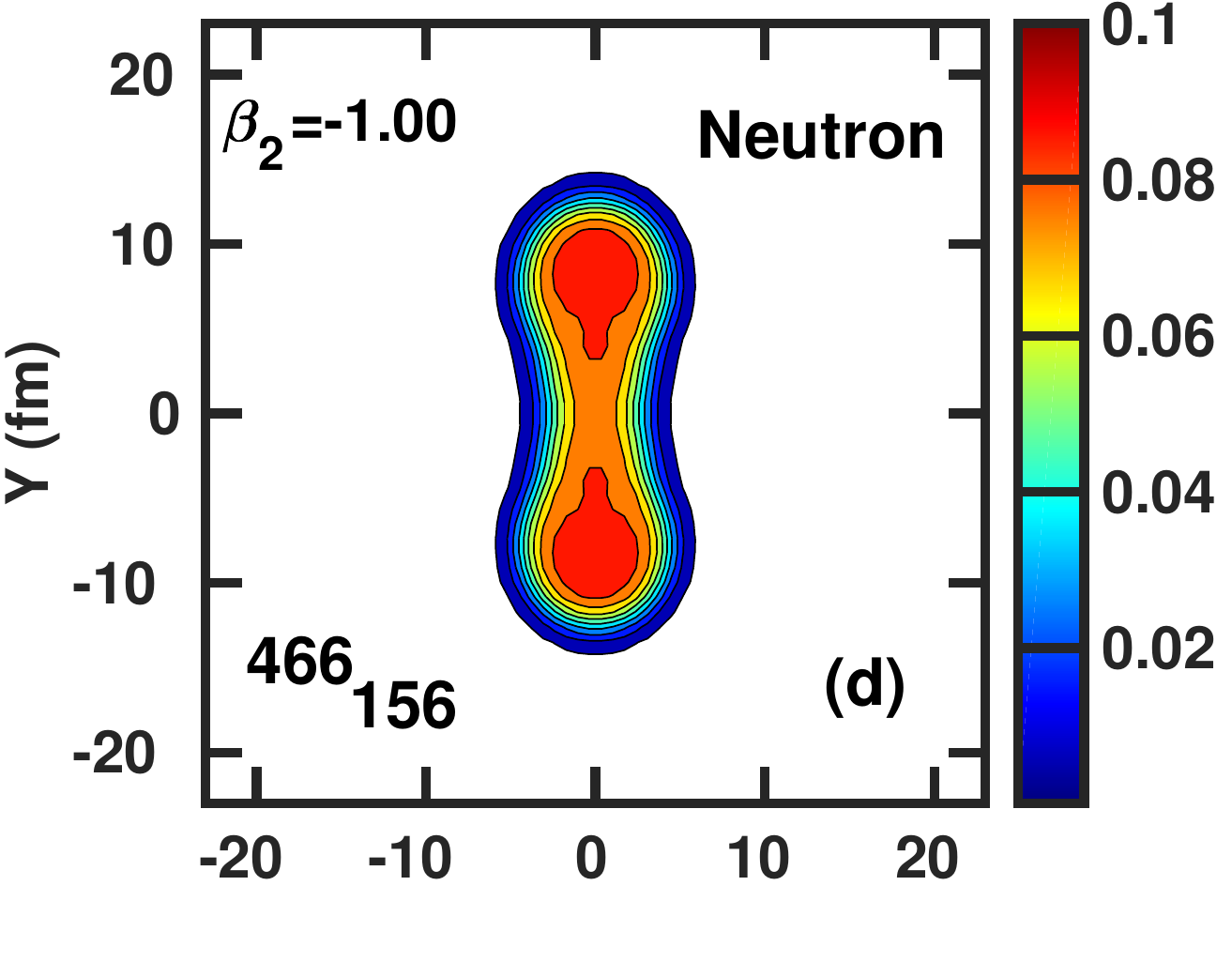}
\includegraphics[angle=0,width=5.9cm]{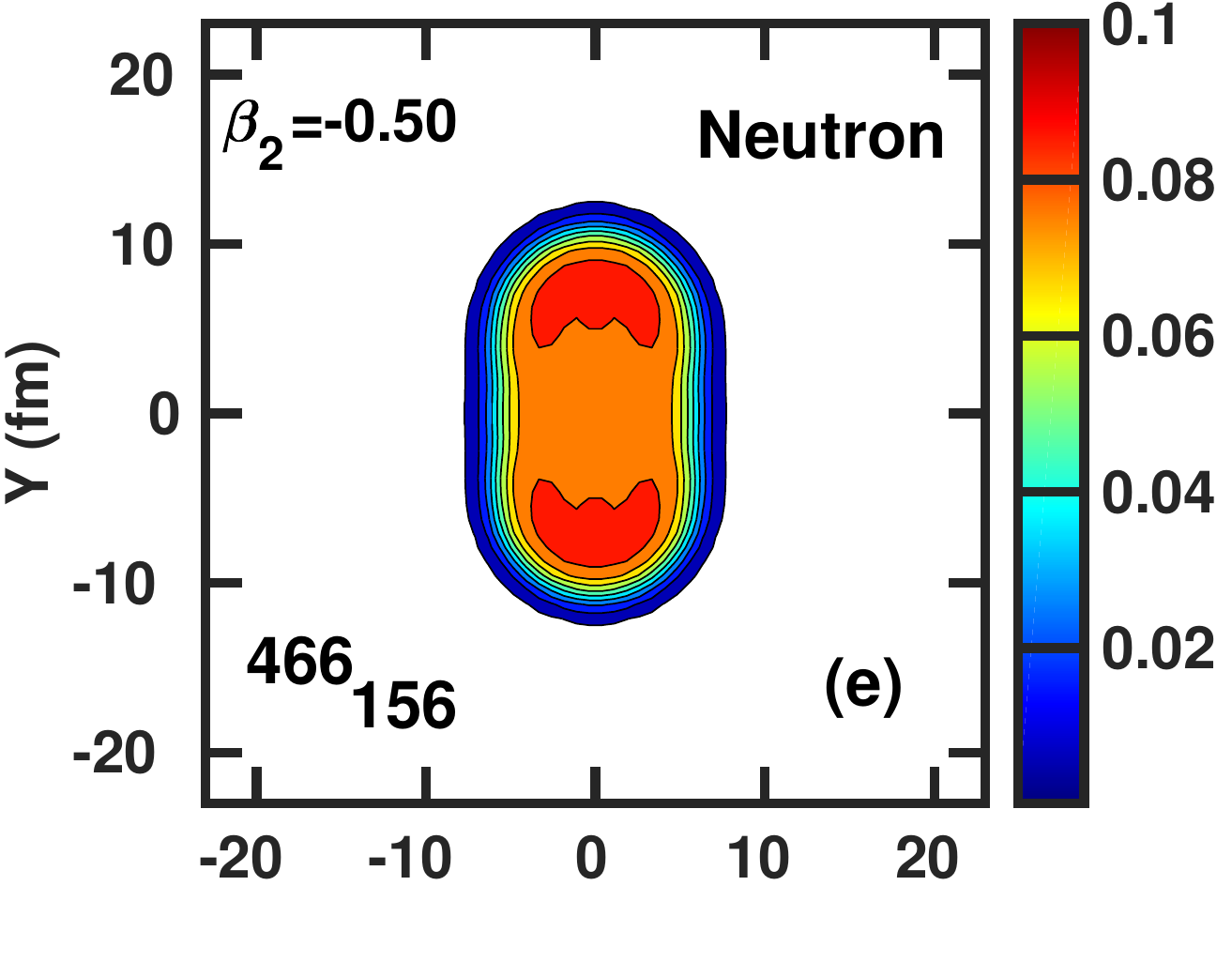}
\includegraphics[angle=0,width=5.9cm]{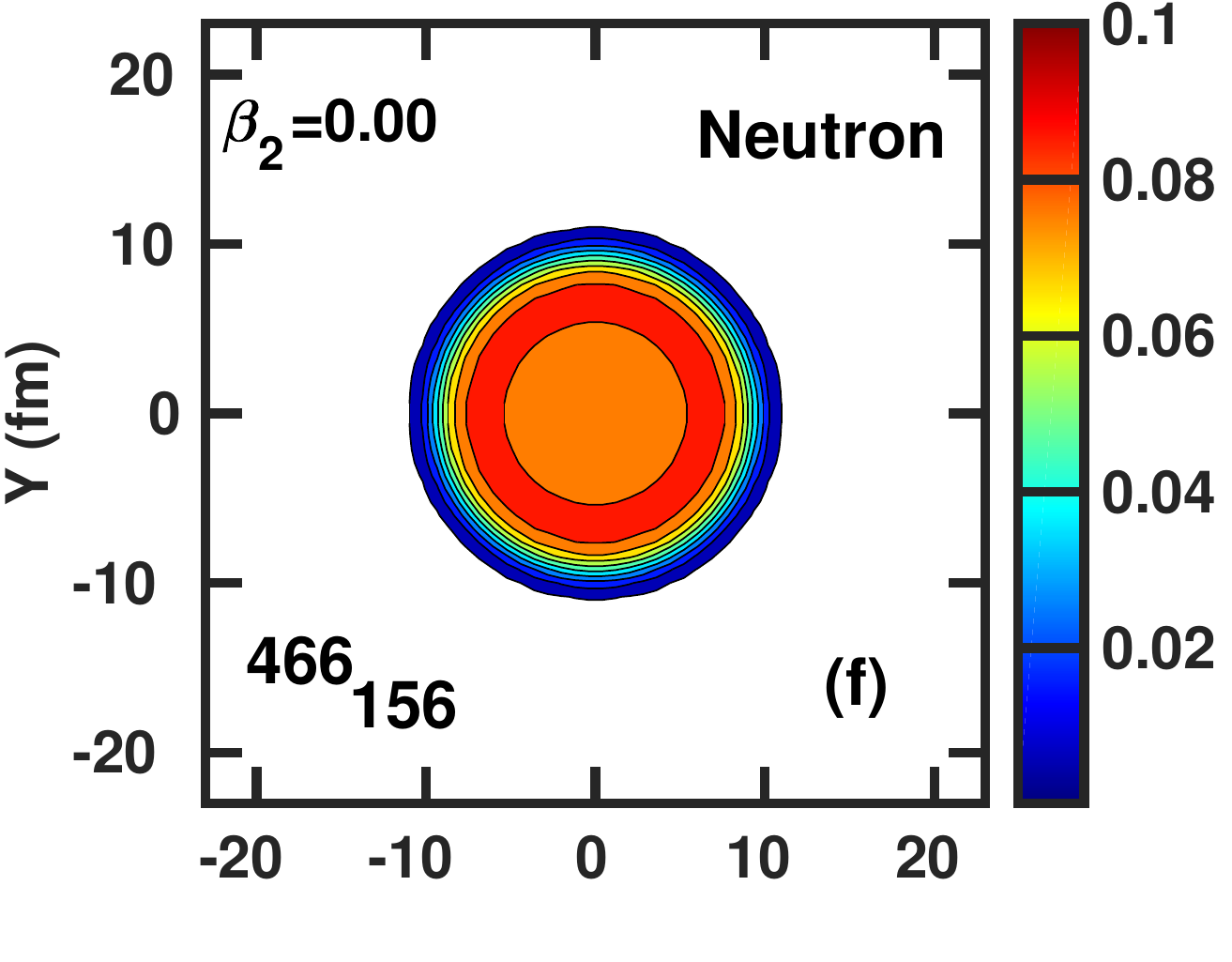}
\includegraphics[angle=0,width=5.9cm]{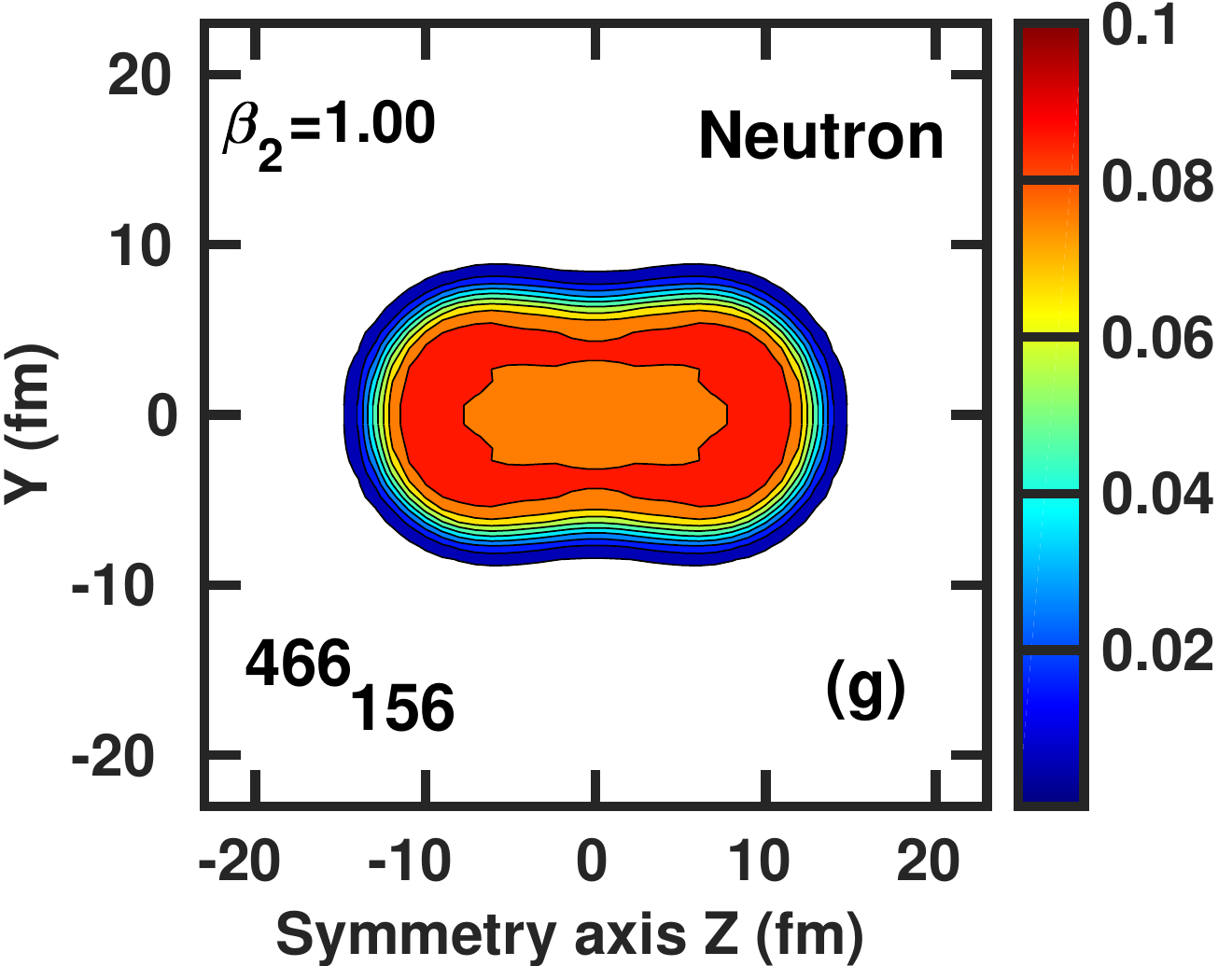}
\includegraphics[angle=0,width=5.9cm]{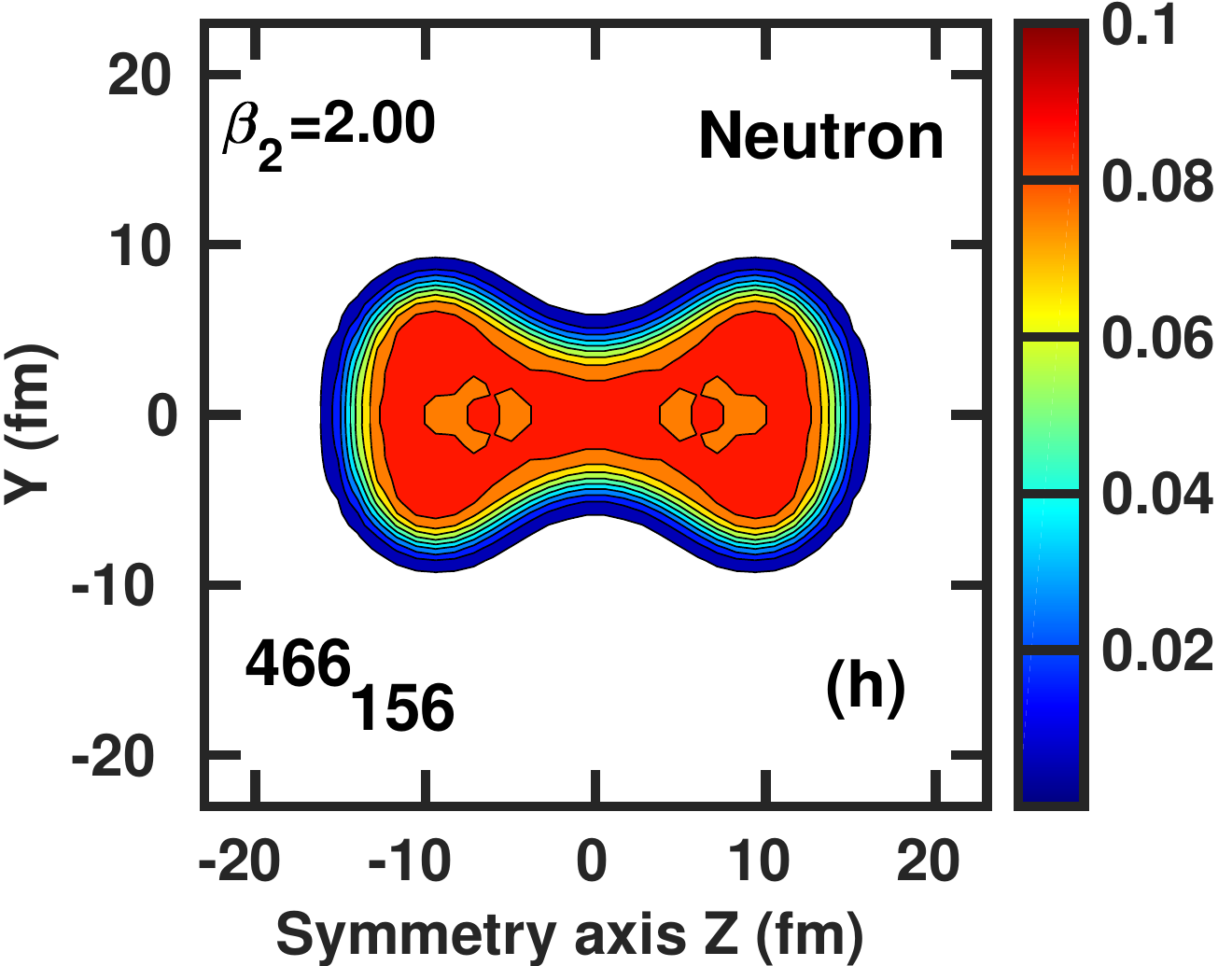}
\includegraphics[angle=0,width=5.9cm]{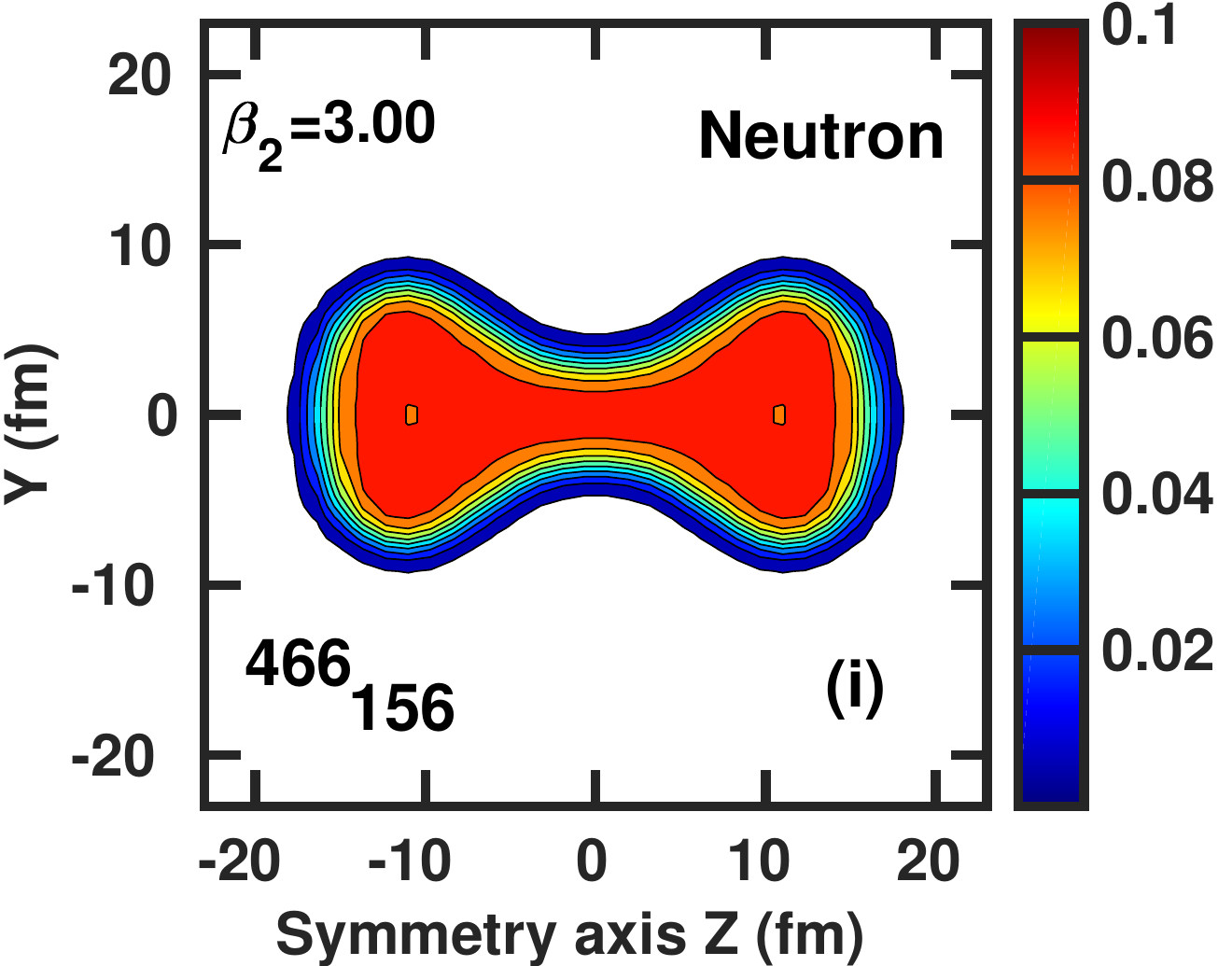}
\caption{(Color online)   Neutron density distributions of the $^{466}$156 
nucleus at the indicated $\beta_2$ values. They are plotted in the $yz$-plane 
at the position of the Gauss–Hermite integration points in the $x$-direction 
closest to zero. The density colormap starts at $\rho_n=0.005$ fm$^{−3}$ 
and shows the densities in fm$^{−3}$. Based on the results of axial RHB
calculations for the lowest in energy solution obtained with $N_F=30$
(see Fig.\ \ref{trunc_basis}b). Note that proton density (not shown here) is 
roughly half of the neutron one.}
\label{densities}
\end{figure*}
  
    The present manuscript aims at the extension of the investigations of 
the properties of hyperheavy nuclei and of nuclear landscape started 
in Ref.\ \cite{AAG.18}.  The topics covered by this investigation are shortly
mentioned in the next paragraph and discussed in details in the sections
below.

   The manuscript is organized as follows. The details of theoretical
calculations are discussed in Sec.\ \ref{theory-details}. Section 
\ref{sect-truncation} is devoted to the analysis of the effects of the 
truncation of the basis on the results of calculations. Density profiles,
charge radii and neutron skins of spherical nuclei located in the
centers of the islands of stability and their dependence on the functional
are considered in Sec.\ \ref{sect-density}.  Sec.\ \ref{shell_closure} 
discusses the shell closures in the islands of stability of spherical 
hyperheavy nuclei.  The stability of spherical nuclei in these islands
with respect of octupole and triaxial distortions is investigated in
Sec.\ \ref{stability_spherical}.  Sec.\ \ref{sect-oblate} analyses 
the impact of octupole deformation on stability of prolate superdeformed 
minima. Systematic analysis of the results of the calculations for the 
$Z=138$ isotopic chain is performed in Sec.\ \ref{Z=138_chain}.
The stability of toroidal shapes in selected nuclei and the evolution
of such shapes along their fission path are considered in Sec.\ 
\ref{toroidal_stability}. Sec.\ \ref{fission-barriers} is devoted to the 
analysis of the impact of triaxial deformation on the fission barriers of 
neutron-rich superheavy nuclei. The extension of nuclear landscape to 
hyperheavy nuclei is discussed in Sec.\ \ref{landscape-extension}. 
Finally, Sec.\ \ref{summary} summarizes the results of our work.

\section{The details of the theoretical calculations}
\label{theory-details}

 The investigations of the properties of super- and hyperheavy nuclei have been 
performed in different theoretical frameworks. Systematic investigation of hyperheavy
nuclei across the nuclear landscape between two-proton and two-neutron drip lines 
is performed within the axial reflection symmetric relativistic Hartree-Bogoliubov  
(RHB) framework (see Ref.\ \cite{AARR.14}).  The stability of prolate minima with 
$\beta_2 \sim 0.5$ of superheavy and low-$Z$ hyperheavy nuclei as well as of 
spherical minima of hyperheavy nuclei with respect of octupole deformation has 
been studied with reflection asymmetric RHB framework using OCT-RHB code of 
Ref.\ \cite{AAR.16}.  Triaxial RHB (TRHB code) \cite{AARR.17} and triaxial relativistic 
mean field + BCS  (TRMF+BCS code)  \cite{AAR.10} frameworks have been employed 
for the study of fission barriers in superheavy nuclei and stability of hyperheavy 
nuclei with respect to triaxial distortions.  Note that the TRHB and TRMF+BCS codes do 
not include octupole deformation.   Considering very time-consuming nature of the 
calculations in the OCT-RHB, TRHB and TRMF+BCS codes, only restricted set of nuclei 
has been investigated in their frameworks.

\begin{figure*}[htb]
\includegraphics[angle=-90,width=16.0cm]{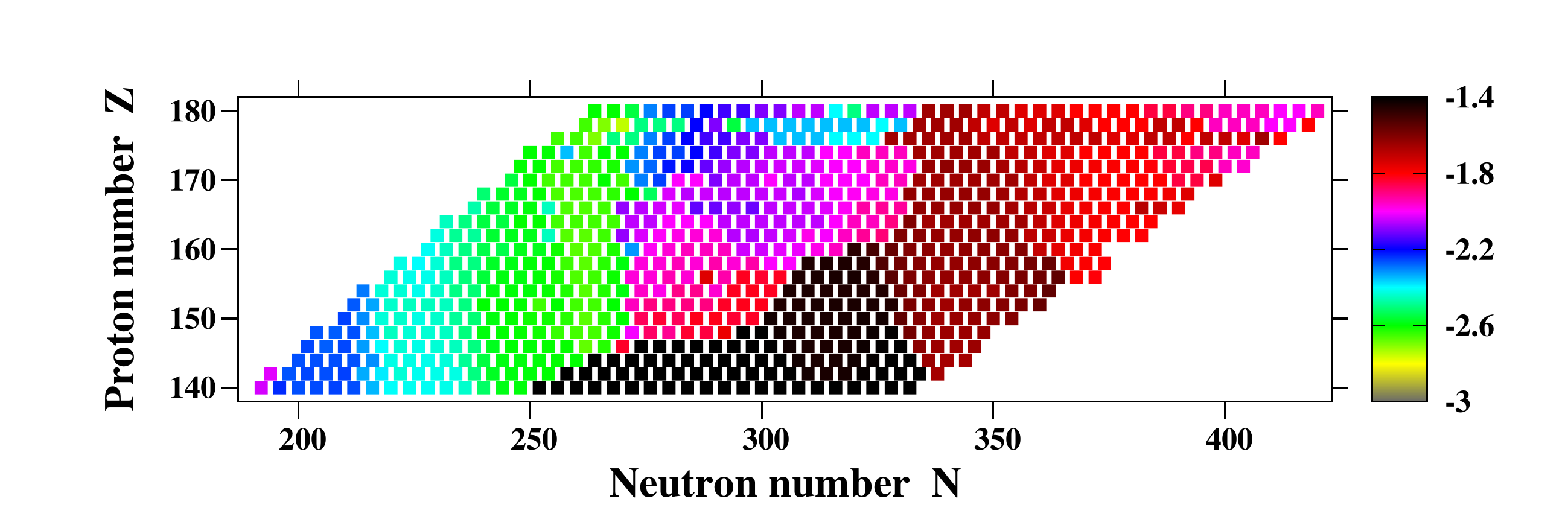}
\caption{(Color online) Proton $\beta_2$ values of the lowest 
in energy solutions of the Z=140-180 nuclei obtained in axial RHB 
calculations with $N_F$=20. The calculations cover the region between
two-proton and two-neutron drip lines.}
\label{beta_2-Z140-180}
\end{figure*}

 The absolute majority of the calculations has been performed with the DD-PC1 
covariant energy density functional \cite{DD-PC1}. This functional is considered 
to be the best relativistic functional today based on  systematic and global studies 
of different physical  observables related to the ground state properties and fission 
barriers \cite{AANR.15,AARR.14,AARR.17,AA.16,PNLV.12,LZZ.12,AAR.16}.  Other 
functionals such as DD-ME2 \cite{DD-ME2}, PC-PK1  \cite{PC-PK1} and NL3* 
\cite{NL3*}), representing other major classes of covariant density functional models 
\cite{AARR.14}, are employed only for the study of some properties of spherical 
nuclei located in the centers of the islands of stability of hyperheavy nuclei (see Fig.\ 6 
in Ref.\ \cite{AAG.18}). This is done for the assessment of systematic theoretical 
uncertainties in the predictions of their properties.

  The constrained calculations in employed codes perform the variation of the
function
\begin{eqnarray} 
E_{RHB/RMF+BCS} + \sum_{\lambda,\mu} C_{\lambda,\mu} (\langle \hat{Q} _{\lambda,\mu} \rangle
                            - q_{\lambda,\mu})^2.
 \label{constr}
\end{eqnarray}
Here $(\lambda,\mu) = (2,0)$, $(\lambda,\mu) = (2,0)$ and $(3,0)$ as well as 
$(\lambda,\mu) = (2,0)$ and $(2,2)$
in the  RHB, OCT-RHB and TRHB (TRMF+BCS) calculations, respectively.  $E_{RHB}$ 
and $E_{RMF+BCS}$ are the total energies in the RHB and RMF+BCS calculations.
$<Q_{\lambda,\mu}>$  stand for the expectation values of the respective multipole
moments which are defined as
\begin{eqnarray}
\hat{Q}_{20} & = & 2z^2 -x^2 -y^2, \\
\hat{Q}_{22} & = & x^2 -y^2, \\
\hat{Q}_{30} & = & z(2z^2 -3x^2 -3y^2).
\end{eqnarray}
$C_{\lambda,\mu}$ in Eq.\ (\ref{constr}) are corresponding stiffness constants 
\cite{RS.80} and $q_{\lambda,\mu}$ are constrained values of respective moments. To 
provide the convergence to the exact value of the desired multipole moment, 
we use the method suggested in Ref.\ \cite{BFH.05}.  Here the quantity $q_{\lambda,\mu}$
is replaced by the parameter $q_{\lambda,\mu}^{eff}$, which is automatically 
modified during the iteration in such a way that we obtain
$\langle \hat{Q}_{\lambda,\mu} \rangle = q_{\lambda,\mu}$
for the converged solution. This method works well in our constrained
calculations. In the OCT-RHB code we also fix the (average) center-of-mass of 
the nucleus at the origin with the constraint
\begin{eqnarray}
\langle \hat{Q}_{10} \rangle =0  
\end{eqnarray}
on the center-of-mass operator $\hat{Q}_{10}$ to avoid a spurious motion
of the center of mass.
   
The deformation parameters $\beta_2$,  $\beta_3$  
and $\gamma$ are extracted from respective multipole moments:
\begin{eqnarray}
Q_{20} &=& \int d^3r \rho({\vec r})\,(2z^2-x^2-y^2),\\
Q_{22} &=& \int d^3r \rho({\vec r})\,(x^2-y^2), \\
Q_{30} &=& \int d^3r \rho({\vec r})\,z(2z^2 -3x^2 -3y^2), \\
\end{eqnarray}
%
%
via
\begin{eqnarray}
\beta_2 &=&  \sqrt{\frac{5}{16\pi}} \frac{4\pi}{3 Z R_0^2} \sqrt{ Q_{20}^2 + 2Q_{22}^2}
\\
\gamma&=& \arctan{\sqrt{2} \frac{Q_{22}}{Q_{20}}}
\\
\beta_3 &=&  \sqrt{\frac{7}{16\pi}} \frac{4\pi}{3ZR_0^3} Q_{30}
\end{eqnarray}
where $R_0=1.2 A^{1/3}$. Note that $Q_{22}=0$ and $\gamma=0$ in
axially symmetric RHB calculations.

\begin{figure*}[htb]
\includegraphics[angle=0,width=18.cm]{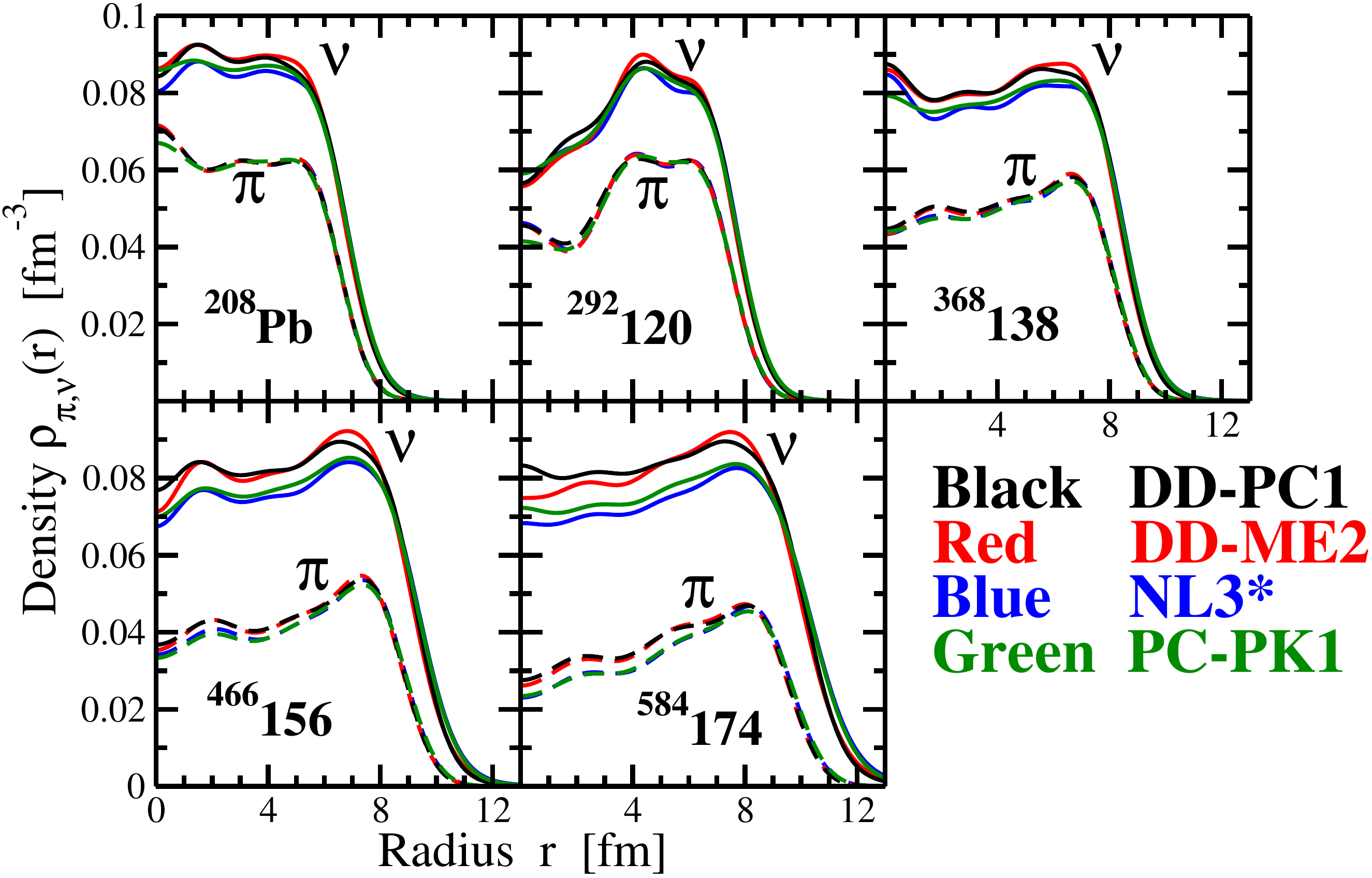}
\caption{(Color online) The evolution of proton and neutron 
densities with the transition from the $^{208}$Pb nucleus to the region 
of hyperheavy nuclei. The figure is based on the results of spherical 
RHB calculations; the employed CEDFs are indicated. The $^{368}$138, 
$^{466}$156 and $^{466}$156 nuclei are located in the centers of the islands 
of stability of spherical hyperheavy nuclei predicted in Ref.\ \cite{AAG.18}. 
Note that it was verified that proton and neutron densities of the nuclei in 
these regions are very similar to the densities of above mentioned central 
nuclei. For comparison, the densities of spherical $^{208}$Pb and $^{292}$120 
nuclei are presented. Note that latter nucleus is an example of substantial 
central depression in density distribution (see Ref.\ \cite{AF.05-dep}) for
details).}
\label{Density-sphere}
\end{figure*}

  The $\beta_2$ and  $\gamma$ values have a standard meaning of the deformations 
of the ellipsoid-like density distributions only for $|\beta_2| \lesssim 1.0$ values. At higher 
$\beta_2$ values they should be treated as dimensionless and particle normalized measures 
of the $Q_{20}$ and $Q_{22}$ moments. This is because of the presence of toroidal 
shapes at large negative $\beta_2$ values and of necking degree 
of freedom at large positive $\beta_2$ values (see Fig.\ \ref{densities} below).

 Note that physical observables are frequently shown as a function of the 
$Q_{20}$,  $Q_{30}$  and $Q_{22}$ moments. However, from our point 
of view such way of presentation has a disadvantage that the physical 
observables of different nuclei related to the shape of the density 
distributions (such as deformations) are difficult to compare because 
the $Q_{20}$, $Q_{30}$ and $Q_{22}$ moments depend on particle 
number(s).

  For each nucleus under study, the deformation energy curves in  
the $-5.0 < \beta_2 < 3.0$ range are calculated in the axial reflection 
symmetric RHB framework \cite{AARR.14}; such large range 
is needed for a reliable definition of the $\beta_2$ value
of the lowest in energy minimum for axial symmetry (LEMAS).
This LEMAS becomes the ground state if the higher order deformations
(triaxial, octupole) do not lead to the instability of these minima.
 
To avoid the uncertainties connected with the definition of the size of the 
pairing window \cite{KALR.10}, we use the separable form of the finite-range 
Gogny pairing interaction introduced in Ref.\  \cite{TMR.09}. Its matrix elements 
in $r$-space have the form
\begin{eqnarray}
\label{Eq:TMR}
V({\bm r}_1,{\bm r}_2,{\bm r}_1',{\bm r}_2') &=& \nonumber \\
= - \,G \delta({\bm R}-&\bm{R'}&)P(r) P(r') \frac{1}{2}(1-P^{\sigma})
\end{eqnarray}
with ${\bm R}=({\bm r}_1+{\bm r}_2)/2$ and ${\bm r}={\bm r}_1-{\bm r}_2$
being the center-of-mass and relative coordinates. The form factor $P(r)$ is of 
Gaussian shape
\begin{eqnarray}
P(r)=\frac{1}{(4 \pi a^2)^{3/2}}e^{-r^2/4a^2}
\end{eqnarray}
The parameters of this interaction have been derived by a mapping of the $^1$S$_0$ 
pairing gap of infinite nuclear matter to that of the Gogny force D1S. The resulting
parameters are: $G=728$ MeV fm$^3$ and $a=0.644$ fm\ \cite{TMR.09}. This pairing 
provides a reasonable description of pairing properties in heaviest nuclei (actinides
and light superheavy nuclei) in which pairing properties can be extracted from 
experimental data \cite{AO.13,AARR.14,DABRS.15}.

\section{The effects of the basis truncation}
\label{sect-truncation}

   Considering that the results published in Ref.\ \cite{AAG.18} and presented in 
   this manuscript represent the first studies of hyperheavy nuclei and toroidal shapes
   in such nuclei within the covariant density functional theory, a special attention has 
   been paid to the investigation of the impact of the truncation of the basis on the 
   results of the calculations at different $\beta_2$ values. The truncation of the basis 
   is performed in such a way that all states 
   belonging to the major shells up to $N_F$ fermionic shells for the Dirac spinors 
   are taken into account. Note also that the results of the calculations depend on
   the deformation $\beta_0$ of the oscillator basis. The detailed investigation of
   the impact of the deformation of the basis on the results of the calculations for the 
   $^{466}$156 nucleus is presented for the $N_F=20$ and $N_F=30$ fermionic 
   shells in Figs.\ \ref{trunc_basis}c and d, respectively. One can see that the 
   deformation of basis $\beta_0 = 0.5 \beta_2$ typically leads to the lowest in energy 
   solutions. Here $\beta_2$ stands for the deformation of the nucleus.
   This truncation scheme is also characterized by the fastest convergence. Similar situation
   has been observed also in other nuclei. Thus, this deformation of the basis has
   been used in all calculations of Ref.\ \cite{AAG.18} and the present manuscript.

  Figs.\ \ref{trunc_basis}a and b show the dependence of the results of 
    calculations on the number of fermionic shells $N_F$ for the $^{208}$Pb
    and $^{466}$156 nuclei. Note that the deformation of basis  
    $\beta_0 = 0.5 \beta_2$ is used for both of these nuclei. In $^{208}$Pb,
    the $N_F=20$ basis provides very accurate description of binding
    energies in the physically interesting range of quadrupole deformations.
    Only at $\beta_2 < -3.5$ there is some difference between the results 
    obtained with $N_F=20$ and $N_F=30$. However, this is not physically
    significant range of the $\beta_2$ values since binding energies at these 
    values exceed binding energy of the ground state by at least 250 MeV.
    
         However, in hyperheavy nuclei the required size of the basis depends 
      both on the nucleus and deformation range of interest. The $N_F=20$ 
      basis is sufficient for the description of deformation energy curves in the 
      region of $-1.8 < \beta_2 < 1.8$  (see Fig.\ \ref{trunc_basis}b).
      The deformation ranges $-3.0 <\beta_2 <-1.8$ and $1.8 <\beta_2 <3.0$ 
      typically require $N_F=24$ (low-$Z$ and  low-$N$ hyperheavy nuclei) or 
      $N_F=26$ (high-$Z$ and high-$N$  hyperheavy nuclei). Even more 
      deformed ground states with $\beta_2 \sim -4.0$ are seen in 
      high-$Z$/high-$N$ hyperheavy nuclei (see Fig.\ \ref{trunc_basis}b
      for the $^{466}156$ results and Fig.\ 1 in Ref.\ \cite{AAG.18} for
      the $^{426}$176 results); their description requires $N_F=30$. Thus, in 
      our studies the truncation of basis is made dependent on the nucleus 
      and typical profile of deformation energy curves or potential energy 
      surfaces.

       Fig.\ \ref{densities}  illustrates the evolution of nuclear shapes 
 along the lowest in energy solution of hyperheavy $^{466}$156 nucleus
 obtained in axial RHB calculations  with  $N_F=30$. Starting from 
 spherical  shape at $\beta_2=0.0$, the increase of prolate deformation 
 leads to the emergence of hyperdeformed shapes at $\beta_2=1.0$, which
  evolve into the shapes consisting of two fragments connected by neck at higher
  $\beta_2$ values.  The separation of the fragments and the size of the neck 
  increases/decreases with increasing $\beta_2$ values. Figs.\ \ref{trunc_basis}b, 
  c and d clearly indicate  increased dependence of the results on the parameters 
  $N_F$ and $\beta_0$ of the basis for these shapes.
      
      The evolution  of the shapes for negative $\beta_2$ values is shown in
   Figs.\ \ref{densities}a-e. Highly deformed oblate shape exists at 
   $\beta_2 = - 0.5$ deformation which transforms into biconcave disk shape
   at $\beta_2 = -1.0$.  Further decrease of the $\beta_2$
   values leads to toroidal shapes. Note that with the increase of absolute value 
   of $\beta_2$ the radius of the toroid increases and the tube radius decreases.
   Total energies and equilibrium deformations of toroidal shapes with 
   $\beta_2 \sim -1.45$,  corresponding 
   to local minimum  seen in deformation energy curves  of Figs.\ \ref{trunc_basis}b-d,
   are rather  well described with $N_F=20$ and show almost no dependence on the
   deformation of basis $\beta_0$. On the contrary, toroidal shapes with larger 
   (in absolute sense) $\beta_2$ values show substantial increase of the dependence 
   of total energies and equilibrium deformations on $N_F$ and $\beta_0$.

   Thus, the present analysis clearly indicates that the $N_F=20$ basis is 
sufficient for a description of classical ellipsoidal shapes and some toroidal 
shapes with relatively low absolute values of $\beta_2$ even in hyperheavy 
$Z<180$ nuclei. On the contrary, significantly  larger basis is required for a 
description of more exotic shapes such as toroidal ones with large absolute 
$\beta_2$ values and two-fragment ones connected by neck.
      
  The equilibrium $\beta_2$ values of the $Z=122-138$ nuclei presented 
in Fig. 2 of Ref.\ \cite{AAG.18} have been calculated with fermionic 
bases including up to $N_F=26$ fermionic shells. The calculations for 
higher $Z$ values require further increase of the size of fermionic 
basis (up to $N_F=30$ in high-$Z$/high-$N$ nuclei). Such calculations are 
extremely time-consuming even in axial RHB framework and thus have not 
been undertaken. On the other 
hand, the type of the LEMAS can be established in the calculations with 
$N_F=20$. This is because even with $N_F=20$ the toroidal shapes 
with $\beta_2 < - 1.4$ represent 
the lowest in energy solutions at axial symmetry in the $Z=140-180$
part of nuclear landscape (see Fig.\ \ref{beta_2-Z140-180}). This
figure clearly shows that classical ellipsoidal shapes are not 
energetically favored in hyperheavy nuclei. However, because of 
the limited size of the basis these $\beta_2$ values have to be 
considered as lower limits (in absolute sense). As illustrated in
Fig.\ \ref{trunc_basis}, further increase of the size of fermionic basis 
will lead to the increase (in absolute sense) of the $\beta_2$ values 
of LEMAS and to more energetically favored status of toroidal shapes 
as compared with ellipsoidal ones.

\begin{figure*}[htb]
\includegraphics[angle=0,width=10.5cm]{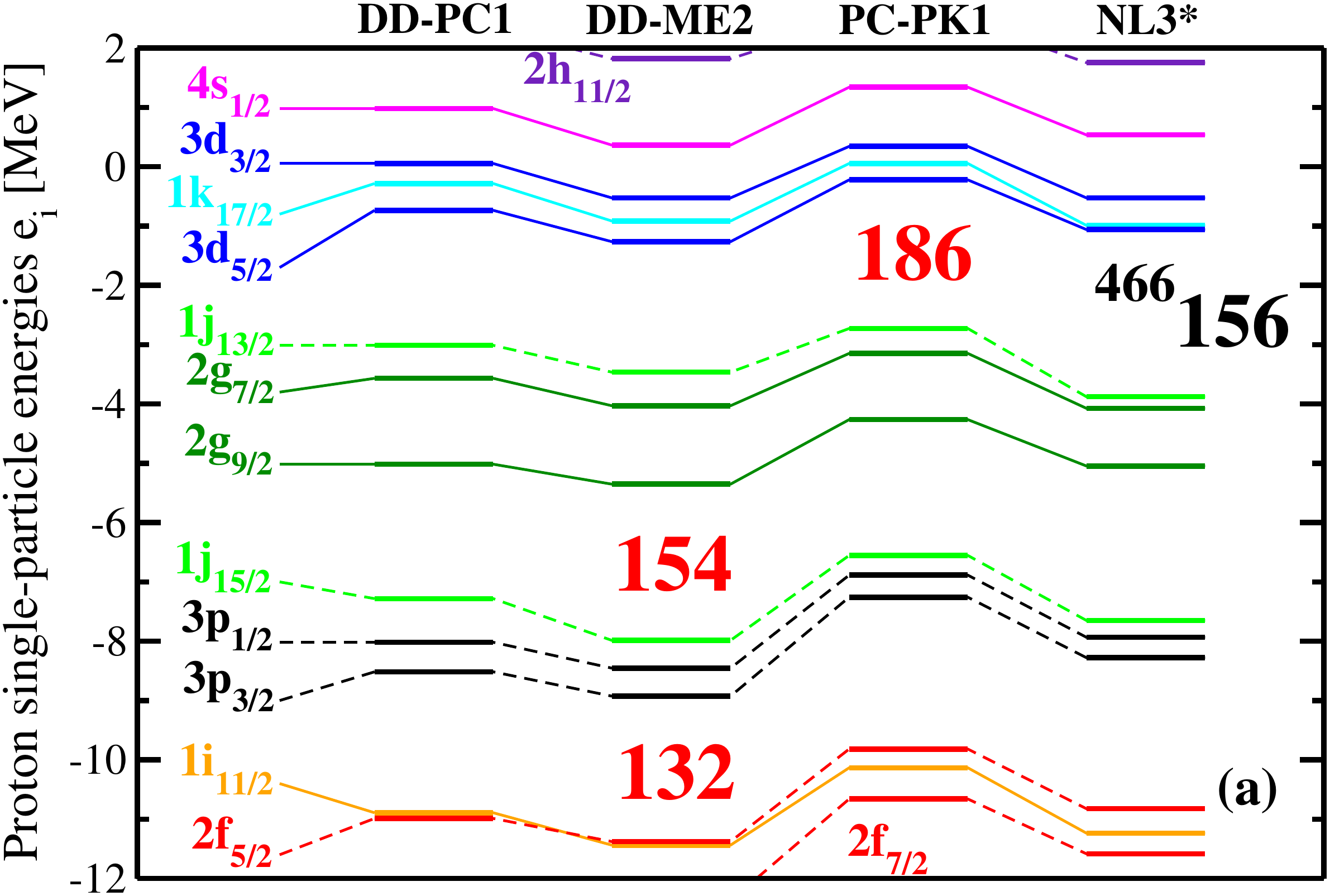}
\includegraphics[angle=0,width=10.5cm]{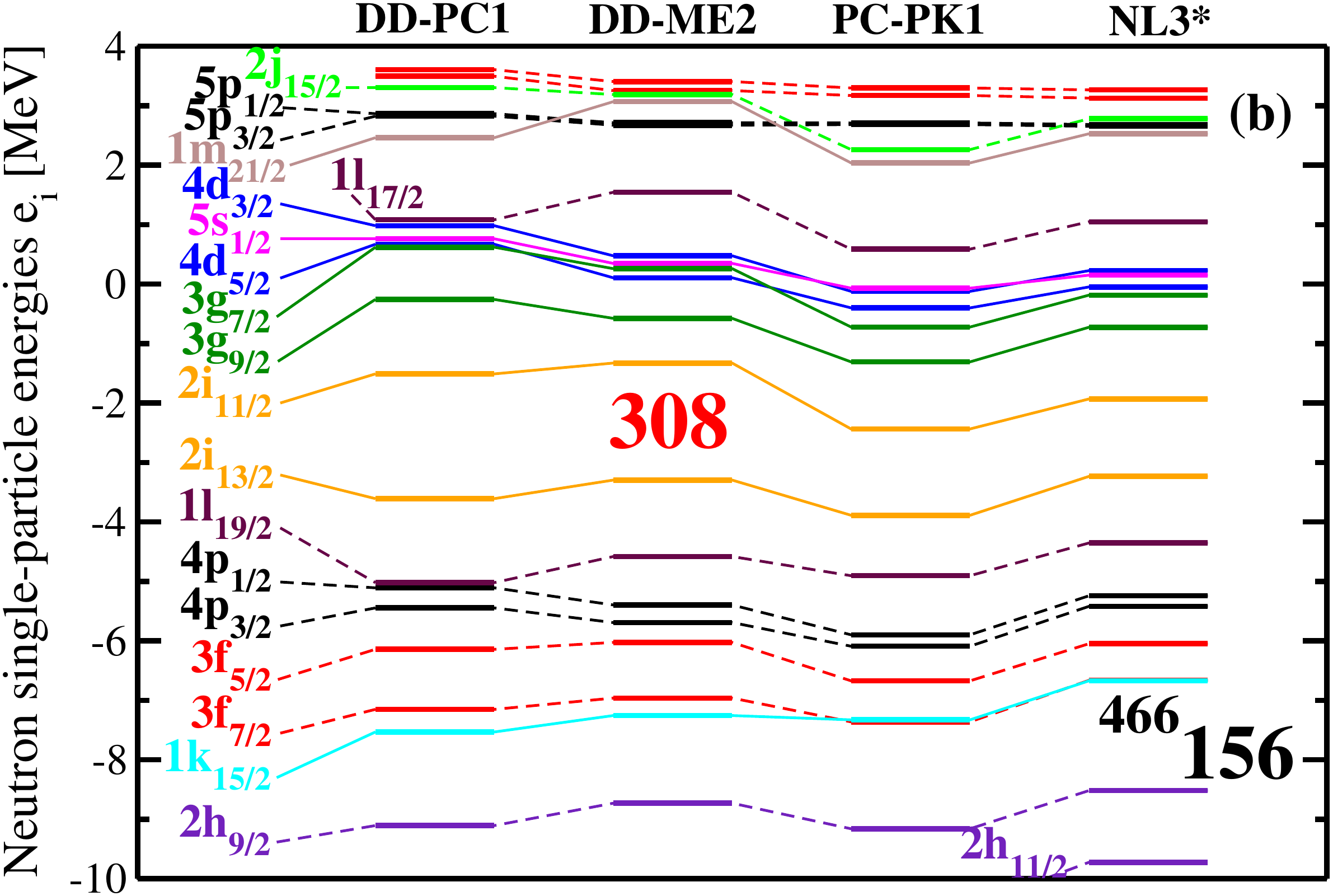}
\caption{(Color online)     Proton and neutron single-particle states at spherical shape in 
the $^{466}$156 nucleus determined with the indicated CEDF’s in the calculations without pairing.
Solid and dashed connecting lines are used for positive and negative parity states. Spherical
gaps are indicated.}
\label{sphere-sp}
\end{figure*}

\section{Density profiles, charge radii and neutron skins of 
spherical hyperheavy nuclei}
\label{sect-density}
  
  Three regions of spherical hyperheavy nuclei centered around 
($Z\sim 138, N\sim 230$),  ($Z\sim 156, N\sim 310$) and 
($Z\sim 174, N\sim 410$), which are expected to be reasonably 
stable against spontaneous fission and $\alpha$-decay, have been 
predicted in Ref.\ \cite{AAG.18}. The largest region with the 
highest fission barriers is centered at $Z\sim 156, N\sim 310$; 
other two regions are smaller with smaller fission barriers (see 
Fig.\ 6 in Ref.\ \cite{AAG.18}). The CEDFs DD-PC1 and DD-ME2 predict
larger regions of stability and substantially higher fission 
barriers (reaching 10 MeV in some nuclei) as compared with the 
NL3* and PC-PK1 functionals.  The large fission barriers obtained 
in the density-dependent functionals will lead to substantial 
stability of spherical hyperheavy nuclei against spontaneous 
fission. This stability is significantly lower for the NL3* and 
PC-PK1 functionals.

\begin{figure*}[htb]
\includegraphics[angle=0,width=12.5cm]{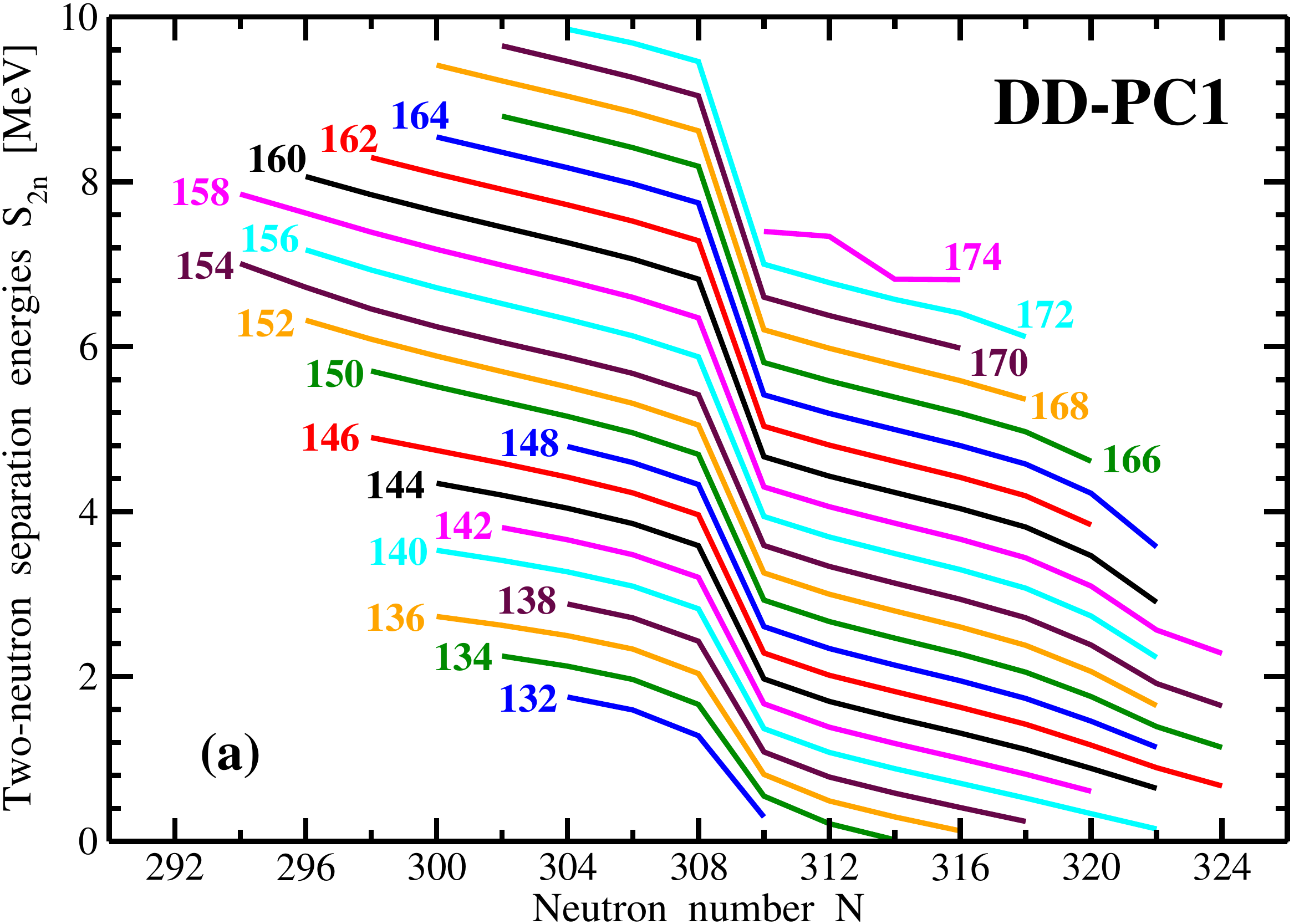}
\includegraphics[angle=0,width=12.5cm]{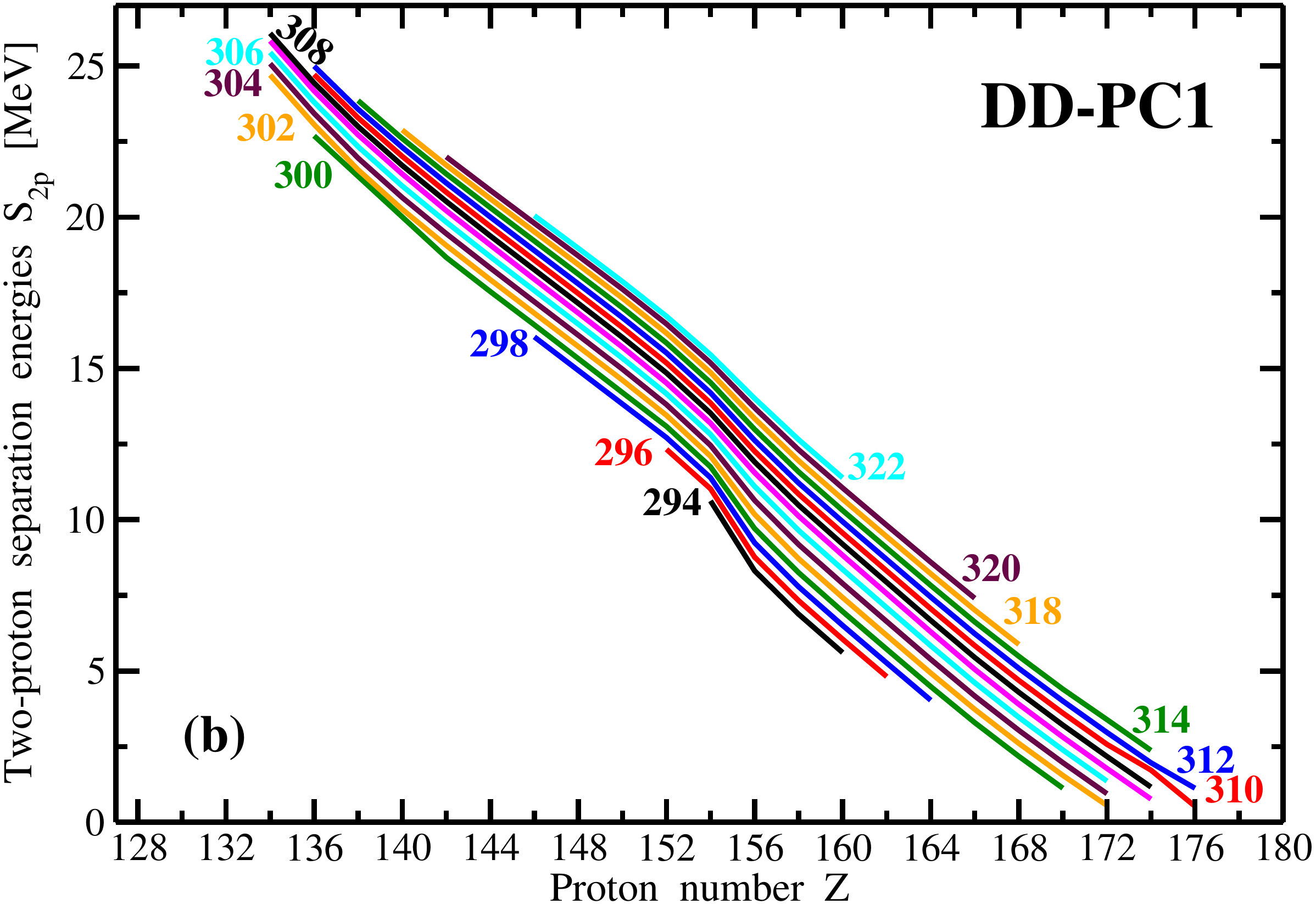}
\caption{(Color online)  Two-neutron and two-proton separation energies
for spherical nuclei located in the $(Z\sim 156, N\sim 310)$ region of
stability of hyperheavy nuclei. They are obtained in the RHB calculations with
the DD-PC1 functional. The lines are labeled by respective proton (panel (a))
and neutron (panel (b)) numbers.}
\label{sphere-sep_energies}
\end{figure*}

   The nuclear matter properties and the density dependence 
are substantially better defined for density-dependent (DD*) 
functionals as compared with non-linear NL3* and PC-PK1 ones 
\cite{AA.16}. As a consequence, in general, they are expected 
to perform better for large extrapolations from known regions. 
In this context, it is also important to look on other features 
which may be critical in the discrimination of the predictions
of different functionals. Thus, this section is dedicated
to the analysis of charge radii, neutron skins and density
distributions of the nuclei located in the centers of this
potential islands of stability of hyperheavy nuclei. These are 
$^{368}138$, $^{466}156$ and $^{584}174$ nuclei. As a benchmark, 
we are using the $^{208}$Pb and $^{292}120$ nuclei. The properties
of latter nucleus were studied in details in Ref.\ \cite{AF.05-dep}.

\begin{figure*}[htb]
\includegraphics[angle=0,width=10.5cm]{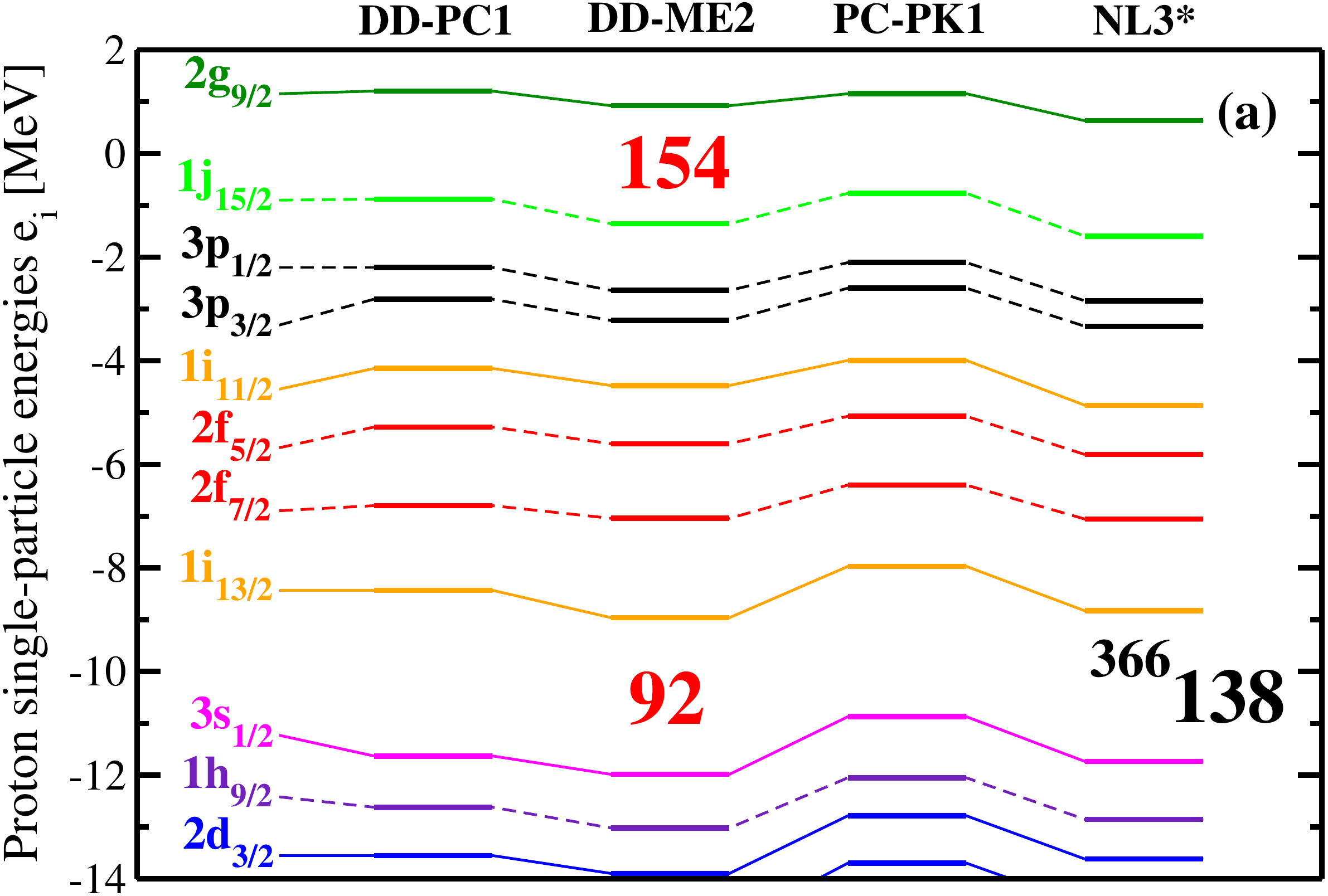}
\includegraphics[angle=0,width=10.5cm]{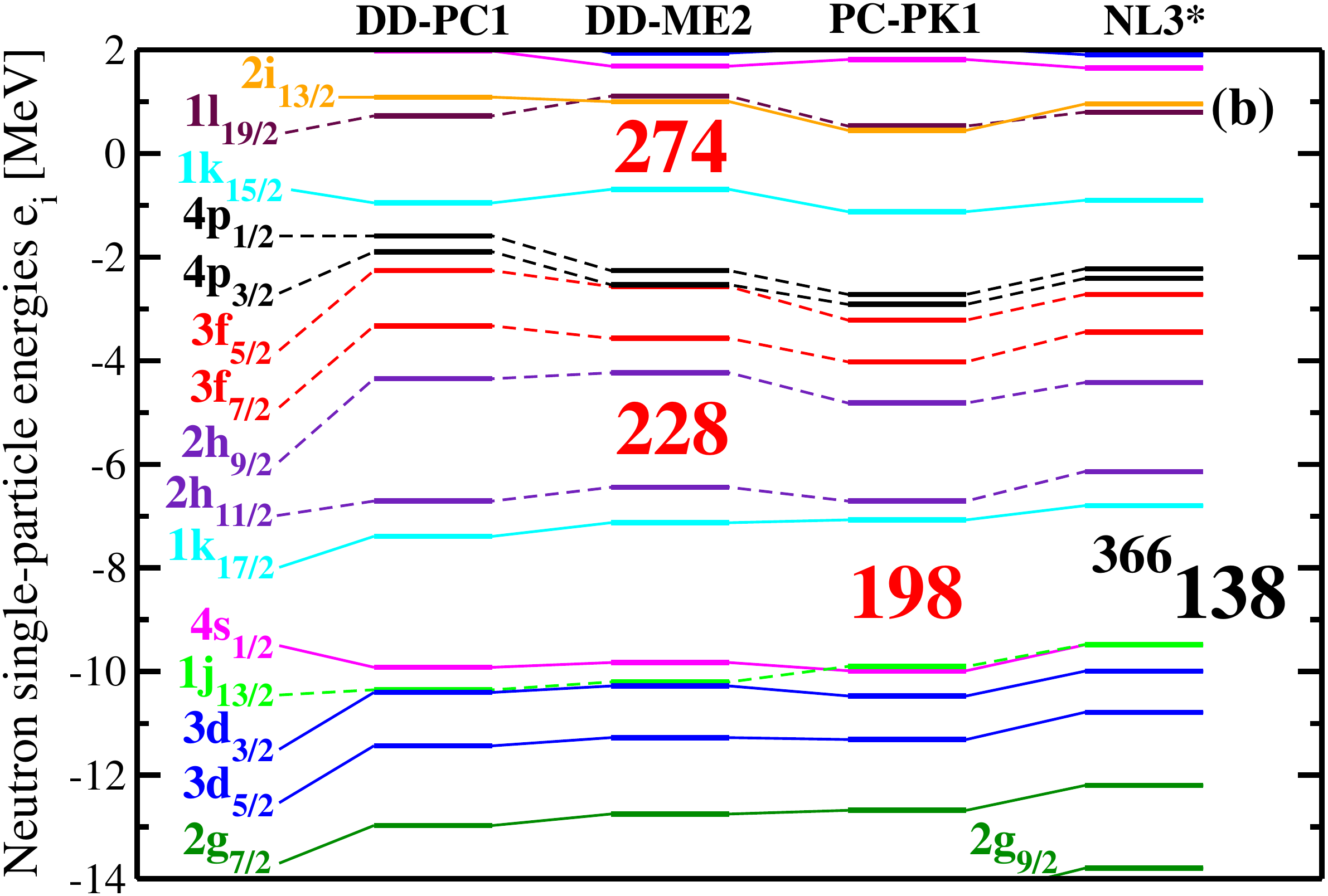}
\caption{(Color online)  The same as Fig.\ \ref{sphere-sp} but for the
$^{366}$138 nucleus.}
\label{sphere-sp-138}
\end{figure*}

\begin{figure*}[htb]
\includegraphics[angle=0,width=10.5cm]{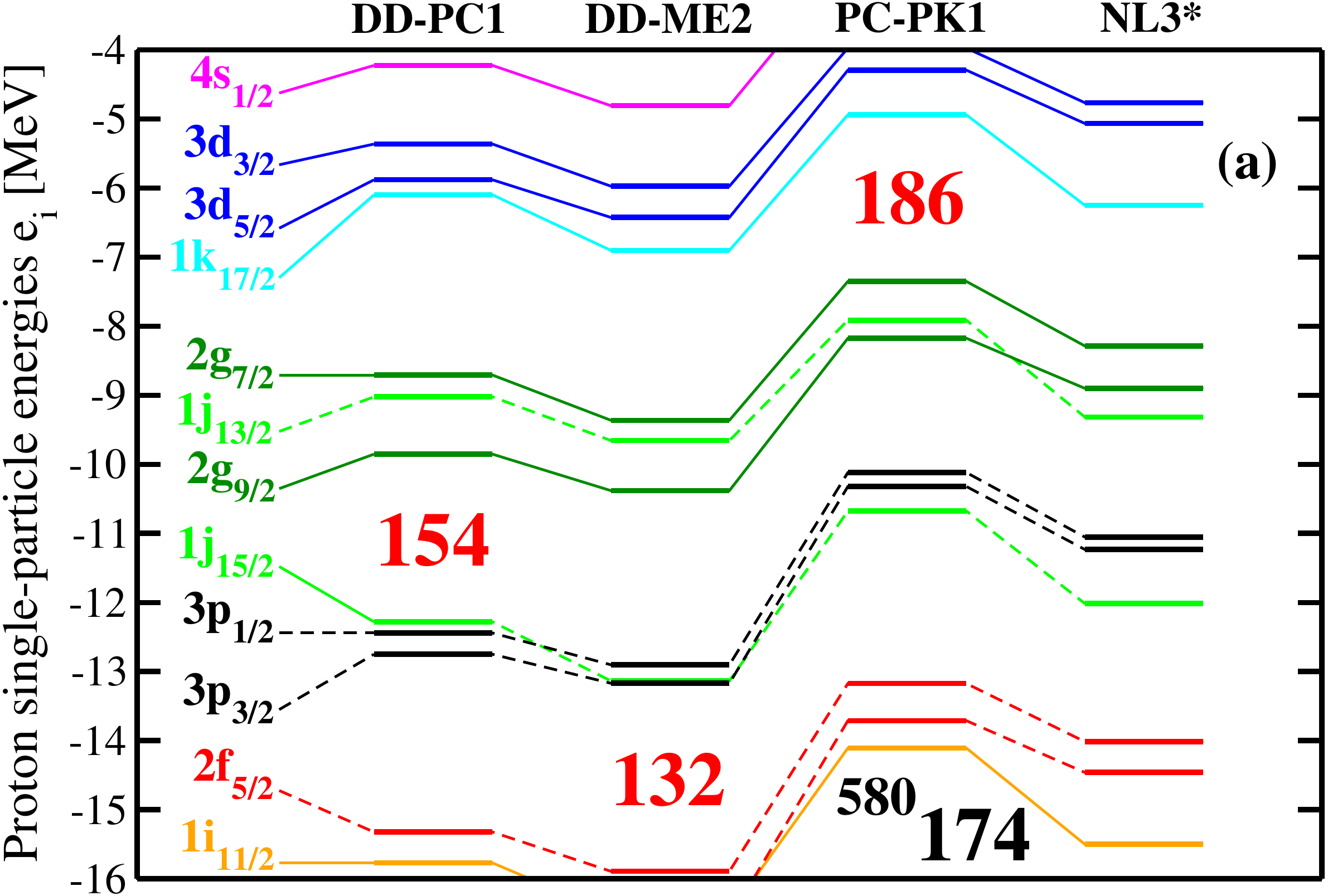}
\includegraphics[angle=0,width=10.5cm]{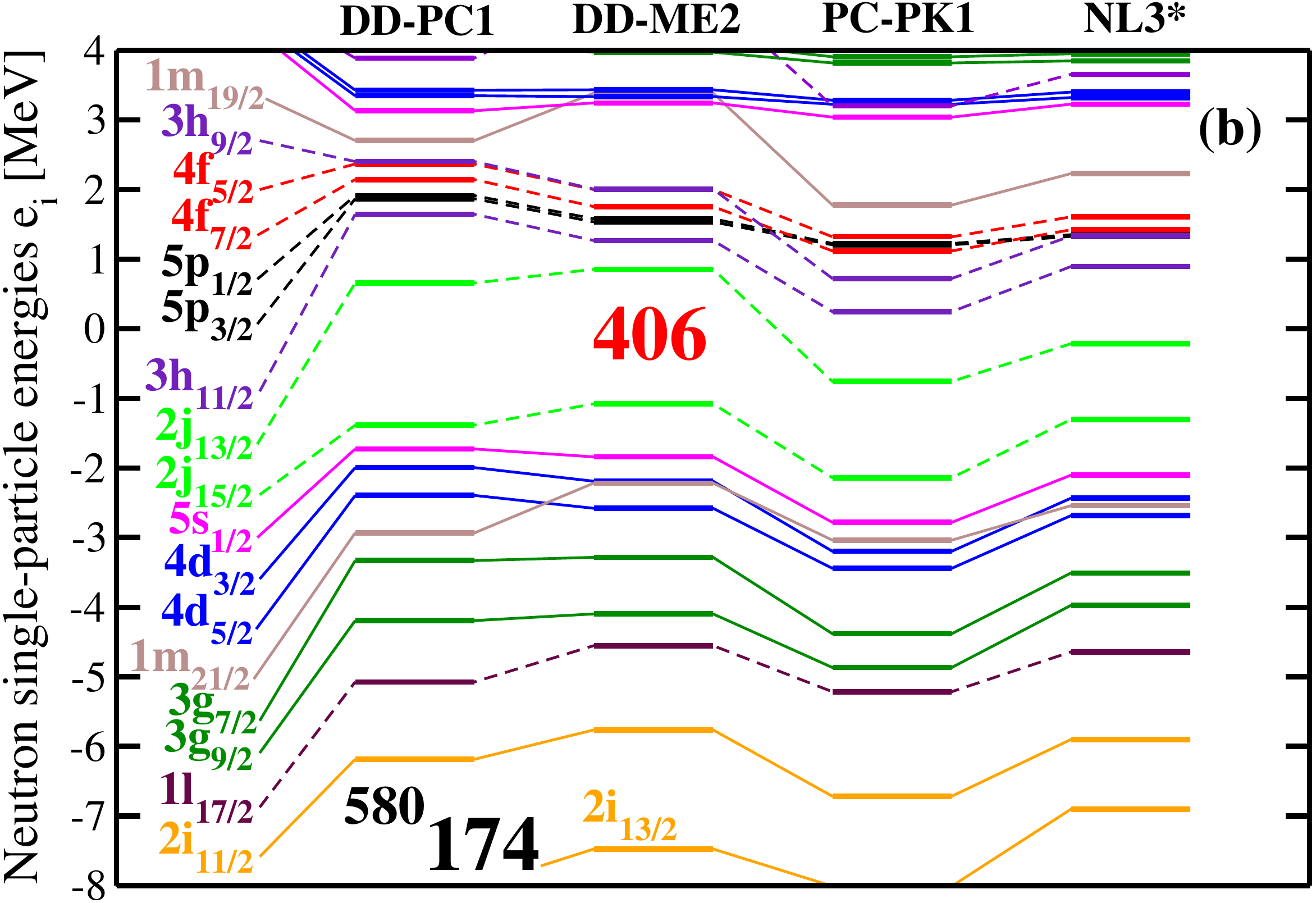}
\caption{ (Color online)  The same as Fig.\ \ref{sphere-sp} but for the $^{580}$174 
nucleus.}
\label{sphere-sp-174}
\end{figure*}

\begin{table}[htb!]
\begin{center}
\caption{Charge radii $r_{ch}$ [in $fm$] and neutron skins $r_{skin}$ 
         [in $fm$] of the density distributions shown in Fig.\ 
         \ref{Density-sphere}.
\label{Table-radii}
}
\begin{tabular}{|c|c|c|c|c|c|c|} \hline 
   Z   &   N   &           &  DD-PC1 &  DD-ME2 &   NL3* &  PC-PK1     \\\hline
  Pb   &  126  & $r_{ch}$   &  5.513  &  5.518  &  5.509 &  5.519   \\\hline
       &       & $r_{skin}$ &  0.202  &  0.193  &  0.288 &  0.257    \\ \hline
 120   &  172  & $r_{ch}$   &  6.272  &  6.282  &  6.276 &  6.286    \\\hline
       &       & $r_{skin}$ &  0.104  &  0.091  &  0.164 &  0.141    \\\hline
 138   &  230  & $r_{ch}$   &  6.759  &  6.765  &  6.799 &  6.811    \\\hline 
       &       & $r_{skin}$ &  0.198  &  0.188  &  0.283 &  0.249     \\\hline
 156   &  310  & $r_{ch}$   &  7.330  &  7.326  &  7.402 &  7.420    \\\hline 
       &       & $r_{skin}$ &  0.290  &  0.295  &  0.427 &  0.364    \\\hline
 174   &  410  & $r_{ch}$   &  7.927  &  7.930  &  8.071 &  8.087    \\\hline 
       &       & $r_{skin}$ &  0.440  &  0.466  &  0.616 &  0.520    \\\hline
    \end{tabular}
  \end{center}
\end{table}

  Charge radii $r_{ch}$ and neutron skins $r_{skin}$ of these 
nuclei are presented in Table \ref{Table-radii} and related
density distributions are shown in Fig.\ \ref{Density-sphere}.
The predictions of different functionals for charge radii
of $^{208}$Pb differ by less than 0.01 fm (see Table 
\ref{Table-radii}); this is also seen in proton density 
distributions (see Fig.\ \ref{Density-sphere}a). On the
contrary, the spread in the predictions of neutron skin is
significant reaching 0.095 fm. Density dependent (DD) functionals
predict the lowest values for the neutron skin, while the NL3* 
functional predicts the highest value and the PC-PK1 results lie 
somewhere in between of the DD and NL3* ones. These features are 
reflected also in the neutron density distributions; they extend 
to higher radii in surface area and have lower density in the 
central region in the NL3* and PC-PK1 functionals as compared 
with the DD ones (see Fig.\ \ref{Density-sphere}a). These differences 
between the functionals are realized when the neutron matter is 
moved from the surface region to the central and middle parts of 
the nucleus.  Comparable features are also seen in the $^{292}120$ 
nucleus (see Table \ref{Table-radii} and  Fig.\ \ref{Density-sphere}b).

  However, these differences between the predictions of the
functionals become enhanced on going to the central nuclei
of the regions of potential stability of spherical hyperheavy 
nuclei. The spreads in the predictions of charge radii increase from
0.014 fm for the $^{292}$120 nucleus to 0.052, 0.094 and 0.16 fm for 
the $^{368}138$, $^{466}156$ and $^{584}174$ nuclei, respectively. 
The largest charge radius is always produced by the PC-PK1 
functional, while the smallest either by DD-PC1 or by DD-ME2. 
Note that latter two functionals give comparable results. 
 
   The spreads in the predictions on going from $^{292}120$ nucleus
to higher $Z$/higher $N$ nuclei increase also for neutron skins (see 
Table \ref{Table-radii}). For example, the difference in neutron skin 
of the $^{584}$174 nucleus calculated with DD-PC1 and NL3* reaches 
0.176 fm. Similar to $^{208}$Pb and $^{292}120$ nuclei, density dependent 
(DD) functionals predict the lowest values for the neutron skin, while 
the NL3* functional predicts the highest value and the PC-PK1 results 
lie somewhere in between of the DD and NL3* ones.

  These results clearly indicate that the accuracy of the 
reproduction of charge radii and neutron skins by the CEDFs 
could be an important criteria in favoring or disfavoring
the predictions of one or another functional for the islands
of stability of spherical hyperheavy nuclei. Among considered
functionals, the DD-ME2 and DD-PC1 functionals provide the best
global description of charge radii (see Sect.\ X in Ref.\ 
\cite{AARR.14}). However, the situation with neutron skins 
is more complex. Even for $^{208}$Pb there is a significant controversy 
in the adopted experimental values of neutron skins (see discussion
in Sect.\ X  of Ref.\ \cite{AARR.14} and in Ref.\ \cite{PREX.12}). 
For example, the experiments based on hadronic probes provide neutron 
skin in $^{208}$Pb around 0.2 fm or slightly smaller. However, these 
experimental data are extracted in model-dependent ways. Alternatively, 
a measurement using an electroweak probe has been carried 
out in parity violating electron scattering on nuclei (PREX) and it 
brings $r_{skin}=0.33\pm 0.17$ \cite{PREX.12}. A central value of 0.33 
fm is particularly intriguing because it is around 0.13 fm higher 
than central values obtained in other experiments. Note that non-linear 
CEDFs typically give $r_{skin} \sim 0.3$ fm (see Table \ref{Table-radii}).
The electroweak probe has the advantage over experiments using hadronic 
probes that it allows a nearly model-independent extraction of the 
neutron radius that is independent of most strong interaction 
uncertainties \cite{PREX-CREX}. Thus, the results obtained in future 
PREX-2 experiment \cite{PREX-CREX} would be quite useful in
helping to discriminate the predictions.

\begin{figure*}[htb]
\centering   
\includegraphics[angle=0,width=8.5cm]{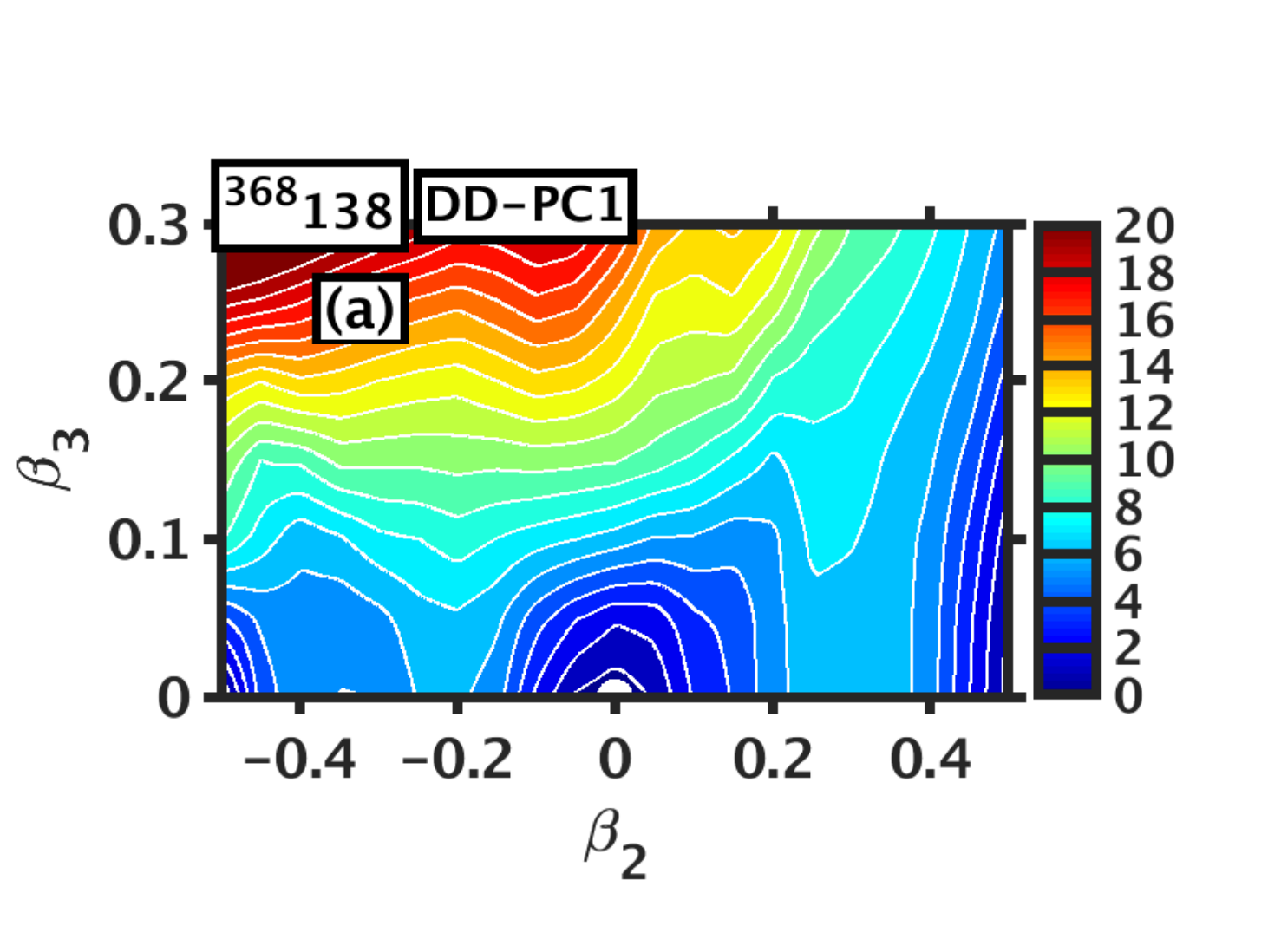}
\includegraphics[angle=0,width=8.5cm]{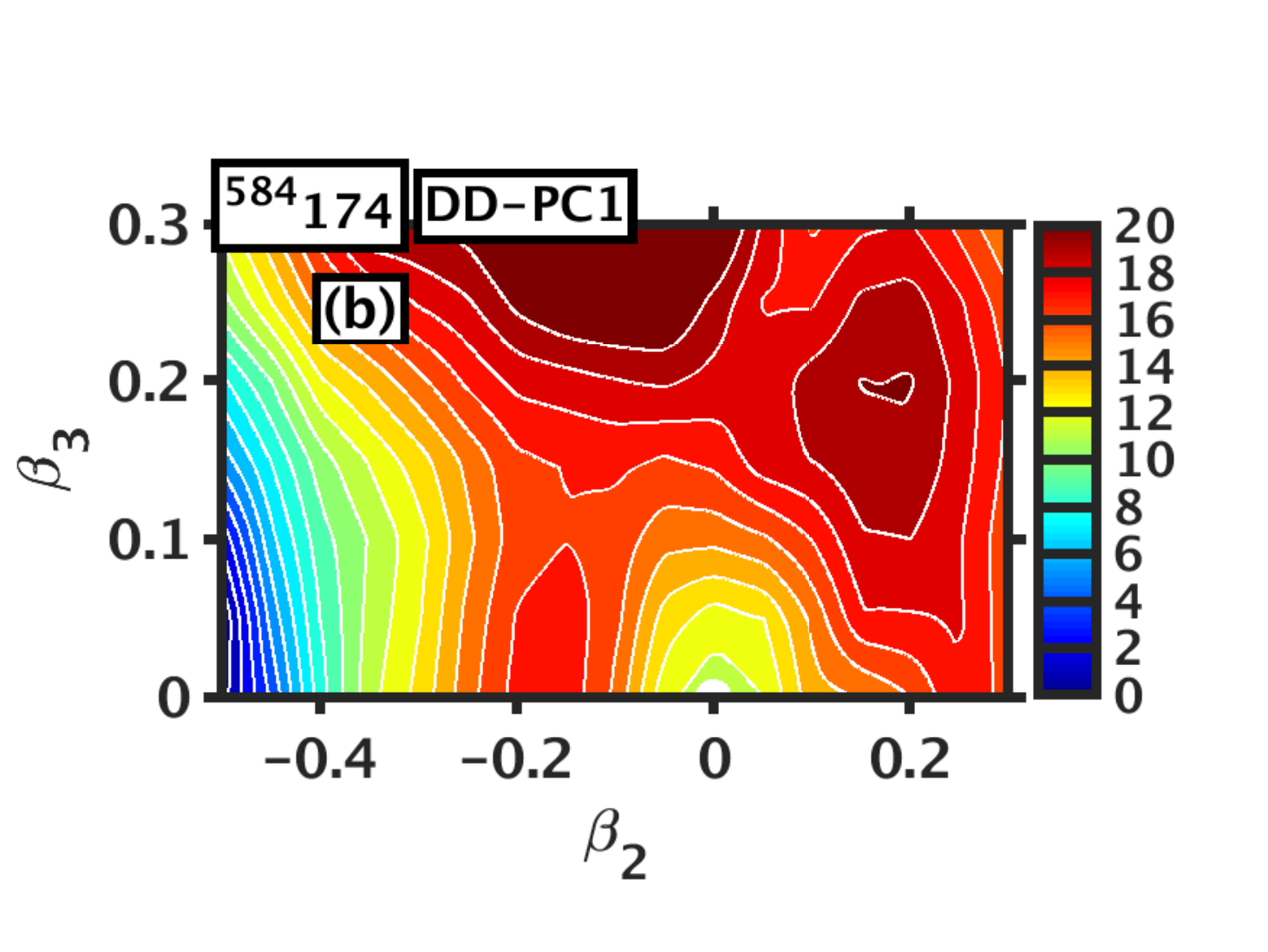}
\includegraphics[angle=0,width=8.5cm]{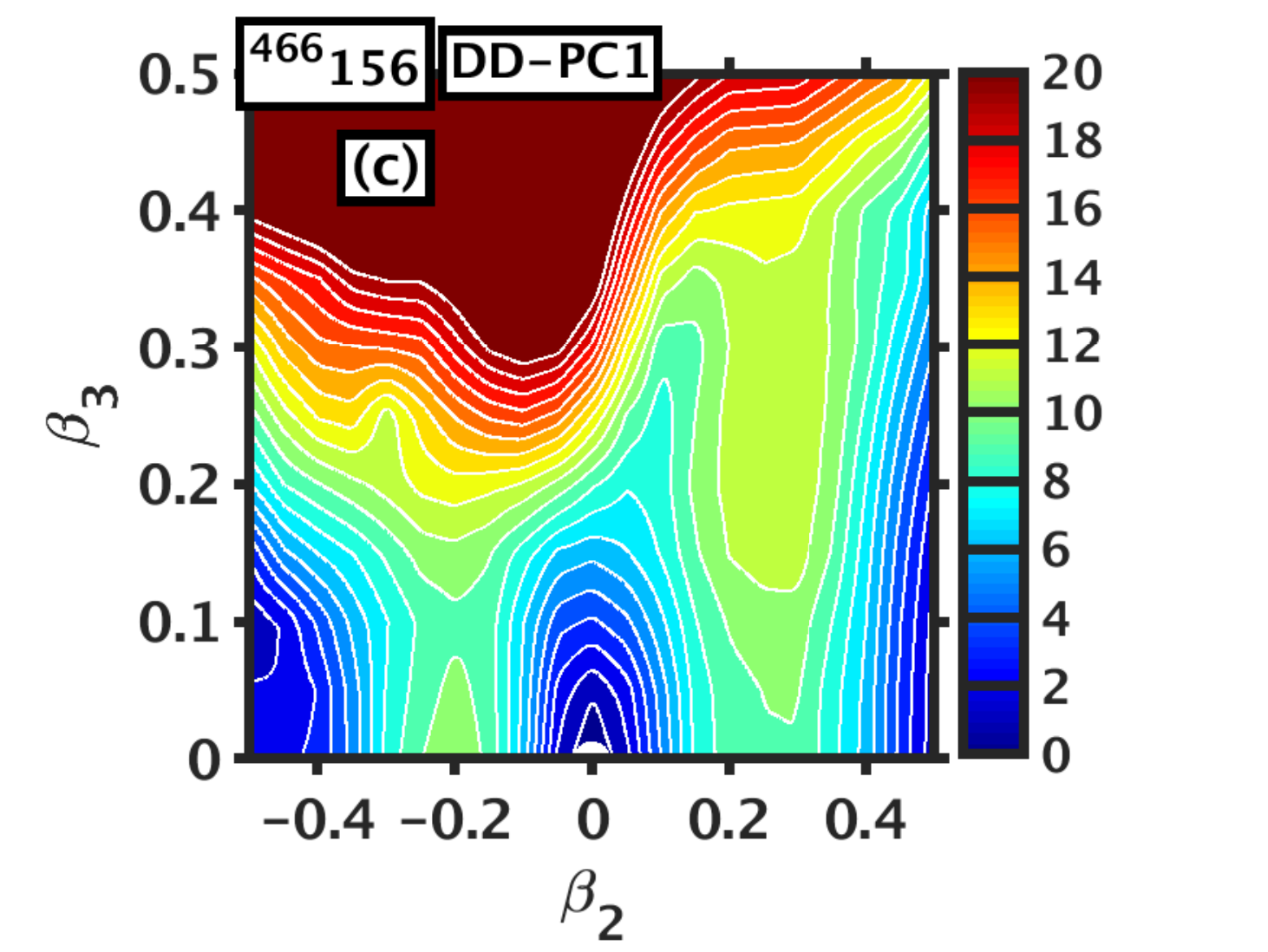}
\caption{(Color online) Potential energy surfaces in the $(\beta_2, \beta_3)$
plane of the central nuclei of the regions of potential stability of spherical
hyperheavy nuclei.  Spherical minimum is indicated by a white semicircle.
Equipotential lines are shown in steps of 1.0 MeV. Note that the results are 
shown in different $(\beta_2, \beta_3$) deformation ranges.}
\label{oct-pes-spher-nuclei}
\end{figure*}

\begin{figure*}[htb]
\centering
\includegraphics[angle=0,width=8.0cm]{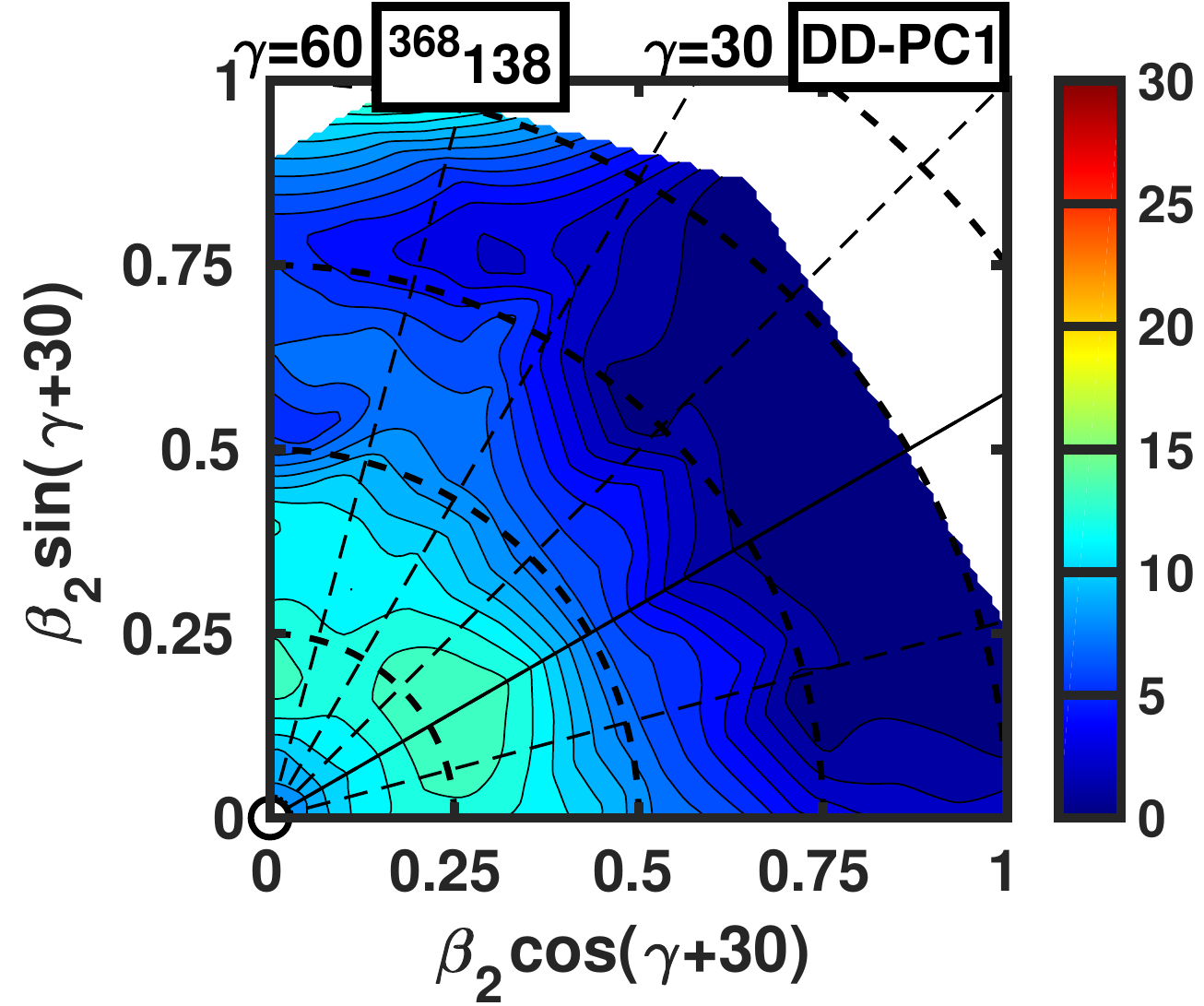}
\includegraphics[angle=0,width=8.0cm]{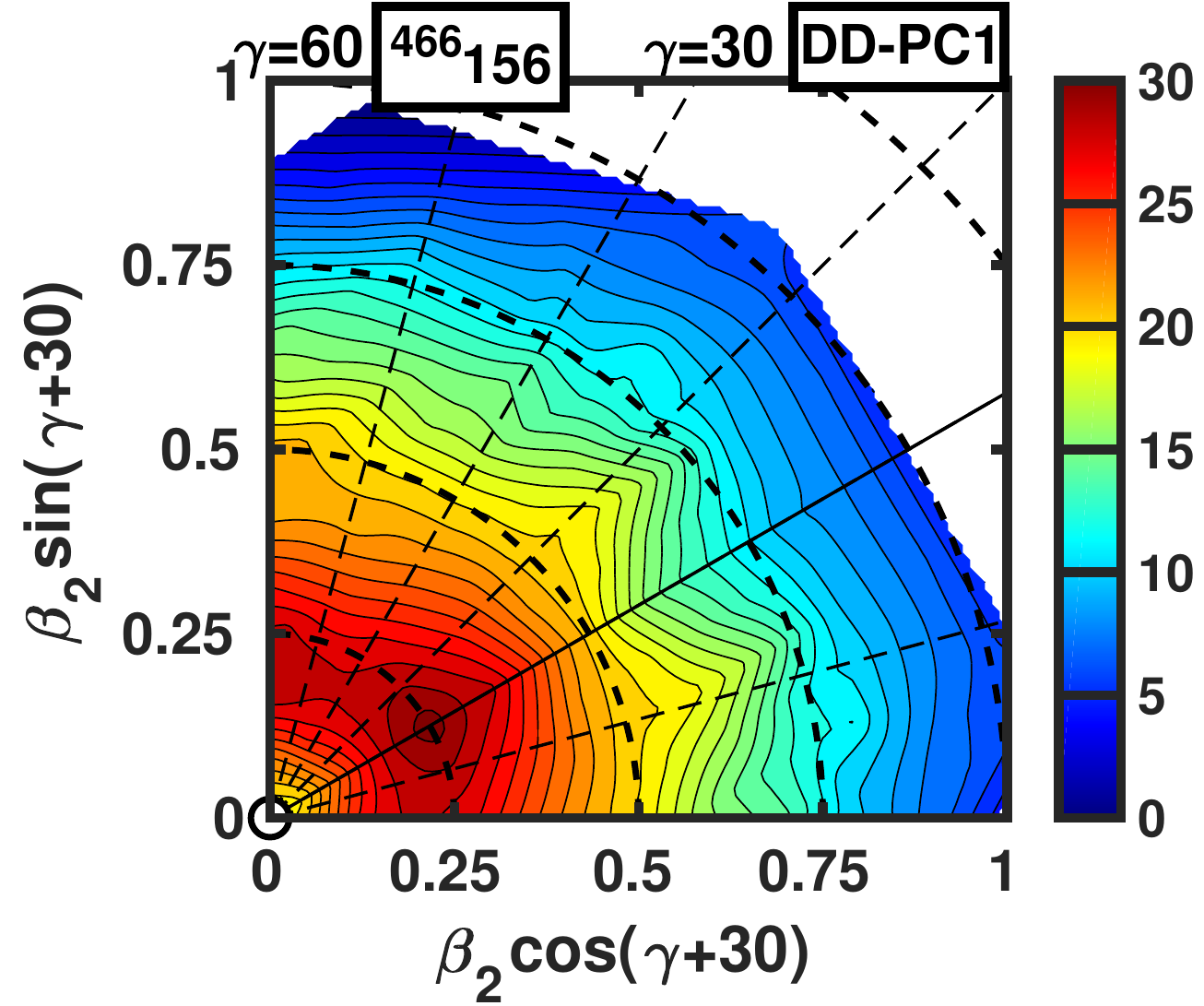}
\includegraphics[angle=0,width=8.0cm]{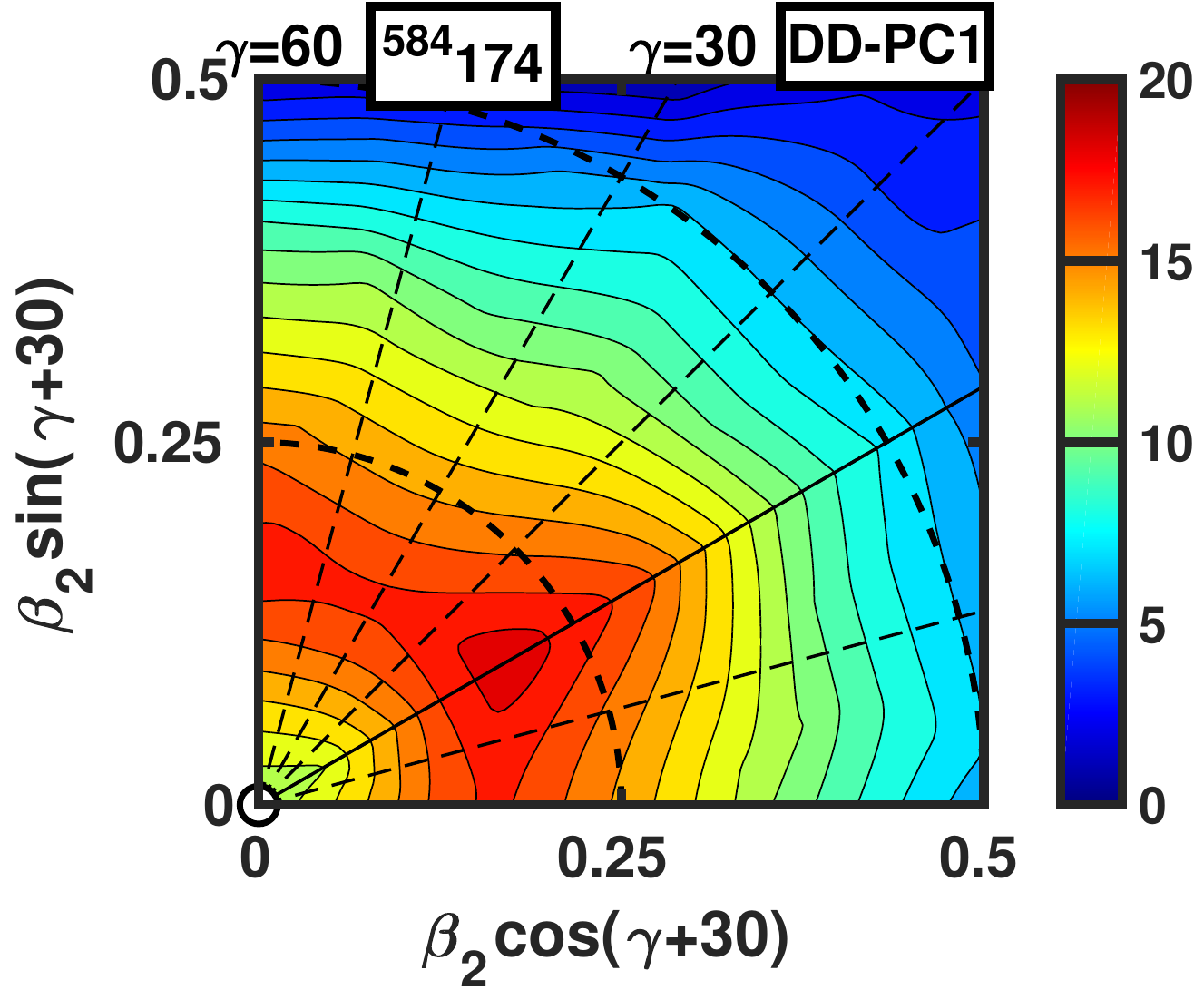}
\caption{(Color online) Potential energy surfaces of the nuclei located in the
centers of the regions of stability of spherical hyperheavy nuclei obtained in 
the TRHB calculations with $N_F=18$.  Note that the topology of potential 
energy surfaces is almost the same in the calculations with $N_F=18$ and
$N_F=20$. Thus, to save computational time these figures are plotted with
$N_F=18$.  The energy difference between two 
neighboring equipotential lines is equal to 1.0 MeV. Spherical minimum is 
indicated by a circle. The colormaps show the 
excitation energies (in MeV) with respect to the energy of the deformation 
point with largest (in absolute value) binding energy. 
Note that the results 
for the $^{548}$174 nucleus are shown in a smaller deformation range (because 
of convergence problems at large $\beta_2$ values) and different colormap is 
used for this nucleus.} 
\label{triaxial-pes}
\end{figure*}

\begin{figure*}[htb]
\includegraphics[angle=0,width=16.0cm]{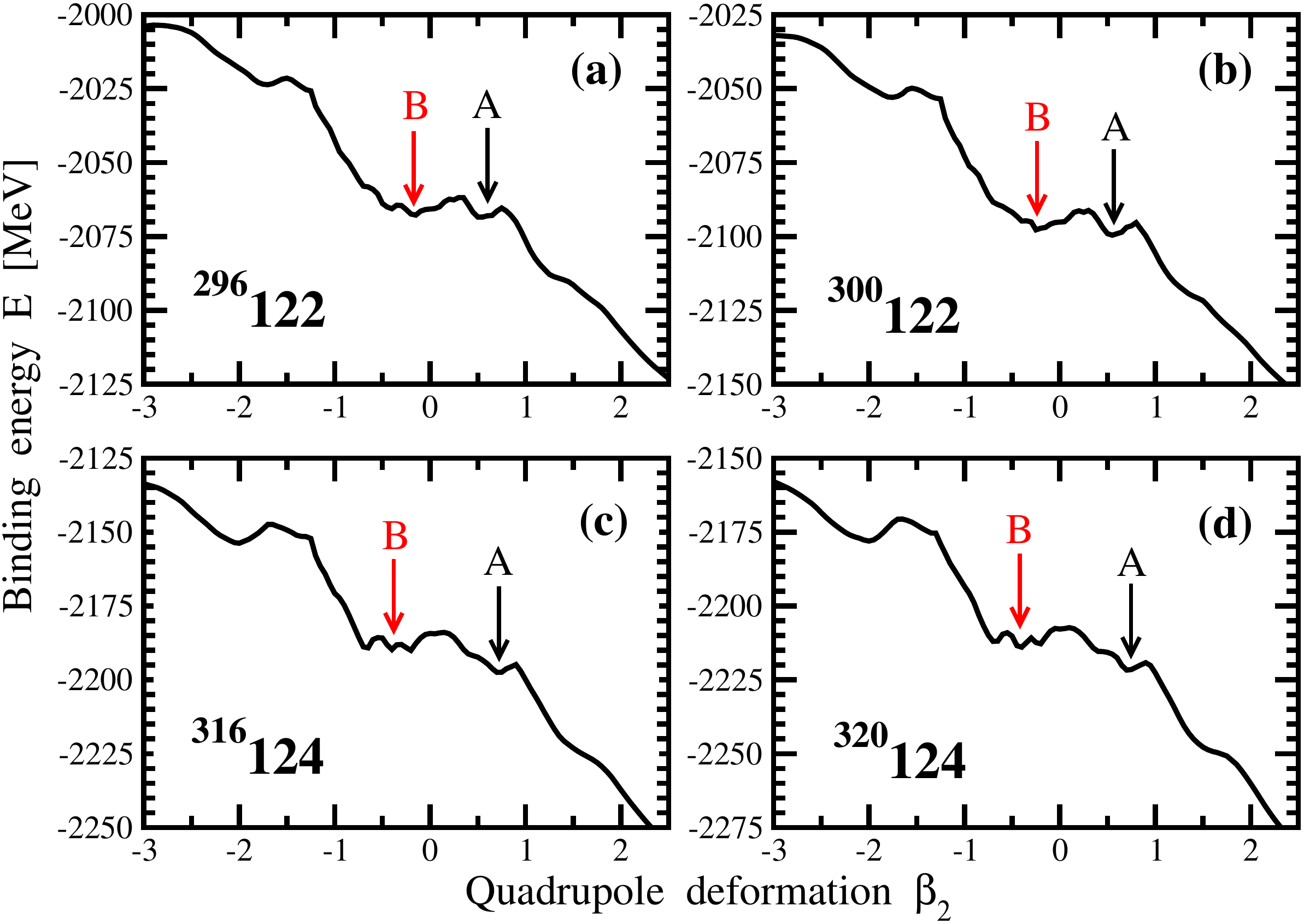}
\caption{(Color online) Deformation energy curves of selected 
even-even superheavy nuclei obtained in axial RHB calculations
performed with $N_F=26$. Arrows A and B indicate prolate 
superdeformed $\beta_2 \sim 0.6$ and oblate $\beta_2 \sim -0.5$ 
local minima, respectively.
\label{def-energy-curves}
}
\end{figure*}

\begin{figure*}[htb]
\includegraphics[angle=0,width=8.5cm]{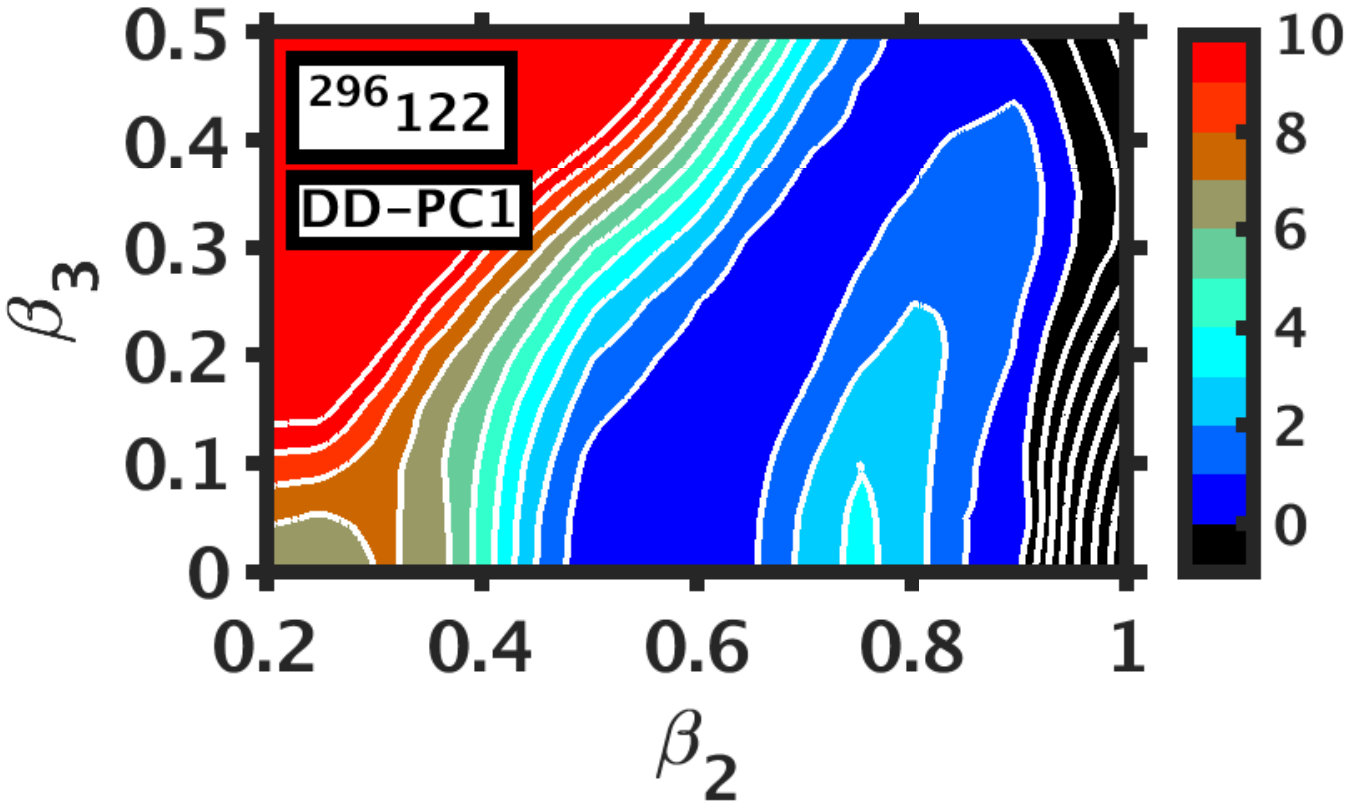}
\includegraphics[angle=0,width=8.5cm]{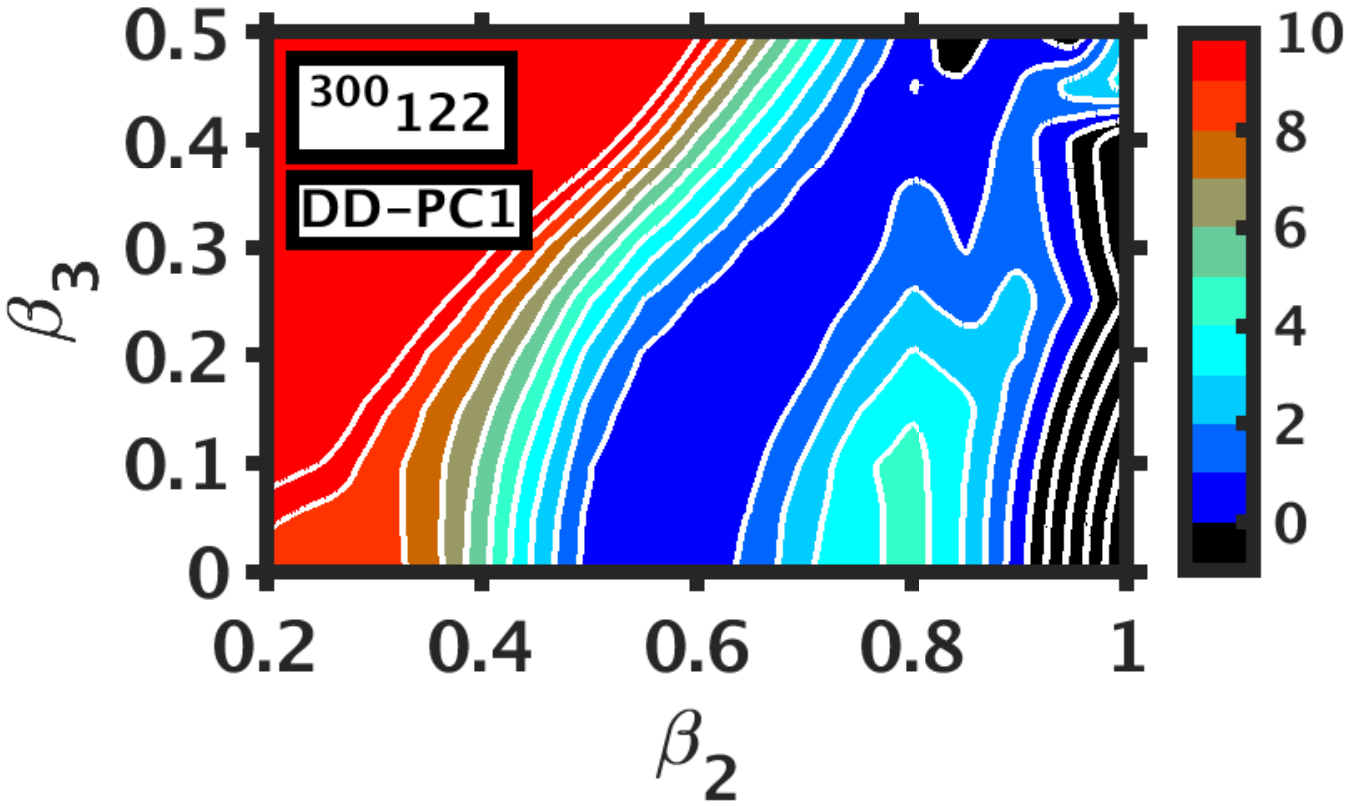}
\includegraphics[angle=0,width=8.5cm]{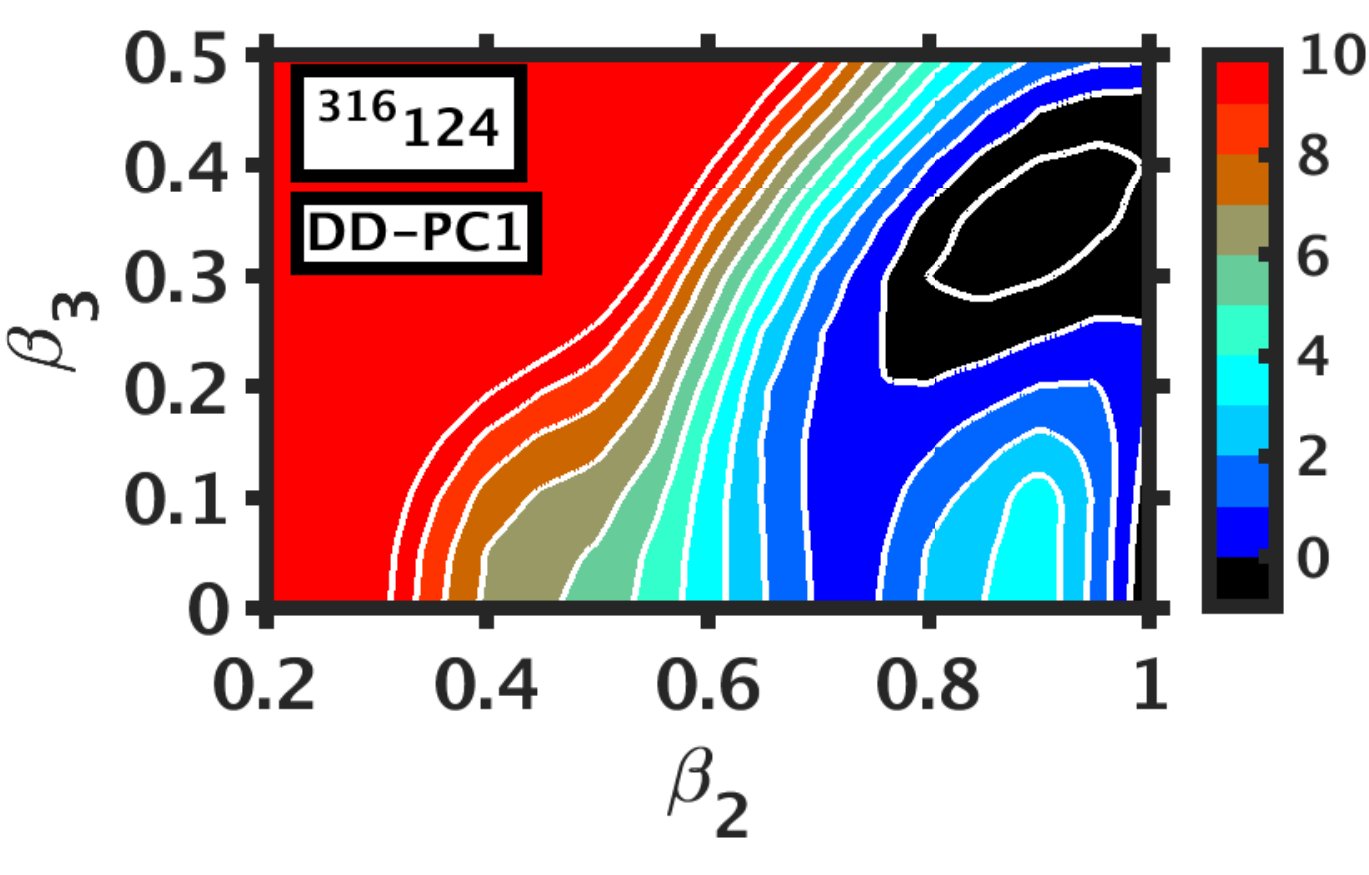}
\includegraphics[angle=0,width=8.5cm]{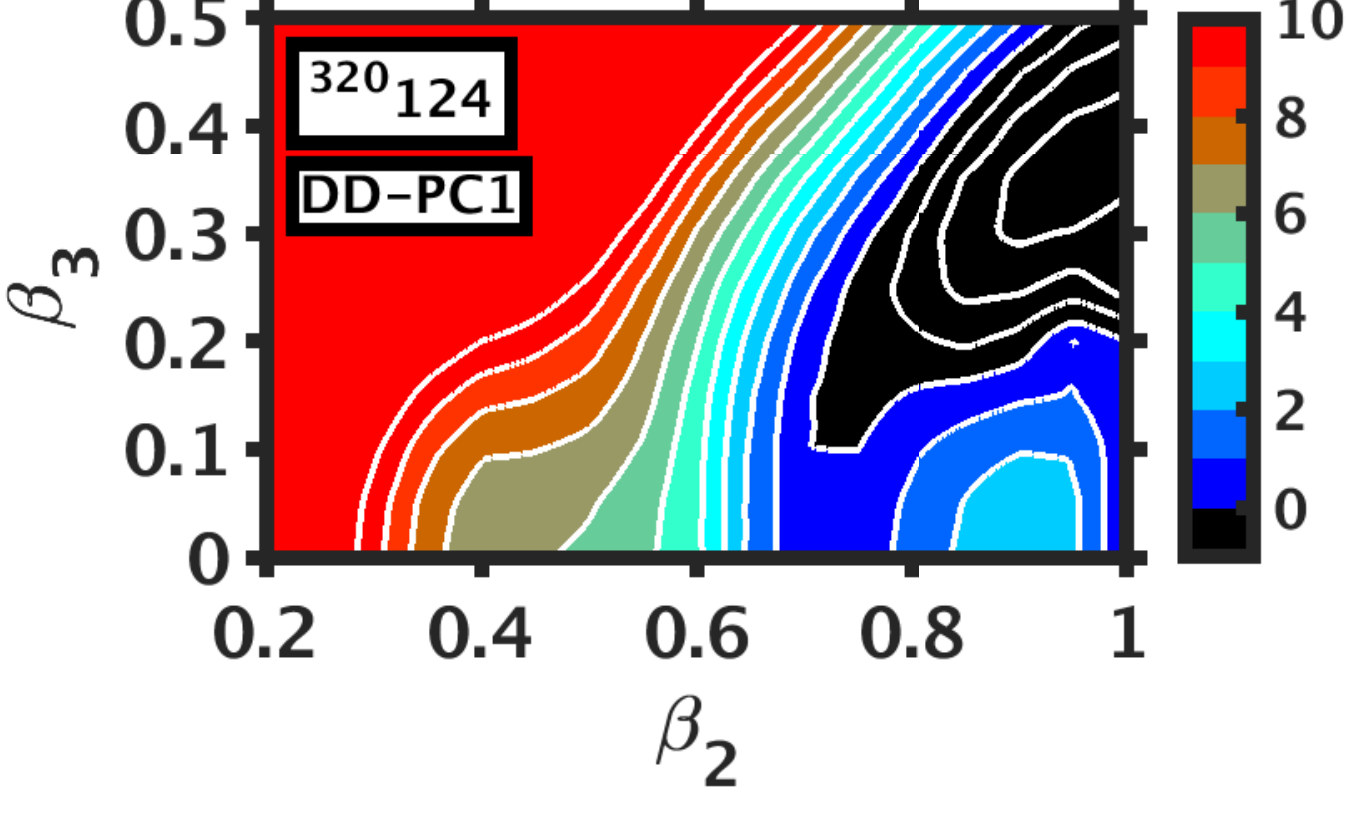}
\caption{(Color online) Potential energy surfaces in the 
$(\beta_2, \beta_3)$ plane for the nuclei shown in Fig.\ 
\ref{def-energy-curves}. Equipotential lines are shown 
in steps of 1.0 MeV. 
\label{SD_minima_octup}
}
\end{figure*}

\begin{figure}[htb]
\includegraphics[angle=0,width=8.5cm]{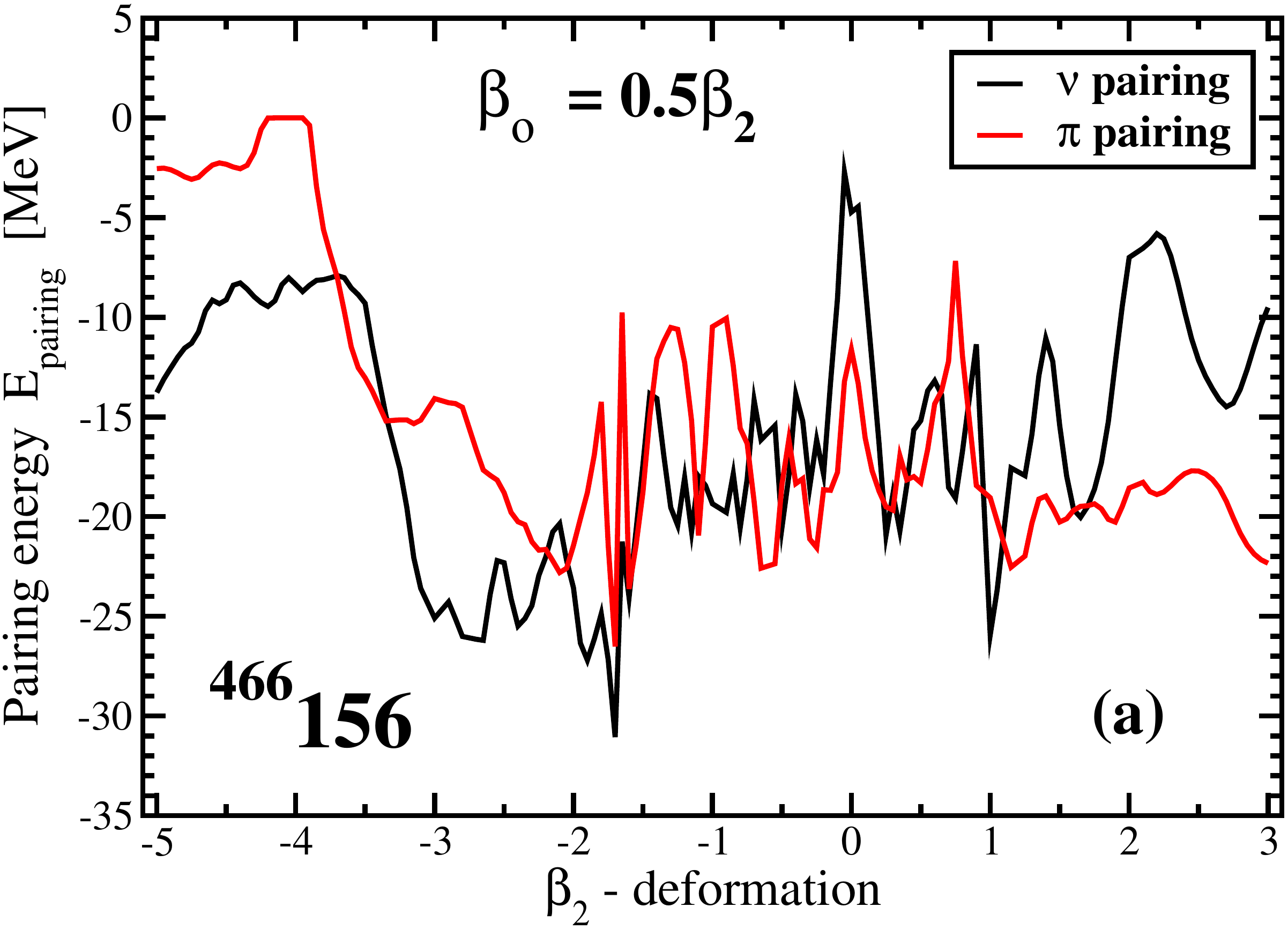}
\includegraphics[angle=0,width=8.5cm]{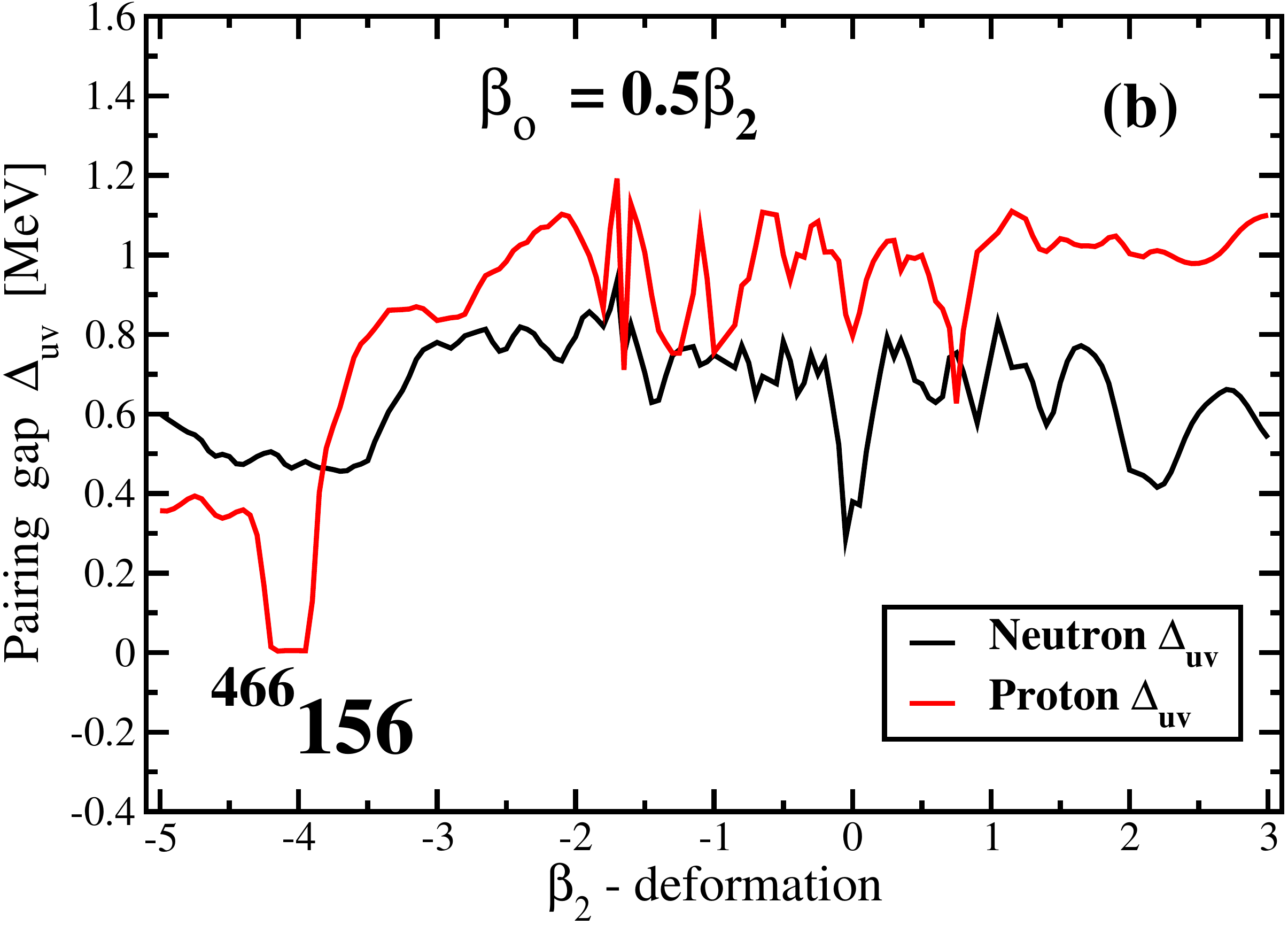} 
\caption{(Color online) Neutron and proton pairing energies $E_{pairing}$
(panel (a)) and pairing gaps $\Delta_{uv}$ (panel (b)) as a function of
$\beta_2$ for the lowest in 
energy solution in the $^{466}$156 nucleus, obtained with $N_F=30$ 
and the deformation of basis $\beta_0 = 0.5 \beta_2$, shown in Fig.\ 
\ref{trunc_basis}d.
\label{Pairing-axial}
}
\end{figure}

\begin{figure*}[htb]
\includegraphics[angle=-90,width=8.5cm]{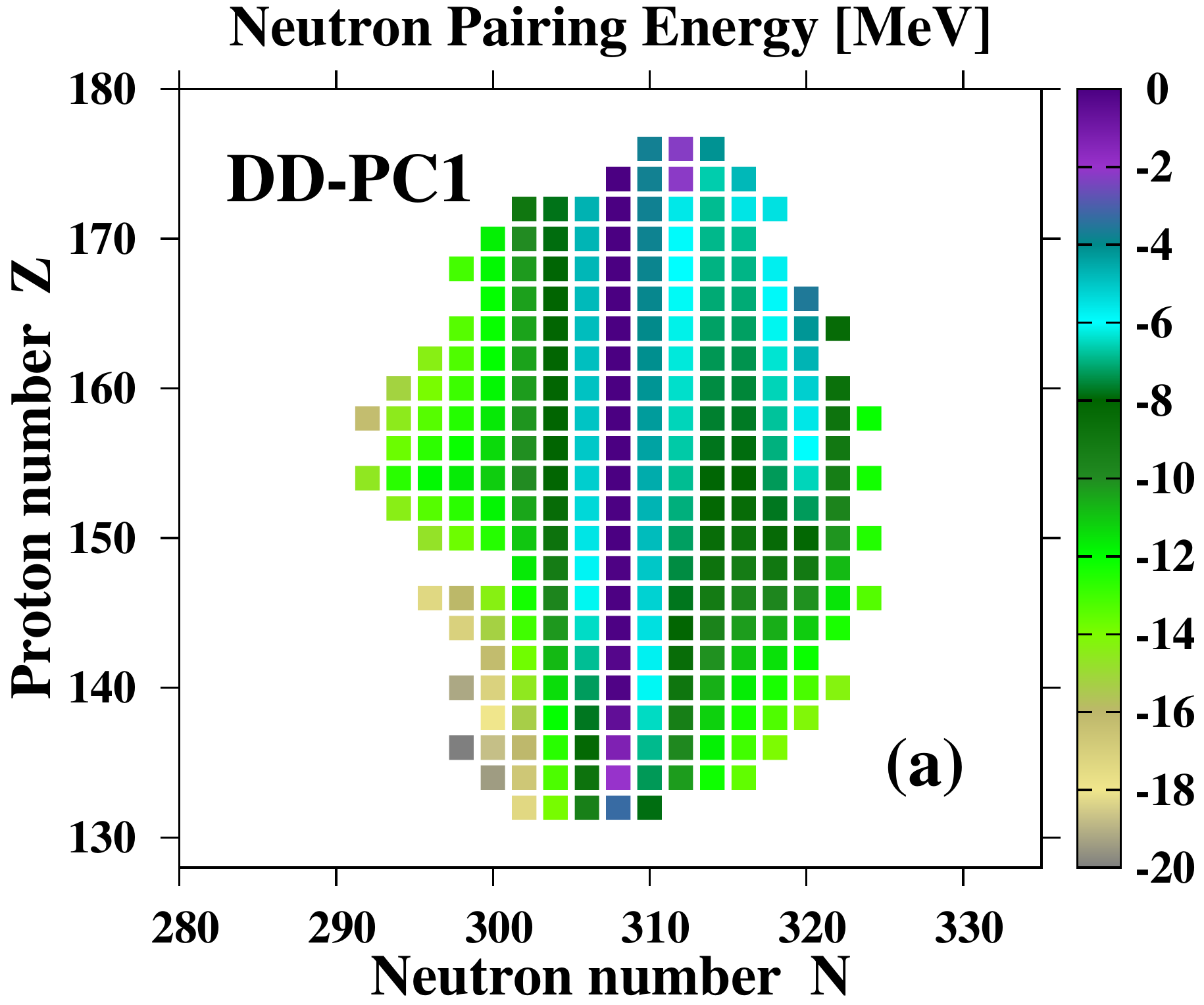}
\includegraphics[angle=-90,width=8.5cm]{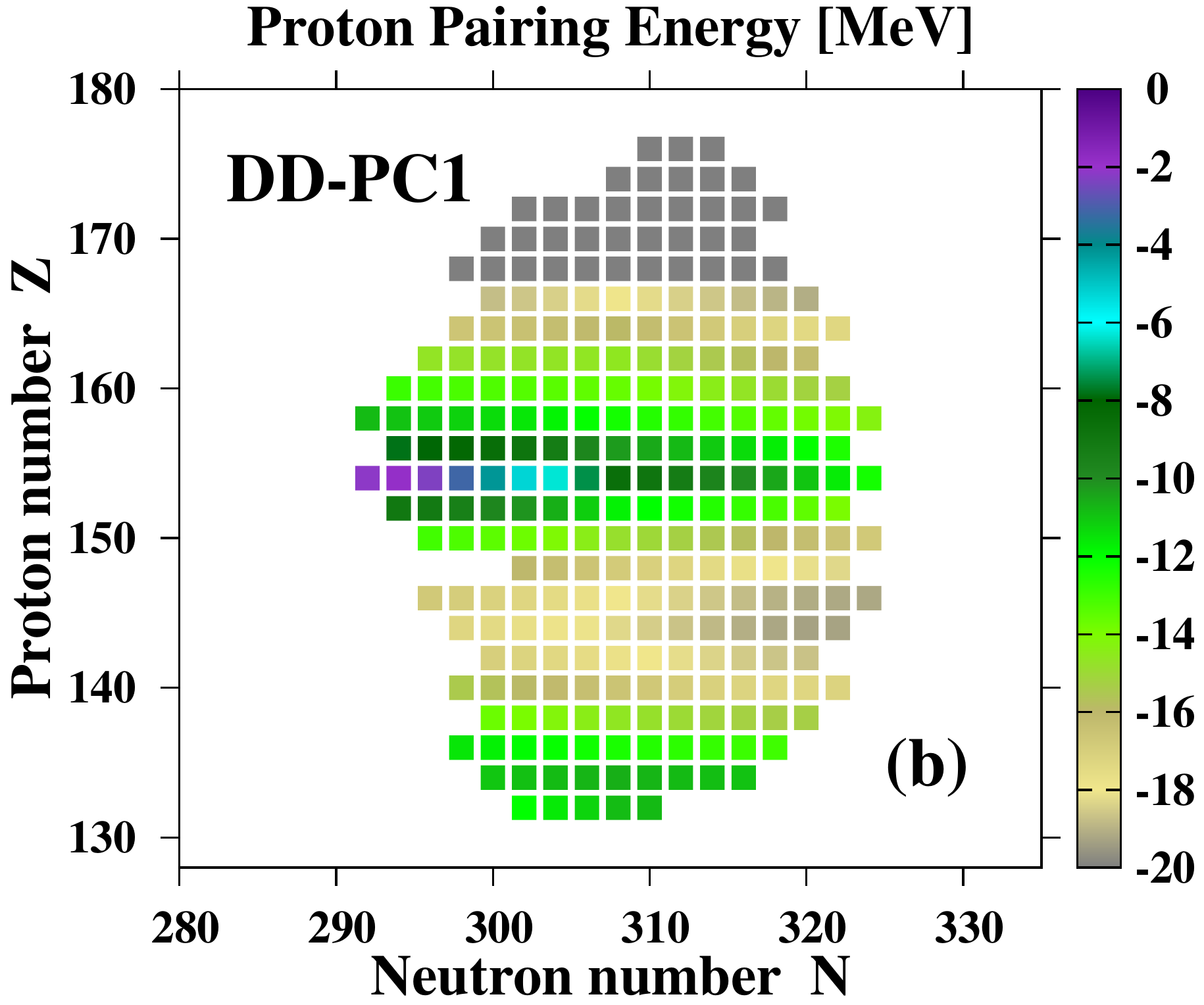}
\caption{(Color online) Neutron (panel (a)) and proton (panel (b)) pairing 
energies $E_{pairing}$ for spherical minima of the nuclei forming the 
$(Z\sim 156, N\sim 310)$ island of stability of spherical hyperheavy 
nuclei.  
\label{Pairing-spherical}
}
\end{figure*}

\begin{figure*}[htb]
\includegraphics[angle=0,width=8.5cm]{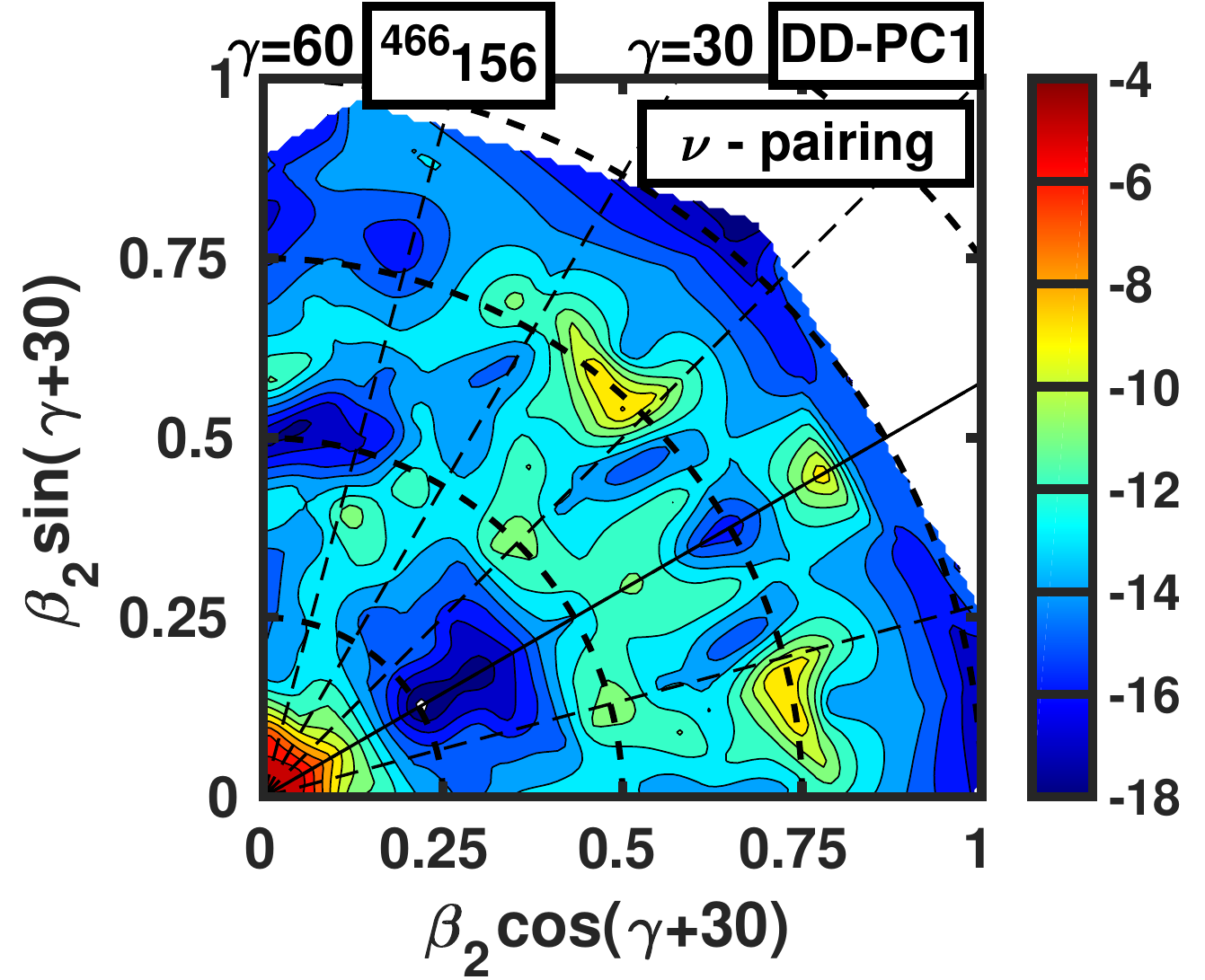}
\includegraphics[angle=0,width=8.5cm]{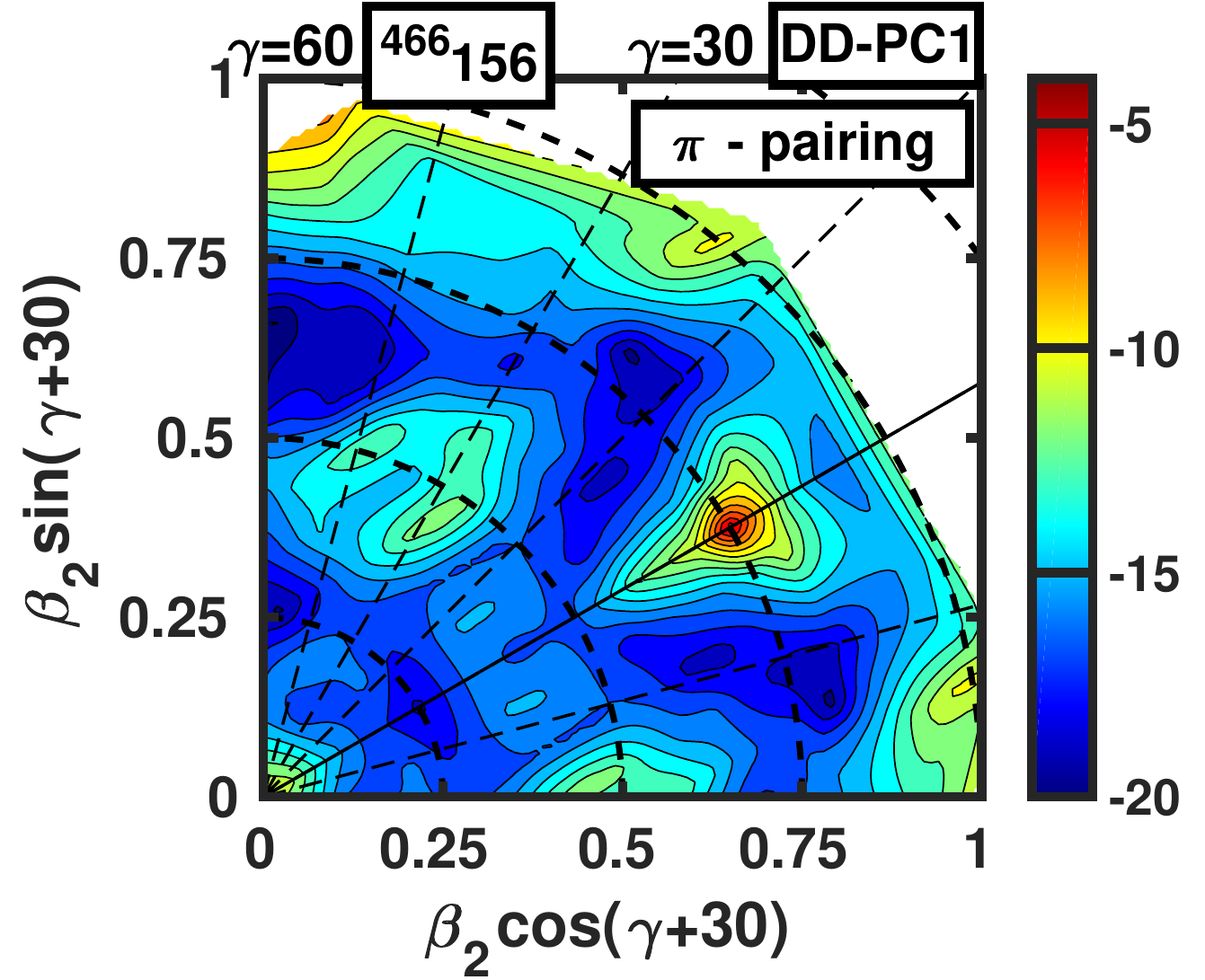}
\caption{(Color online) The evolution of neutron (panel (a)) and proton 
(panel (b)) pairing energies $E_{pairing}$ as a function of the $\beta_2$ 
and $\gamma$ deformations in the $^{466}$156 nucleus.
}
\label{Pairing-triaxial}
\end{figure*}

\begin{figure*}[htb]
\includegraphics[angle=90,width=18.cm]{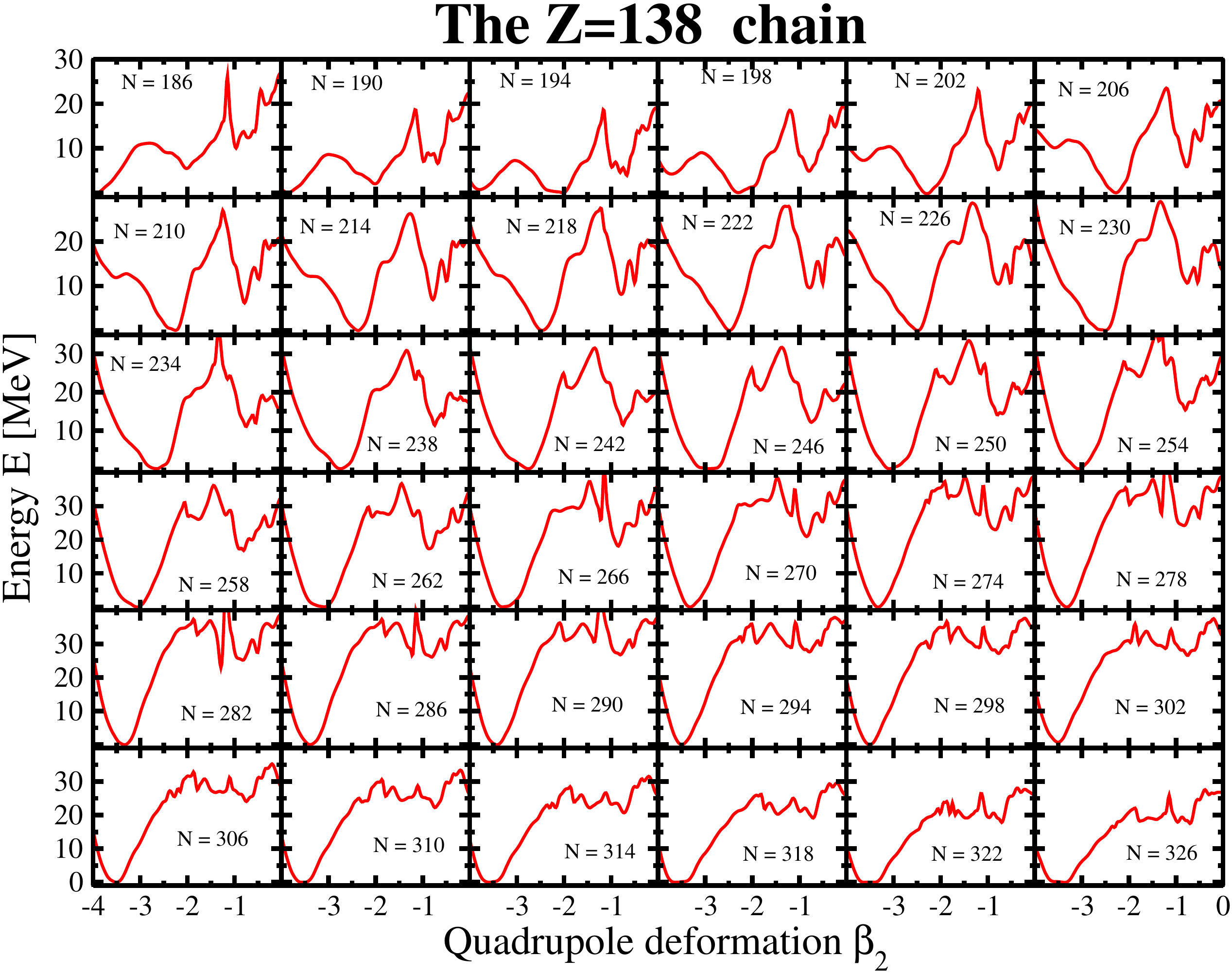}
\caption{(Color online) Deformation energy curves of even-even
$Z=138$ hyperheavy nuclei obtained in axial RHB calculations 
with DD-PC1 functional and the $N_F=26$ basis. The nuclei located 
between two-proton and two-neutron drip lines are shown in step of 
$\Delta N=4$. The curves are plotted only for negative $\beta_2$ 
values since prolate solutions are unstable. The deformation energy 
curves are normalized in such a way that the minimum of total 
energy for negative $\beta_2$ values corresponds to zero 
energy. 
\label{Z138-E}
}
\end{figure*}

\begin{figure*}[htb]
\includegraphics[angle=90,width=18.cm]{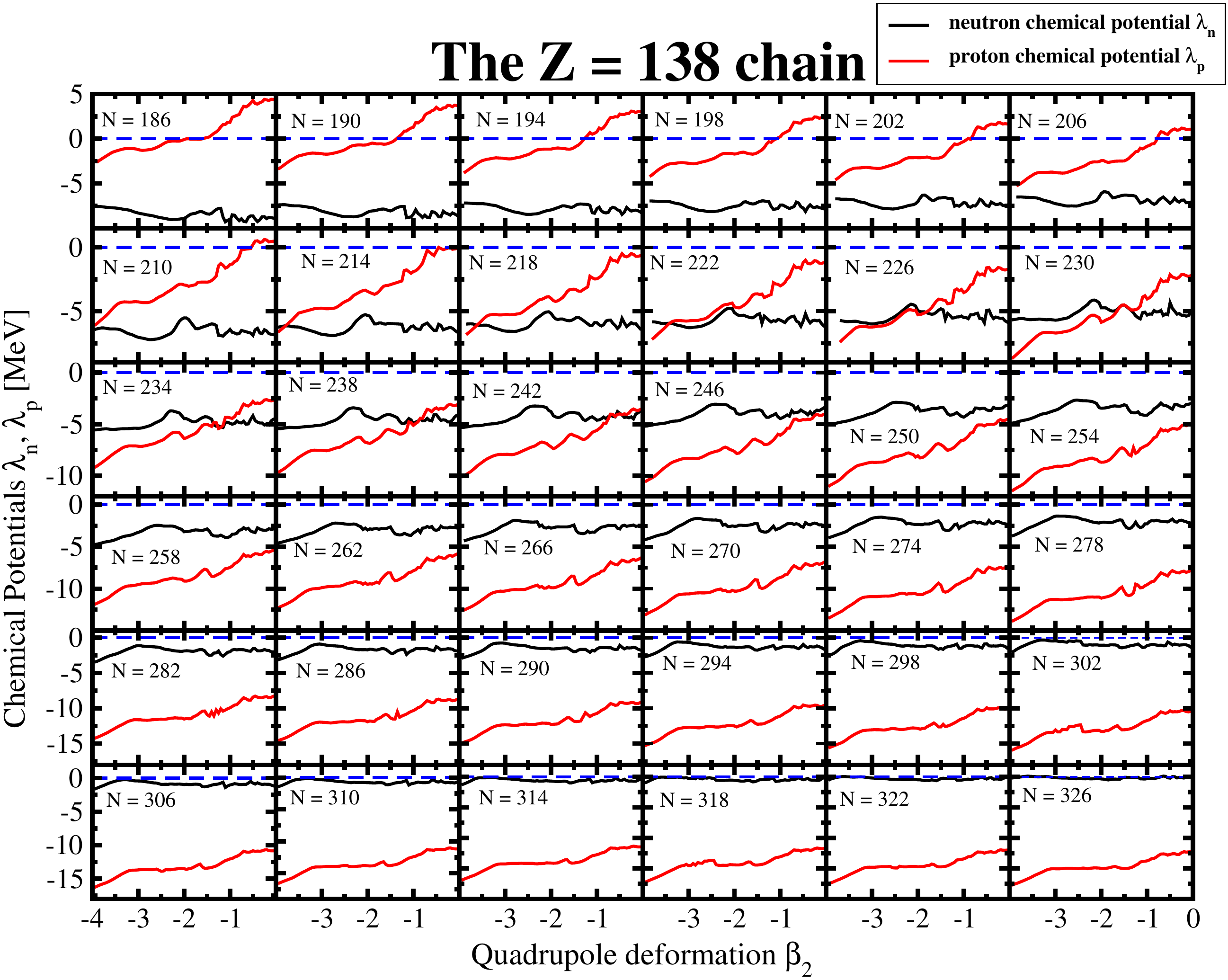}
\caption{(Color online) Proton and neutron chemical potentials
as a function of the  $\beta_2$ values for the solutions 
displayed in Fig.\ \ref{Z138-E}. Blue dashed lines show the 
continuum threshold.
\label{Z138-lambda}
}
\end{figure*}

  General features of proton and neutron density 
distributions of hyperheavy nuclei seen in Fig.\ \ref{Density-sphere} 
do not depend on employed functional. Apart of neutron density
in $^{368}138$ nucleus (which is almost the same in the center of nucleus
and at its surface), both types of  densities are characterized by the 
density depression in the central part of the nucleus. 
Here we use the ratio $\omega = \rho^{cent}/\rho^{surf}$ of the 
density at the center $\rho^{cent}$ to the maximum
density at the surface $\rho^{surf}$ averaged over the set of
employed functionals to characterize this depression.
The central density depressions in neutron subsystems of the 
$^{466}156$ and $^{584}$174 nuclei are rather modest with 
$\omega_{\nu}= 0.814$ and 0.86, respectively. Thus, neutron
densities are close to flat density distributions and could not be 
characterized as specific for semi-bubble 
nuclei (in the language of Ref.\ \cite{DBDW.99}).  However,  central 
depressions are significantly more pronounced in proton subsystems 
of hyperheavy nuclei with $\omega_{\pi}=0.753$, 0.651 and 0.534 for 
the $^{368}138$, $^{466}156$ and $^{584}$174 nuclei, respectively. 
Thus, they are close to the ones expected for semi-bubble nuclei 
(see Ref.\ \cite{DBDW.99}). Note that in a given nucleus the proton  
density is roughly half of the neutron one.

\begin{figure*}[htb]
\includegraphics[angle=0,width=8.5cm]{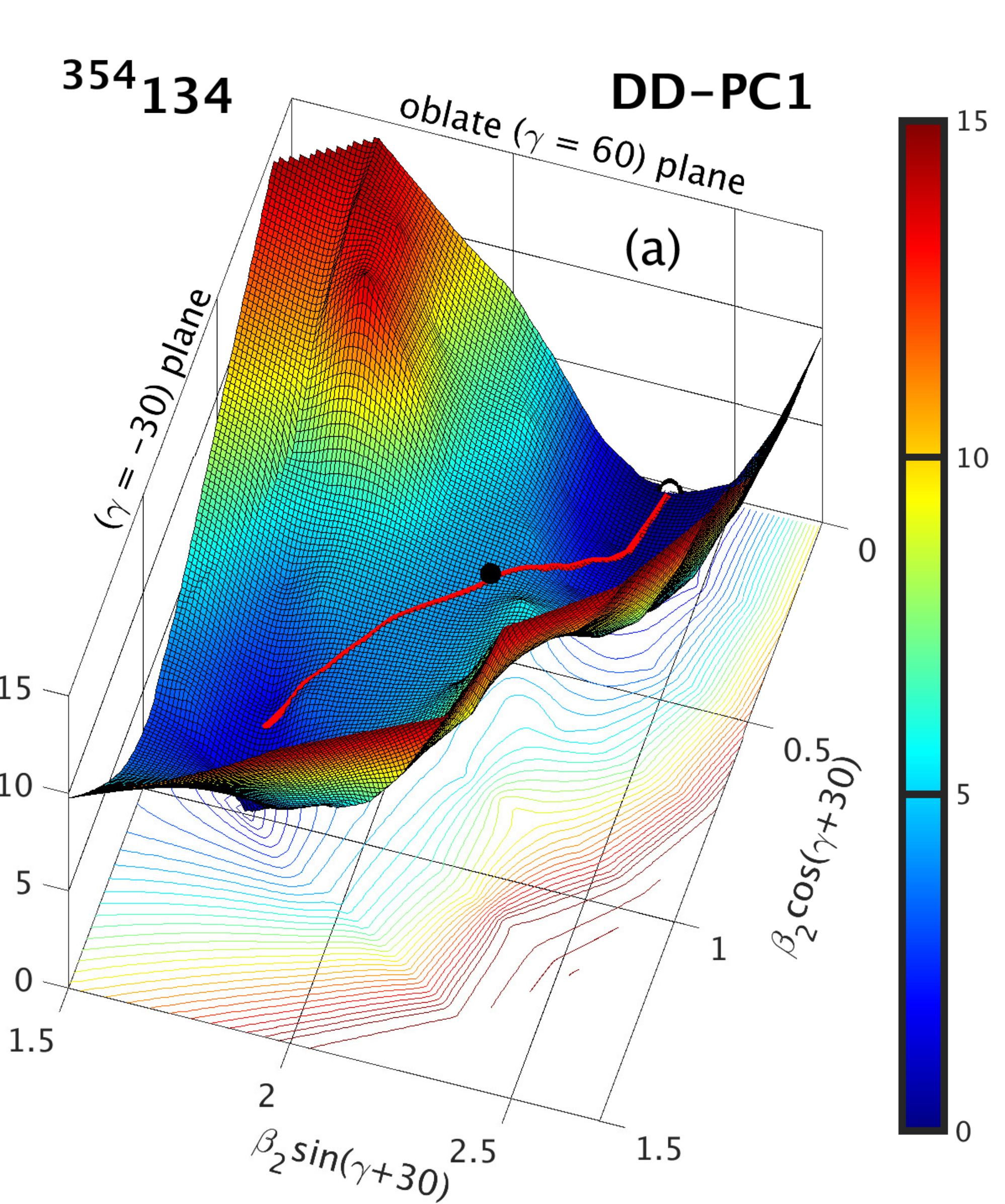}
\includegraphics[angle=0,width=8.5cm]{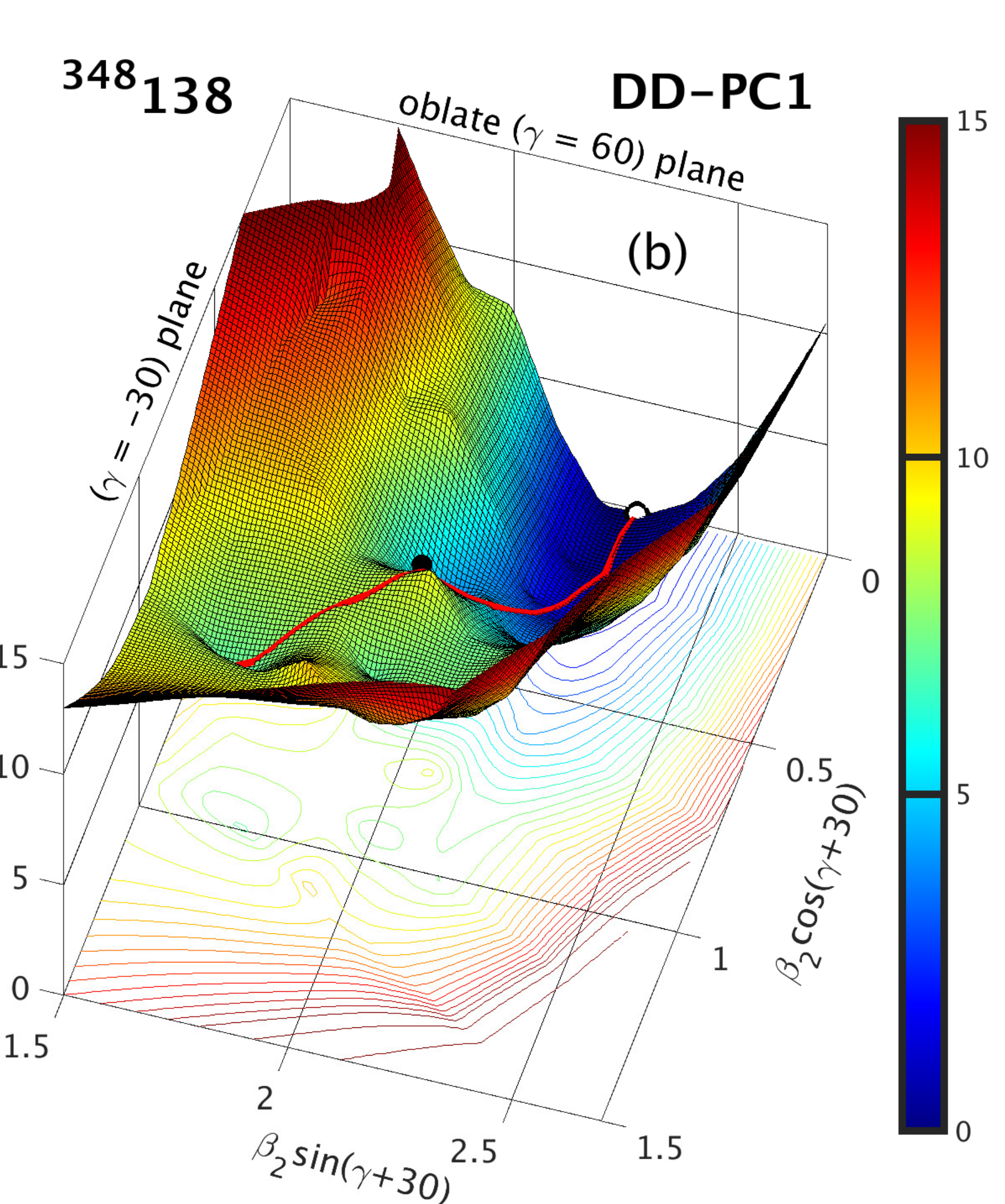}
\caption{(Color online) Three-dimensional potential energy surfaces with 
their two-dimensional projections (contour plots) for the solutions with 
minimum at $\beta_2 \sim 2.3, \beta_4 \sim +1.5, \gamma=60^{\circ}$
in indicated nuclei. Based on the results of the TRMF+BCS calculations
of Ref.\ \cite{AAG.18}. These solutions are excited ones in axial 
calculations, but they are the lowest in energy stable solutions in 
triaxial calculations. The red line shows static fission path from the 
minimum indicated by open white circle; the saddle point is shown 
by black solid circle. The energy difference between two neighboring 
equipotential lines in contour plot is 0.5 MeV.}
\label{fission-path}
\end{figure*}

\begin{figure*}[htb]
\includegraphics[angle=0,width=5.8cm]{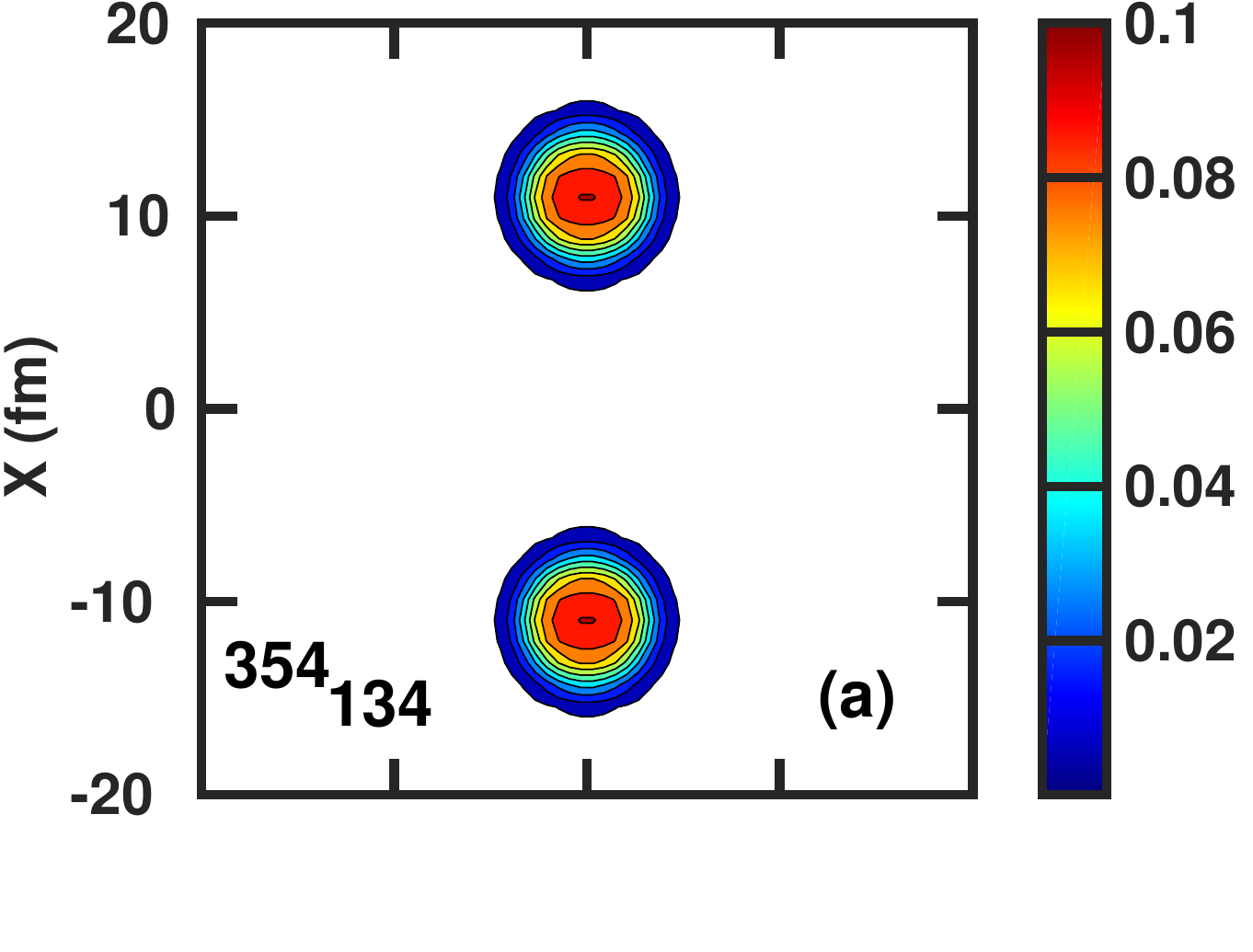}
\includegraphics[angle=0,width=5.8cm]{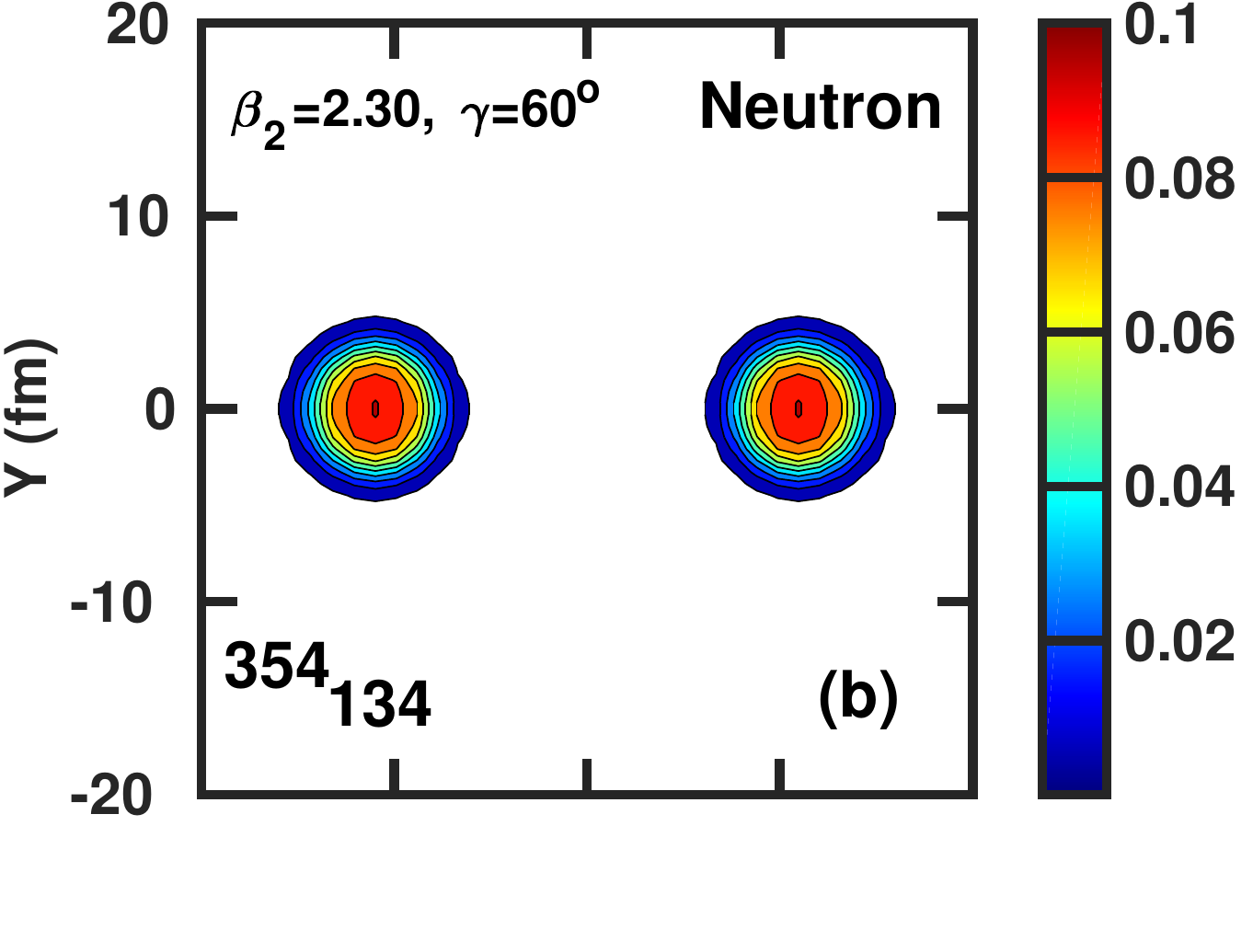}
\includegraphics[angle=0,width=5.8cm]{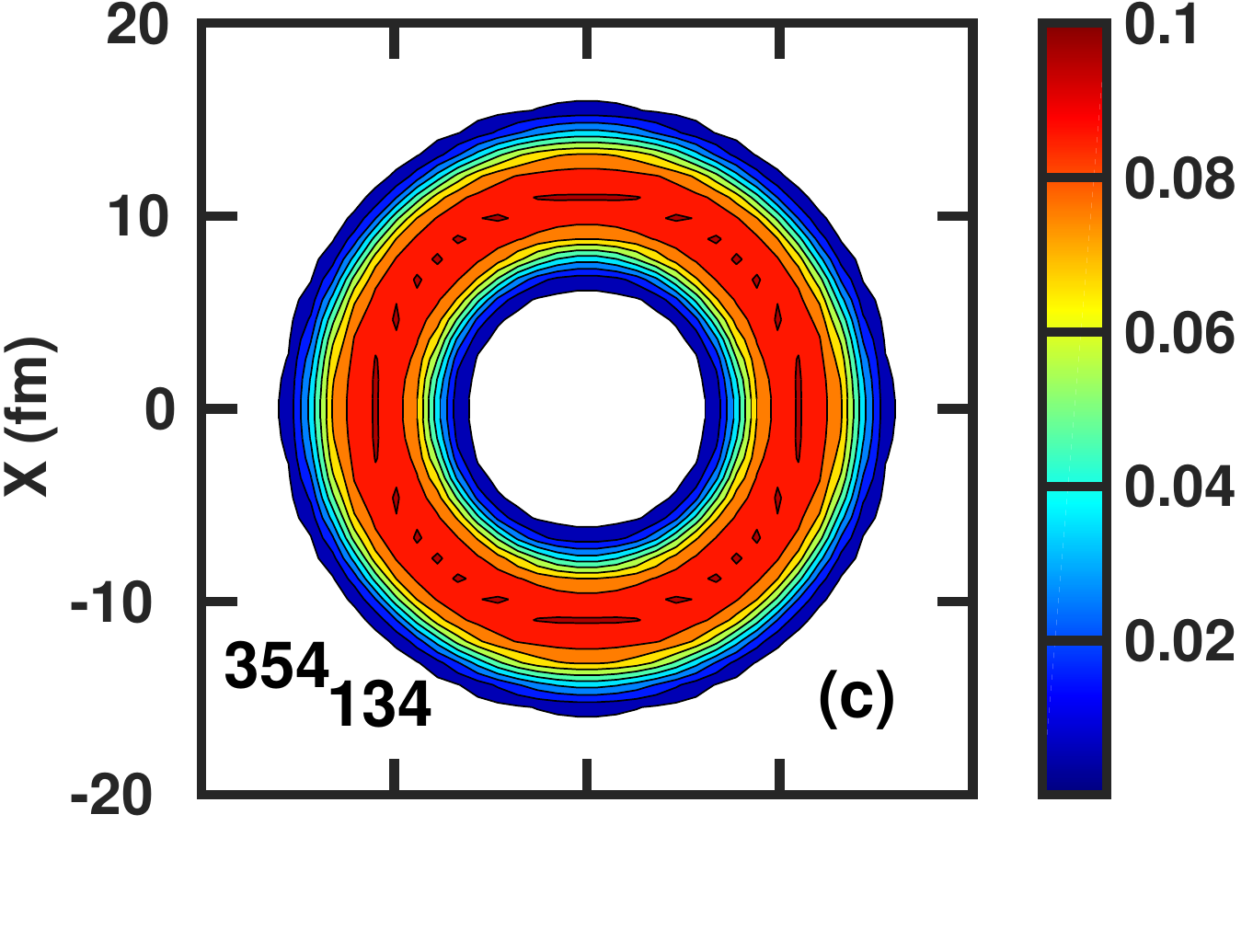}
\includegraphics[angle=0,width=5.8cm]{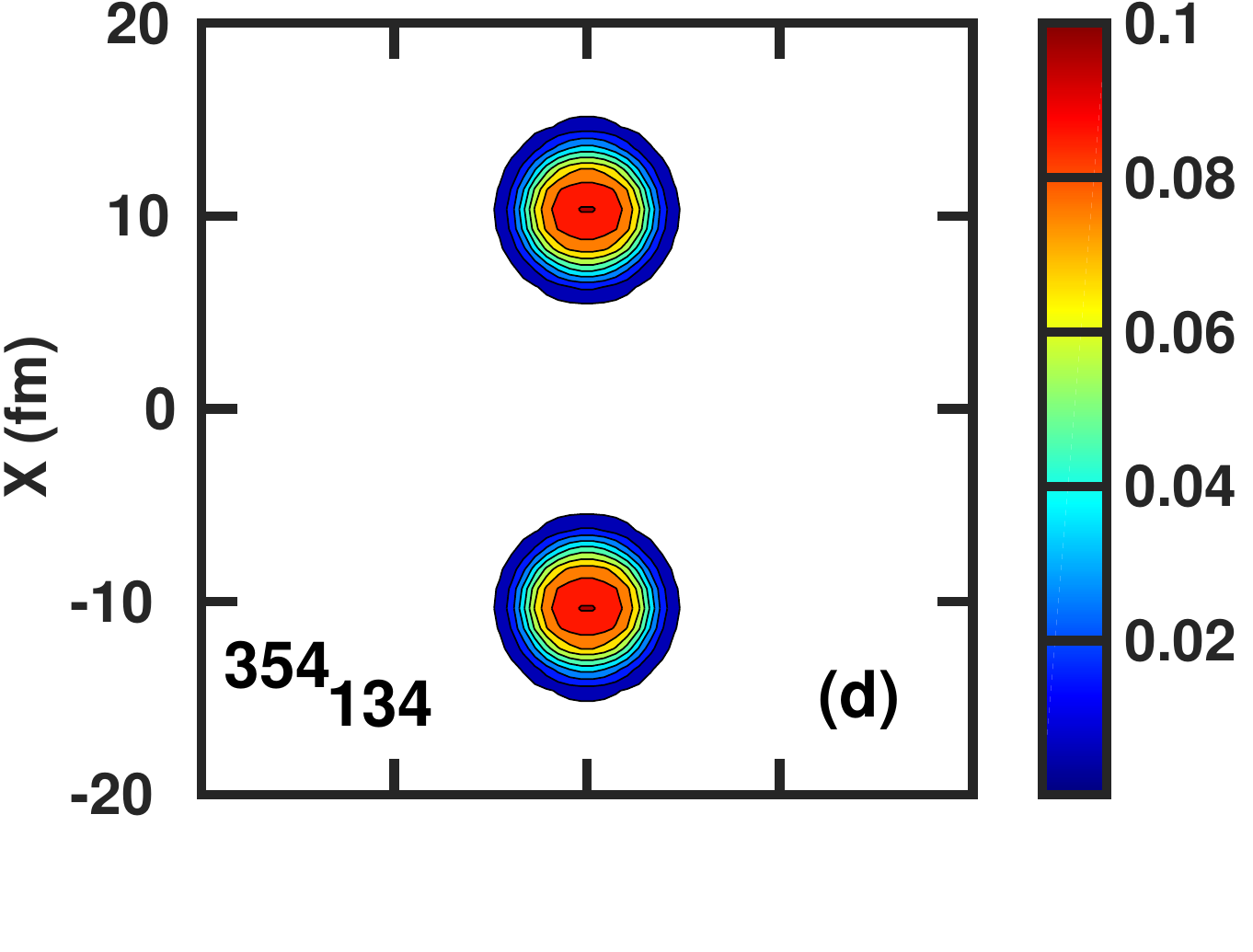}
\includegraphics[angle=0,width=5.8cm]{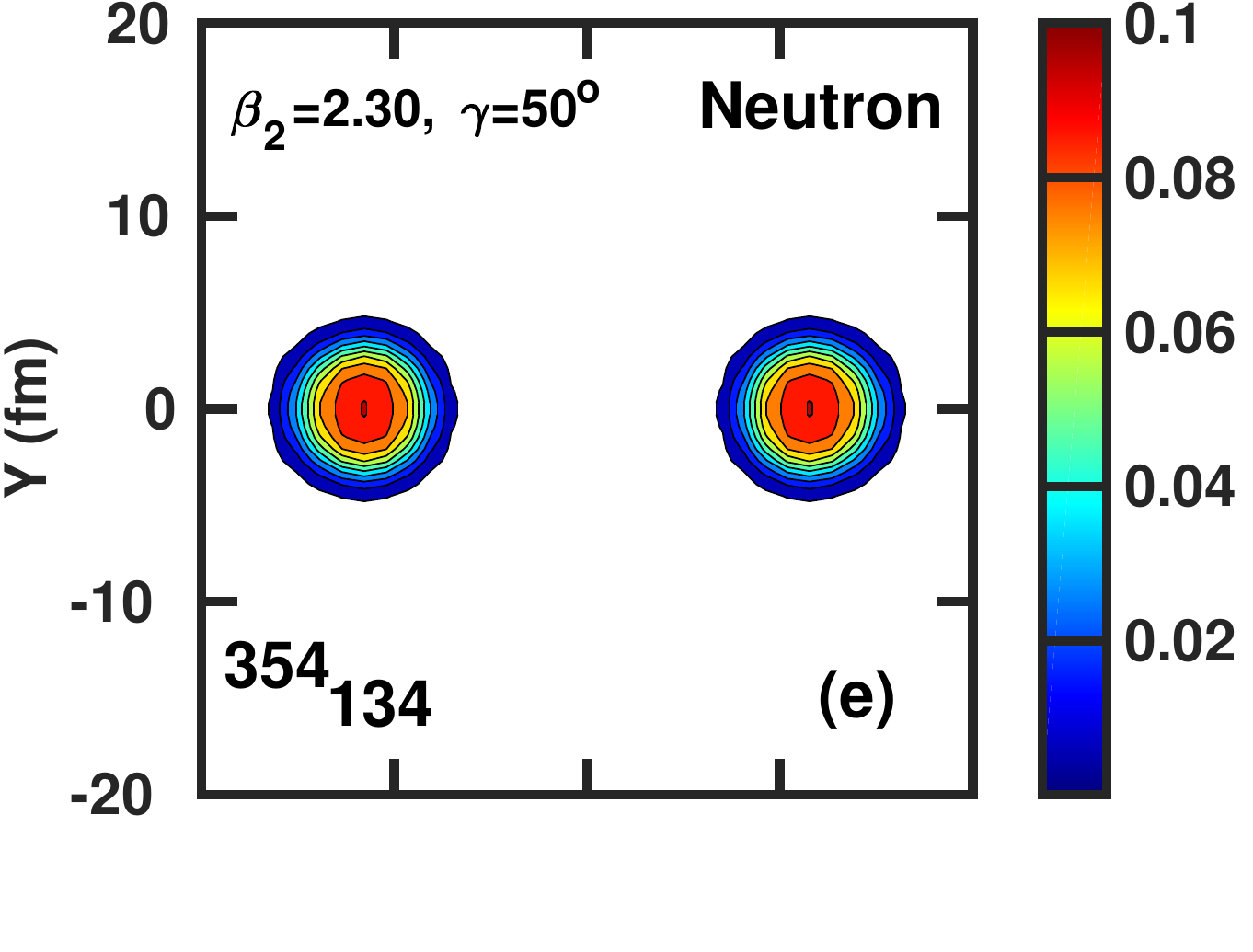}
\includegraphics[angle=0,width=5.8cm]{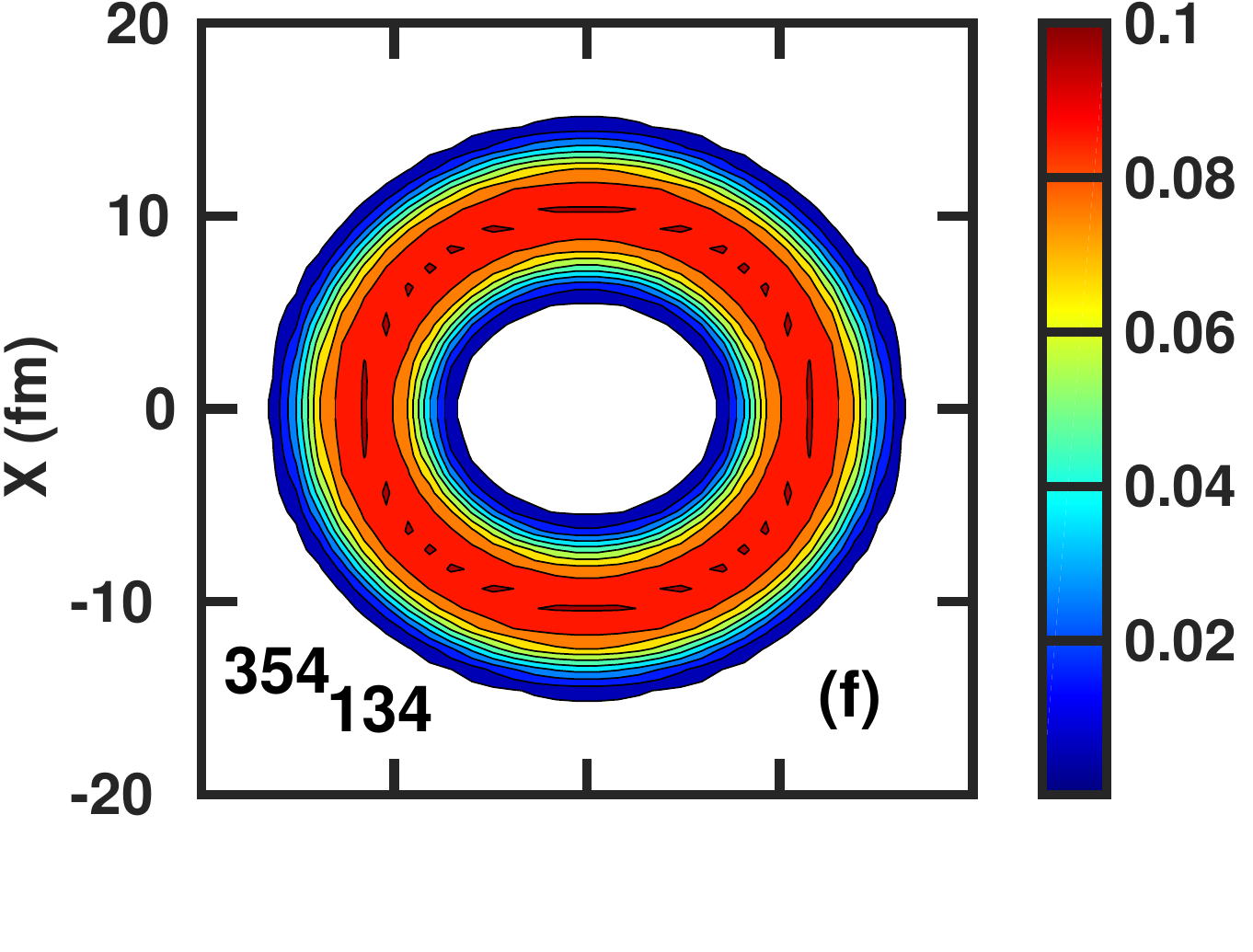}
\includegraphics[angle=0,width=5.8cm]{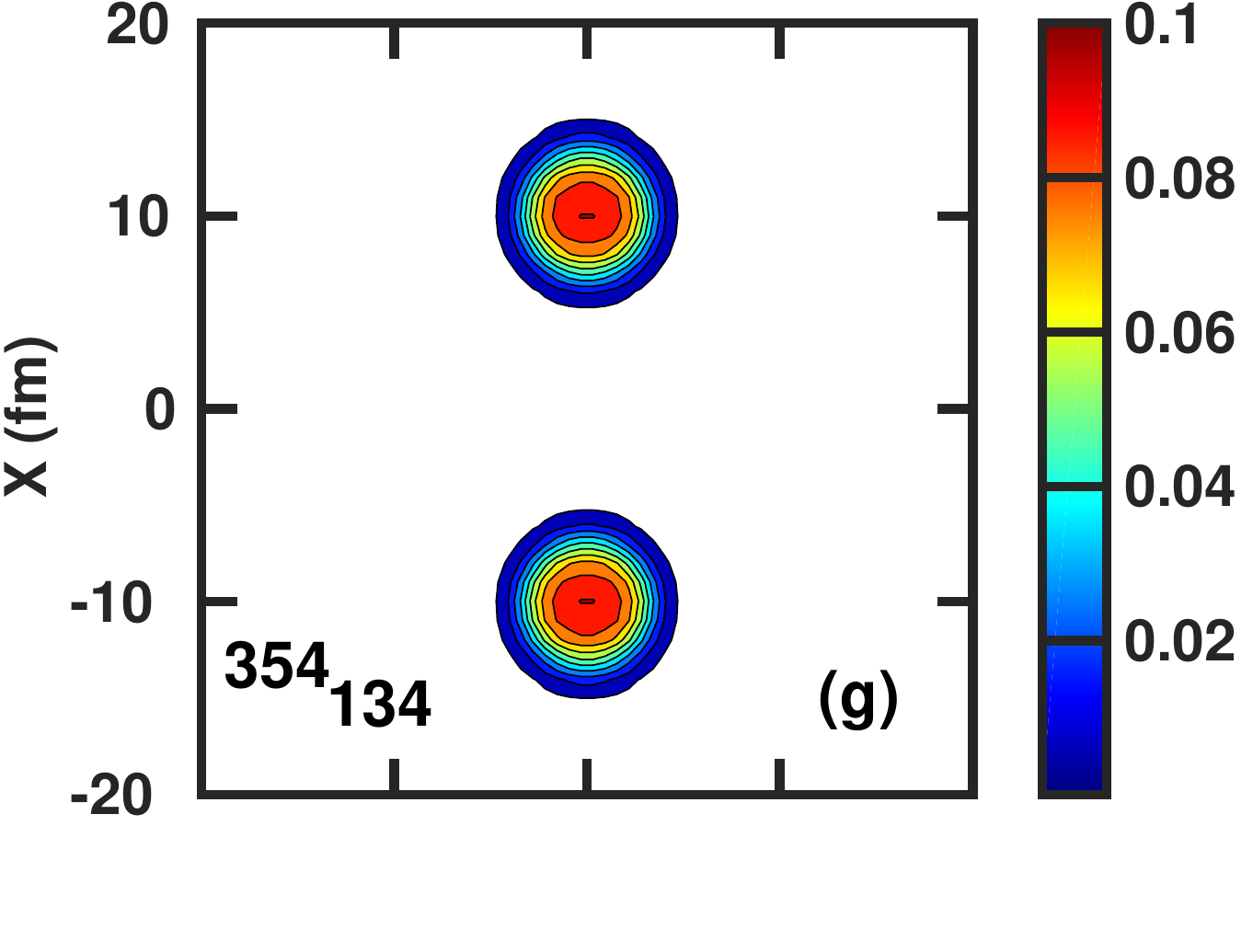}
\includegraphics[angle=0,width=5.8cm]{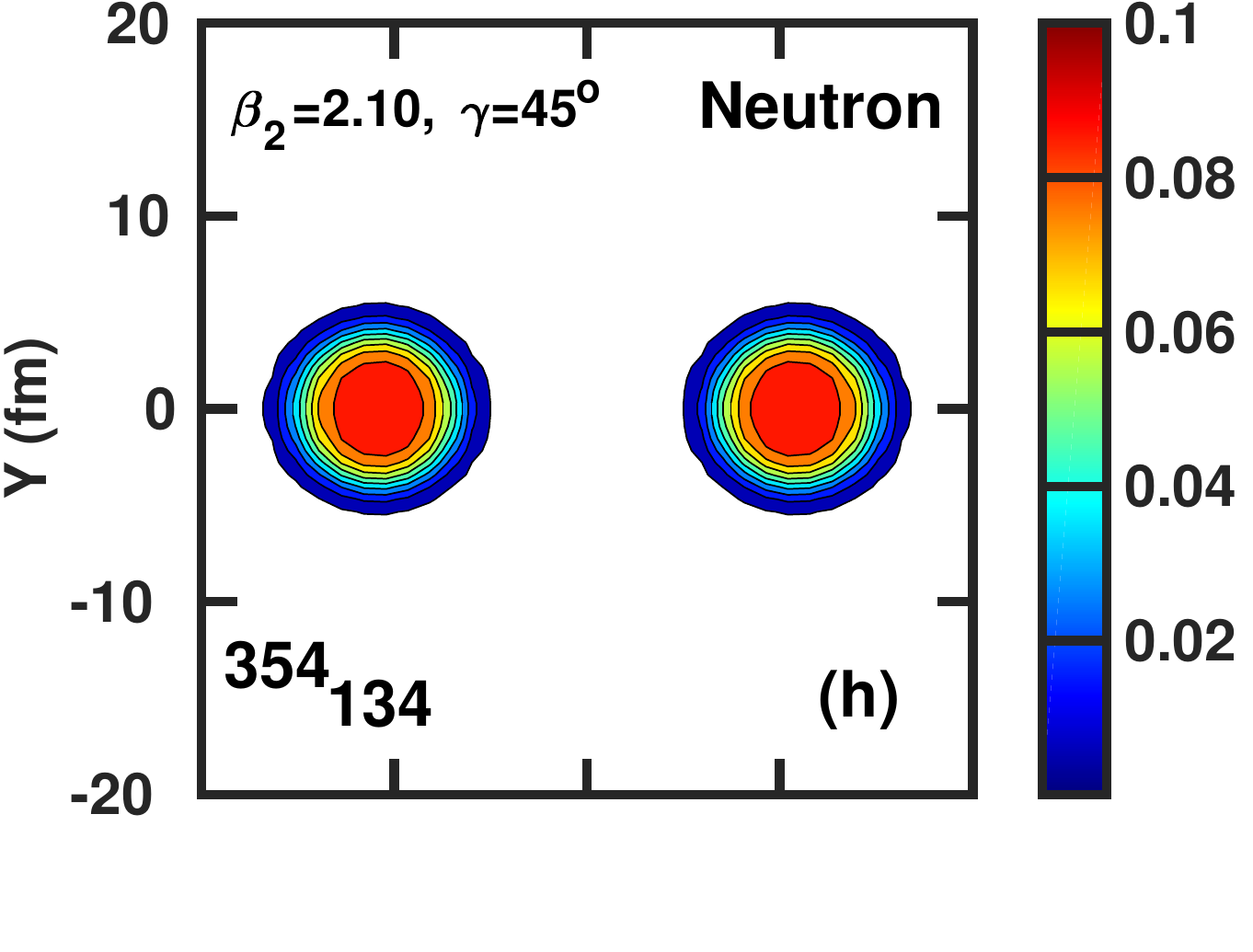}
\includegraphics[angle=0,width=5.8cm]{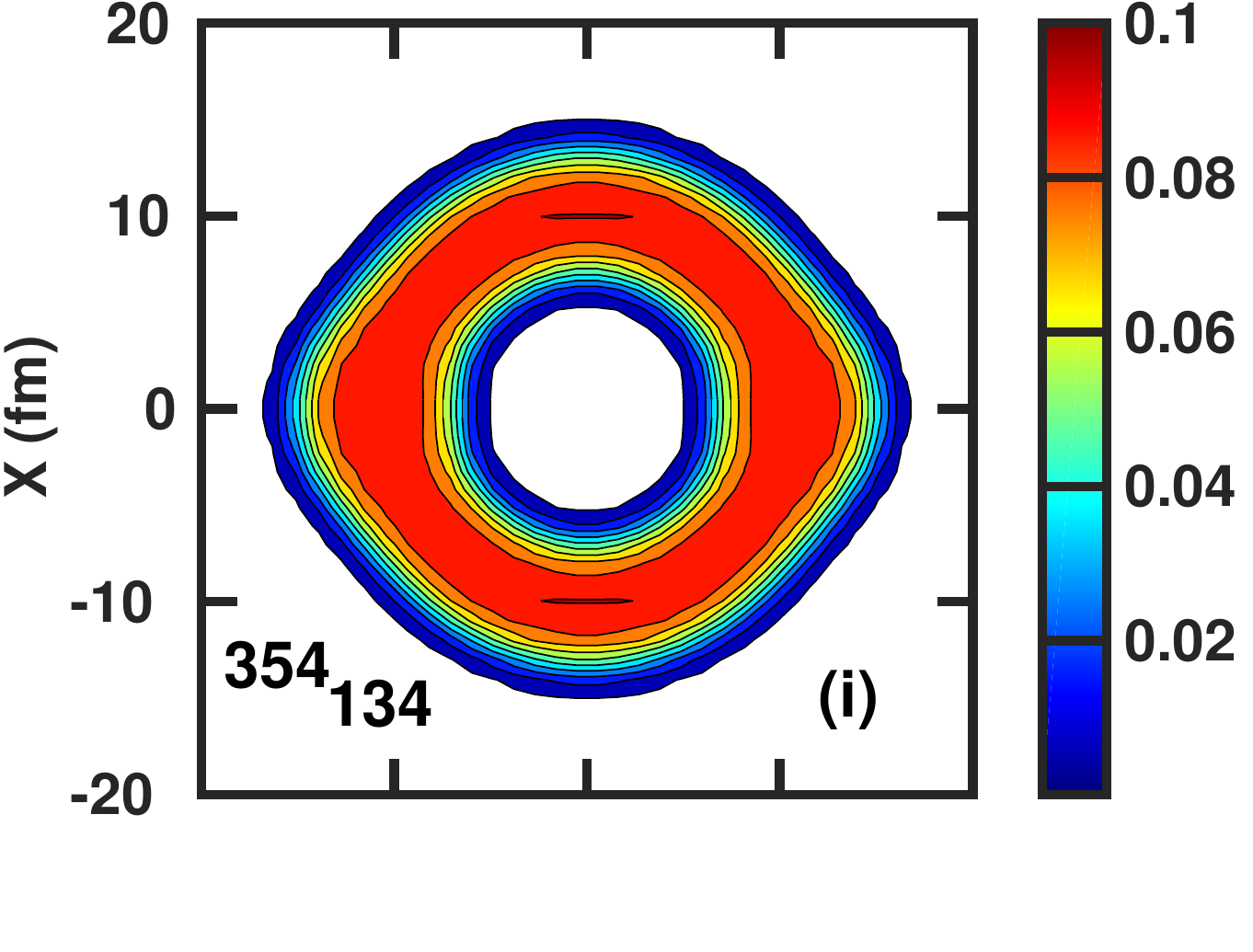}
\includegraphics[angle=0,width=5.8cm]{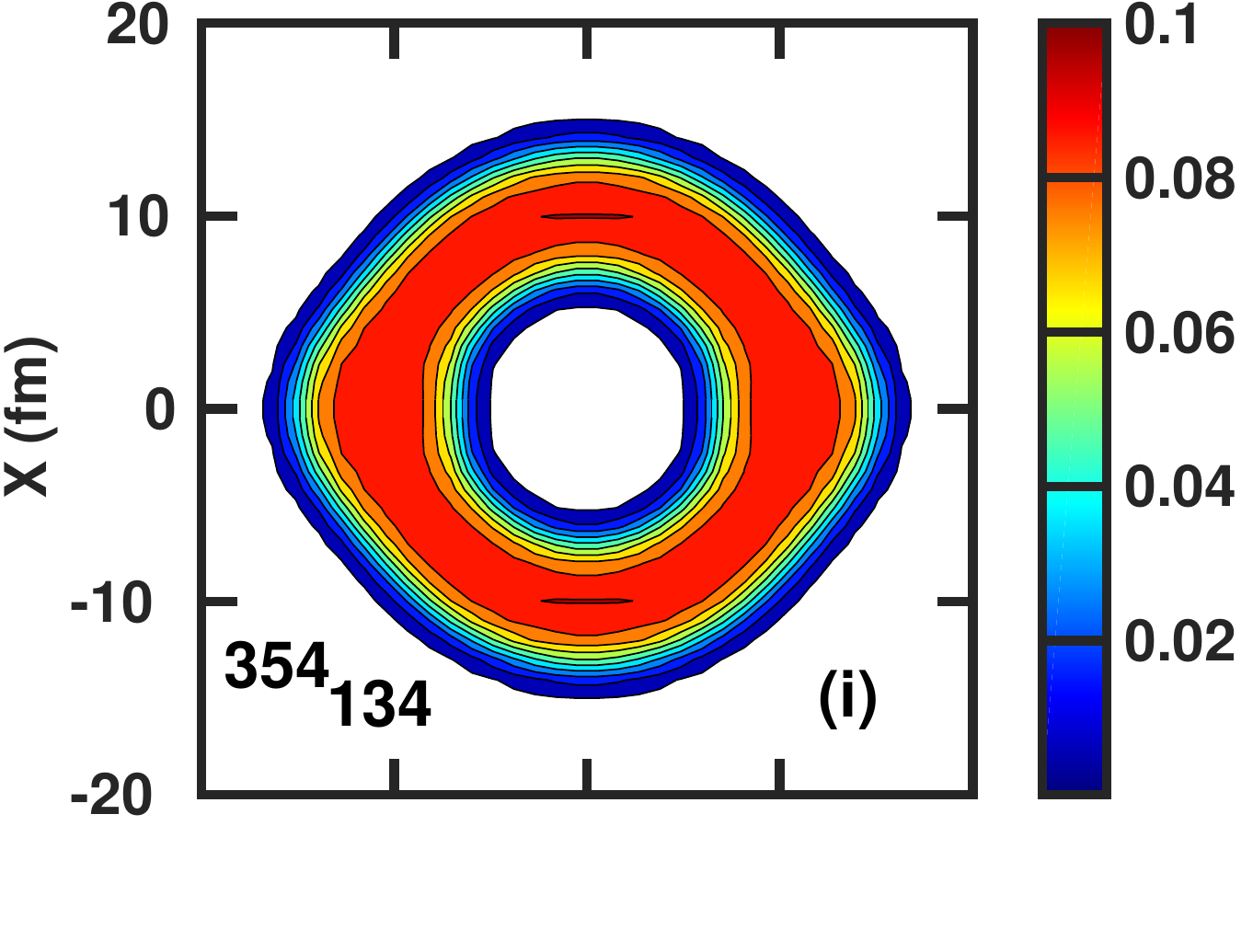}
\includegraphics[angle=0,width=5.8cm]{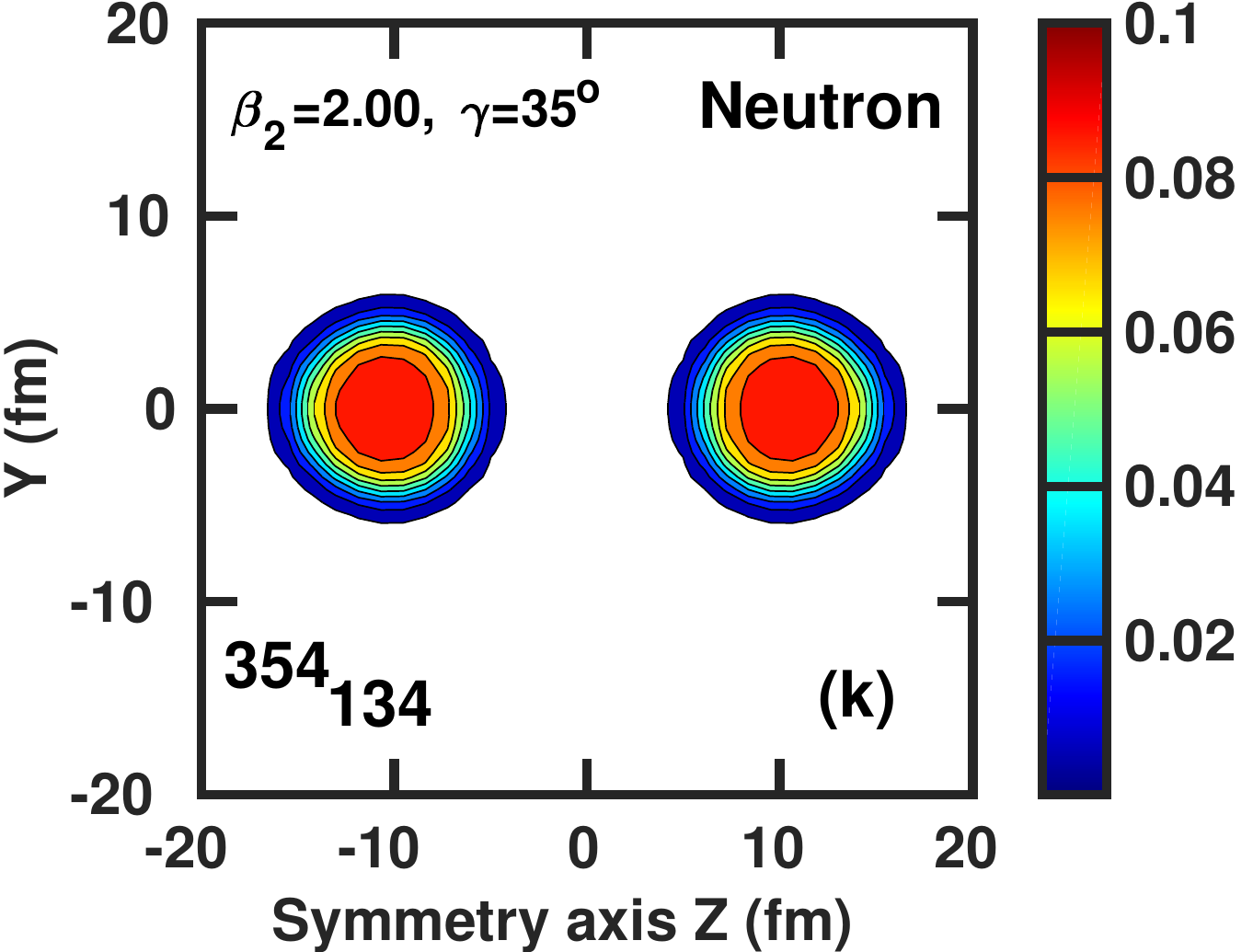}
\includegraphics[angle=0,width=5.8cm]{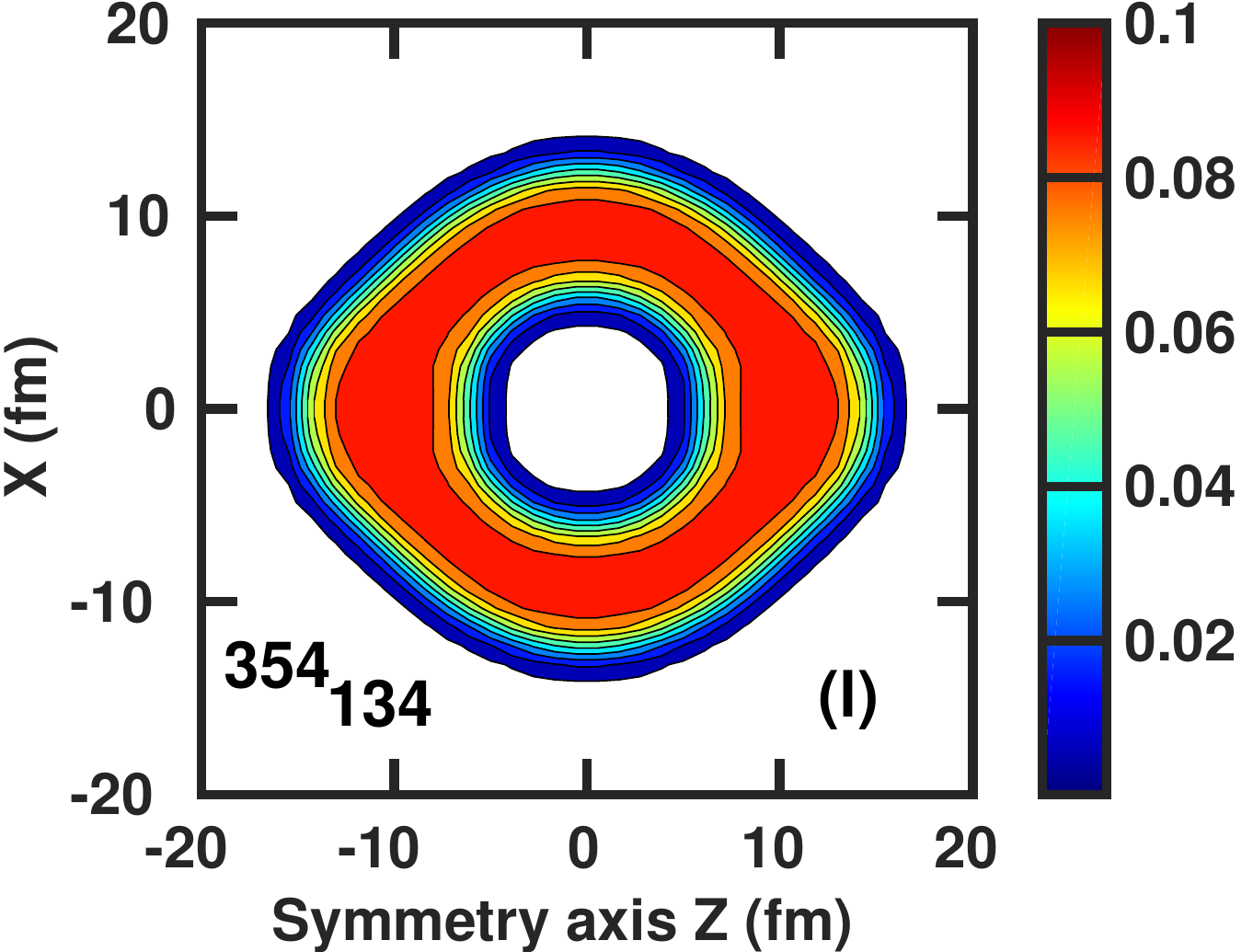}
\caption{(Color online)  The evolution of toroidal shapes along the fission path in the $^{354}$134 
              nucleus shown on left panel of Fig.\ \ref{fission-path}. Neutron density distributions 
              are shown at indicated ($\beta_2, \gamma$)-deformations along this fission  path.
              To give a full three-dimensional representation of the density distributions, they are 
              plotted in the $xy$, $yz$ and $xz$ planes at the positions of the Gauss-Hermite 
              integration points in the $z$, $x$ and $y$ directions closest to zero, respectively. 
              The density colormap starts at $\rho_n=0.005$ fm$^{-3}$ and shows the densities 
              in fm$^{-3}$.        
         }
         \label{shape-evol}
\end{figure*}

\begin{figure*}[htb]
\includegraphics[angle=0,width=8.5cm]{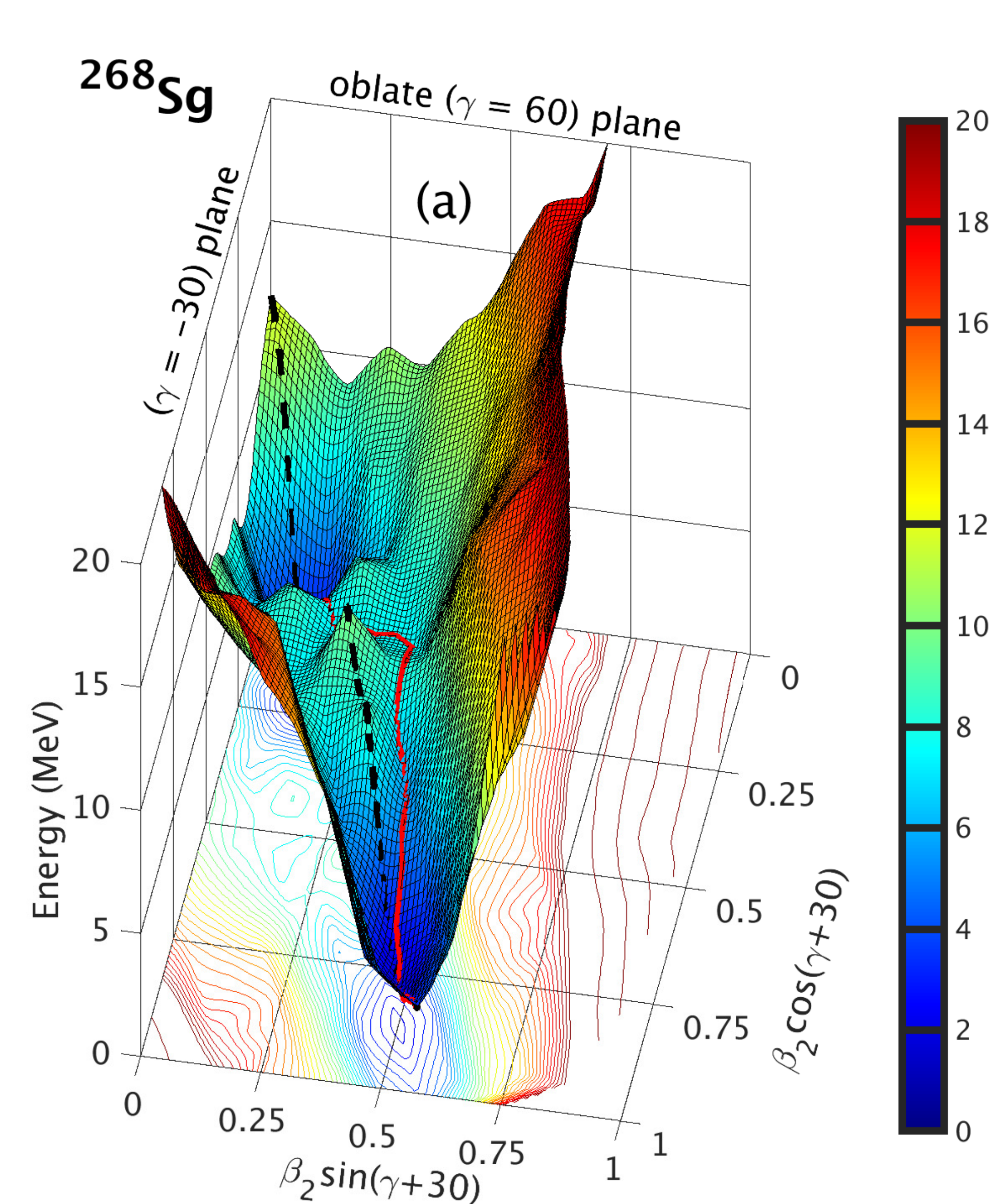}
\includegraphics[angle=0,width=8.5cm]{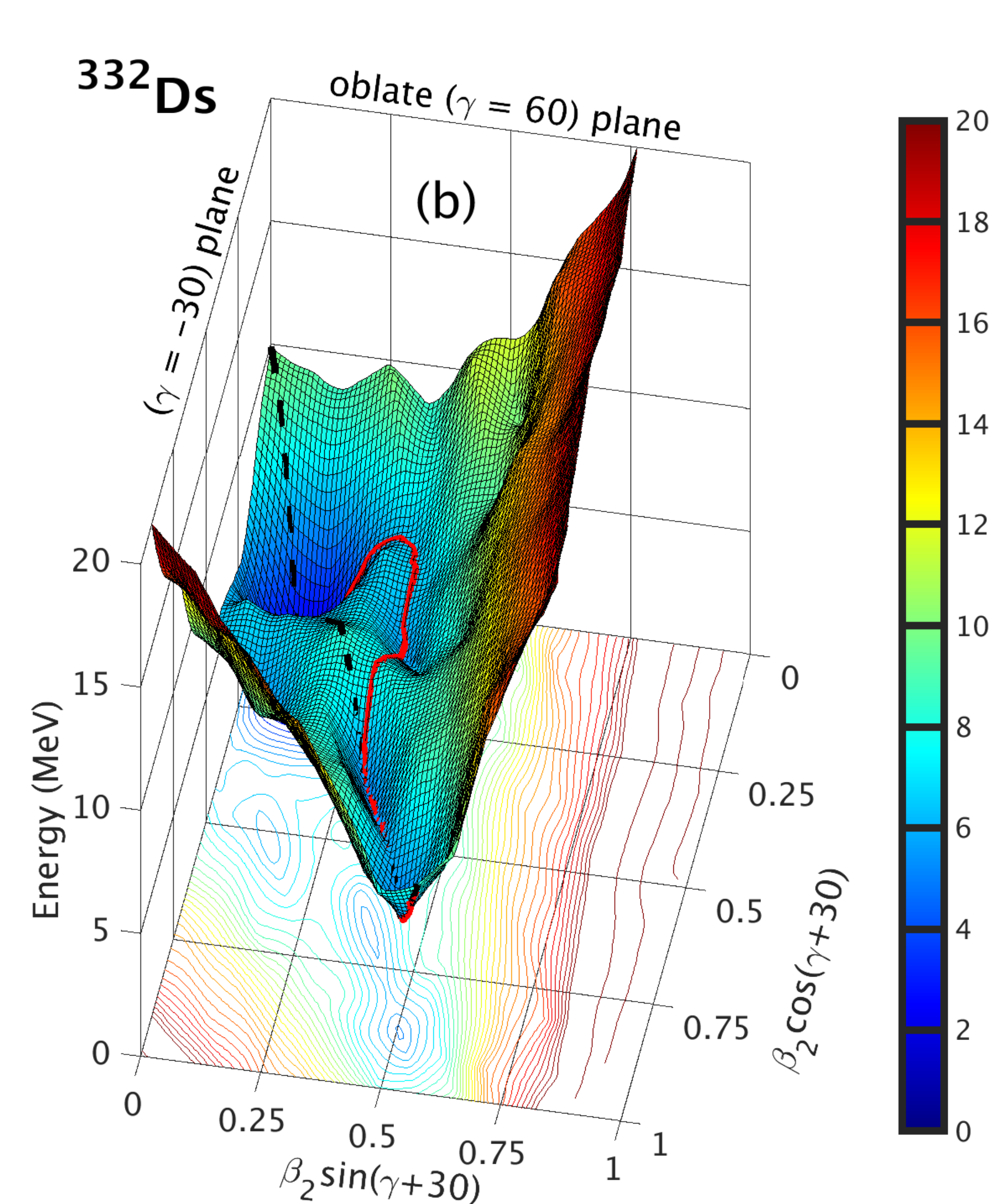}
\includegraphics[angle=0,width=8.5cm]{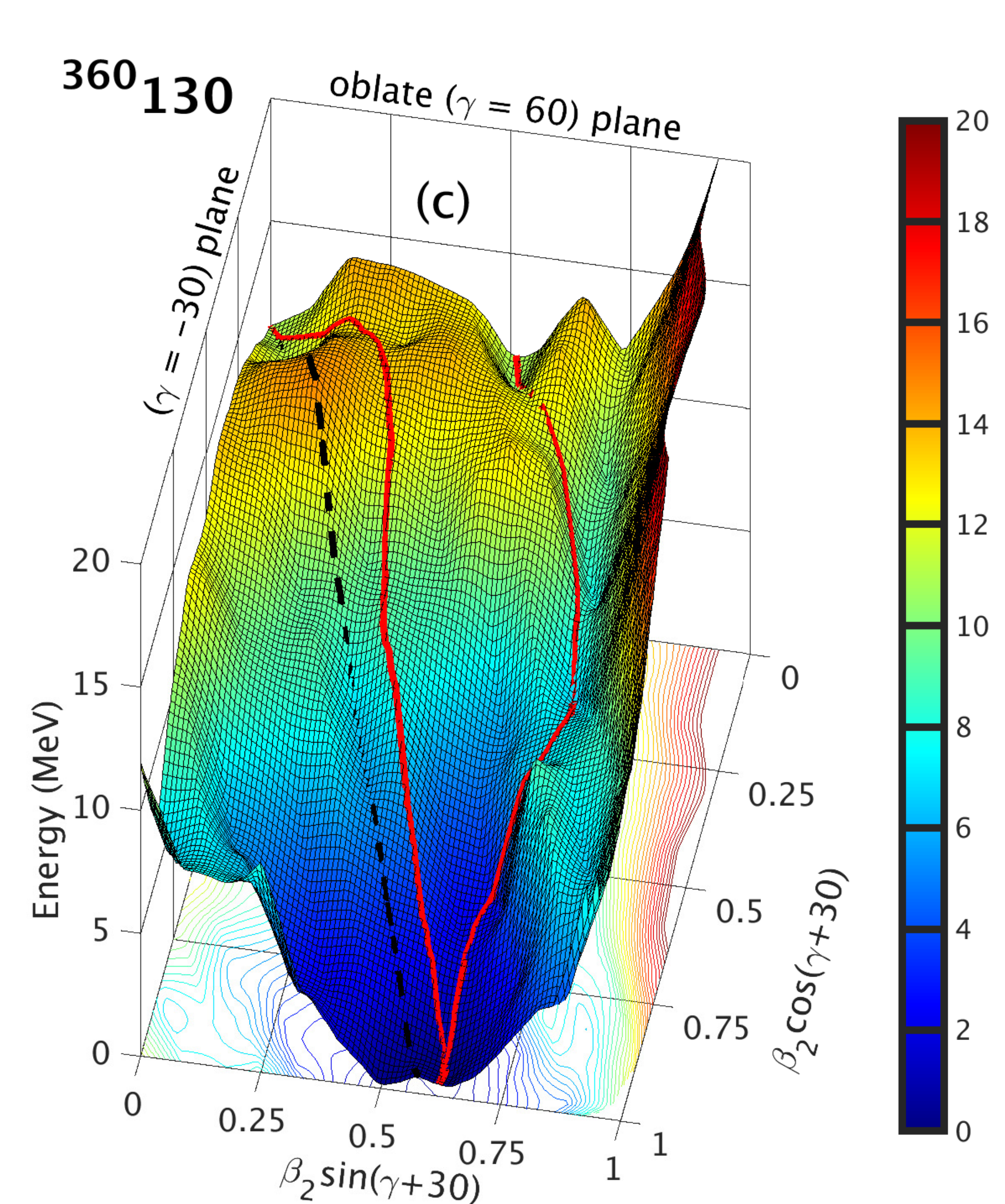}
\includegraphics[angle=0,width=8.5cm]{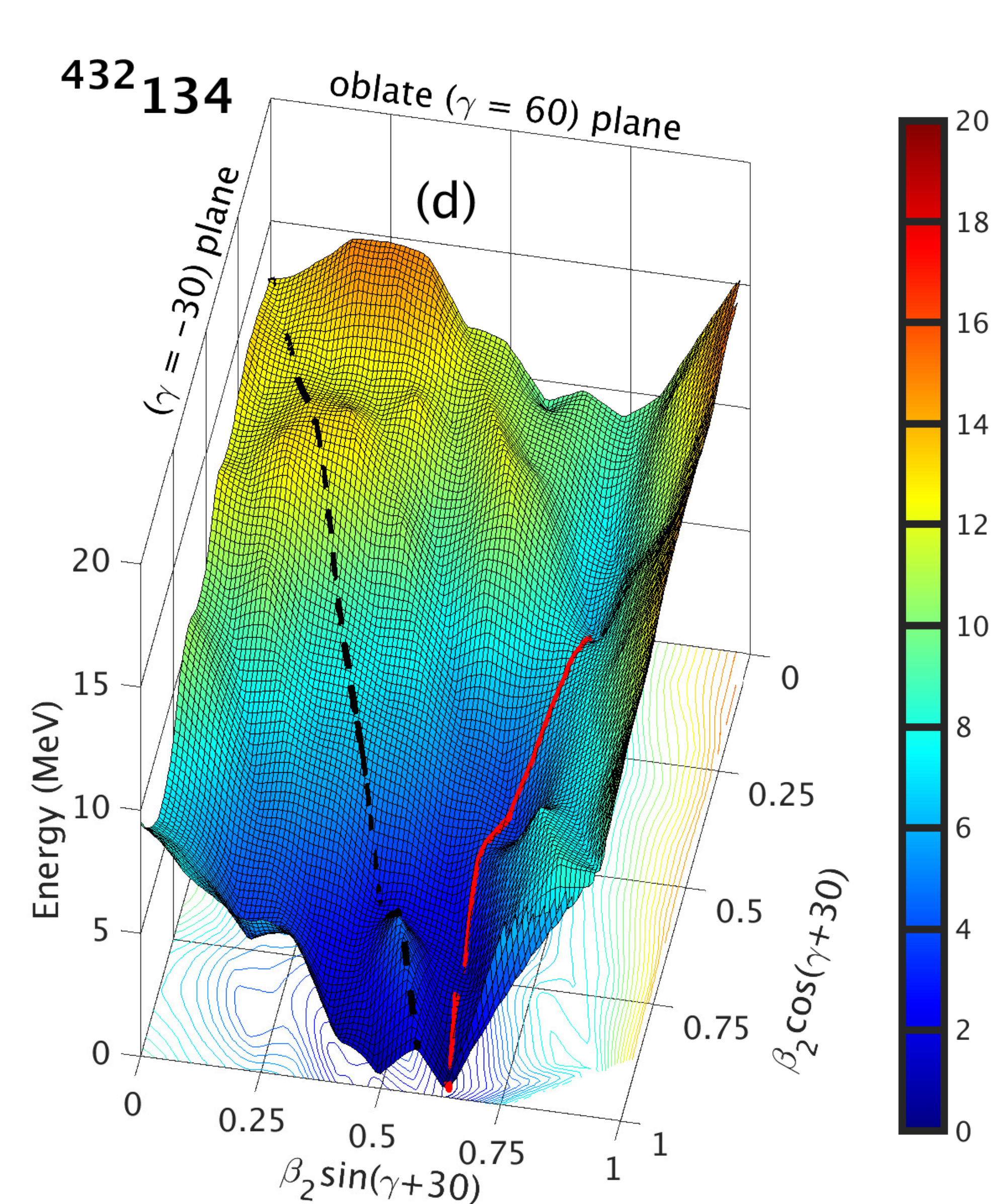}
\caption{(Color online) Three-dimensional potential energy surfaces 
with their two-dimensional projections (contour plots) for the nuclei 
with the ground states having ellipsoidal shape. They have been obtained 
in the TRHB calculations with $N_F=18$. The red line shows static 
fission path from respective  minimum, while black dashed line the 
$\gamma=0^{\circ}$ axis. The energy difference between two neighboring 
equipotential lines in contour plot is 0.5 MeV.
\label{PES_3D_landscape}
}
\end{figure*}

   It is interesting that central depression is more pronounced in the 
$^{292}120$ superheavy nucleus (with $\omega_{\pi} \sim \omega_{\nu} \sim 0.65$)
as compared with higher-$Z$ hyperheavy nuclei (see Fig.\ \ref{Density-sphere}); 
the only exception is proton subsystem of the  $^{584}$174 nucleus. The detailed 
analysis of Ref.\  \cite{AF.05-dep} strongly suggests that such central depression 
in the density distributions of the $^{292}120$ nucleus is mostly due to underlying 
shell structure: it emerges due to the occupation of specific high-/low-$j$ 
orbitals which place matter mostly in the surface/central region of the nucleus.  Indeed, the 
occupation of the neutron $3d_{5/2}$, $3d_{3/2}$ and $4s_{1/2}$ orbitals on going
from $N=172$ to $N=184$ (the $^{304}$120 nucleus) destroys this central
depression (see Fig.\ 2 in Ref.\ \cite{AF.05-dep}). However, on going to hyperheavy
nuclei the role of strong Coulomb force (which pushes the matter to surface
region) in creation of central depression in density distributions is expected to 
become dominant \cite{SNR.18}). The fact that the densities of the nuclei are 
similar within the regions of potentially stable spherical hyperheavy nuclei 
suggests reduced role of shell effects.

\begin{figure}[htb]
\includegraphics[angle=0,width=8.5cm]{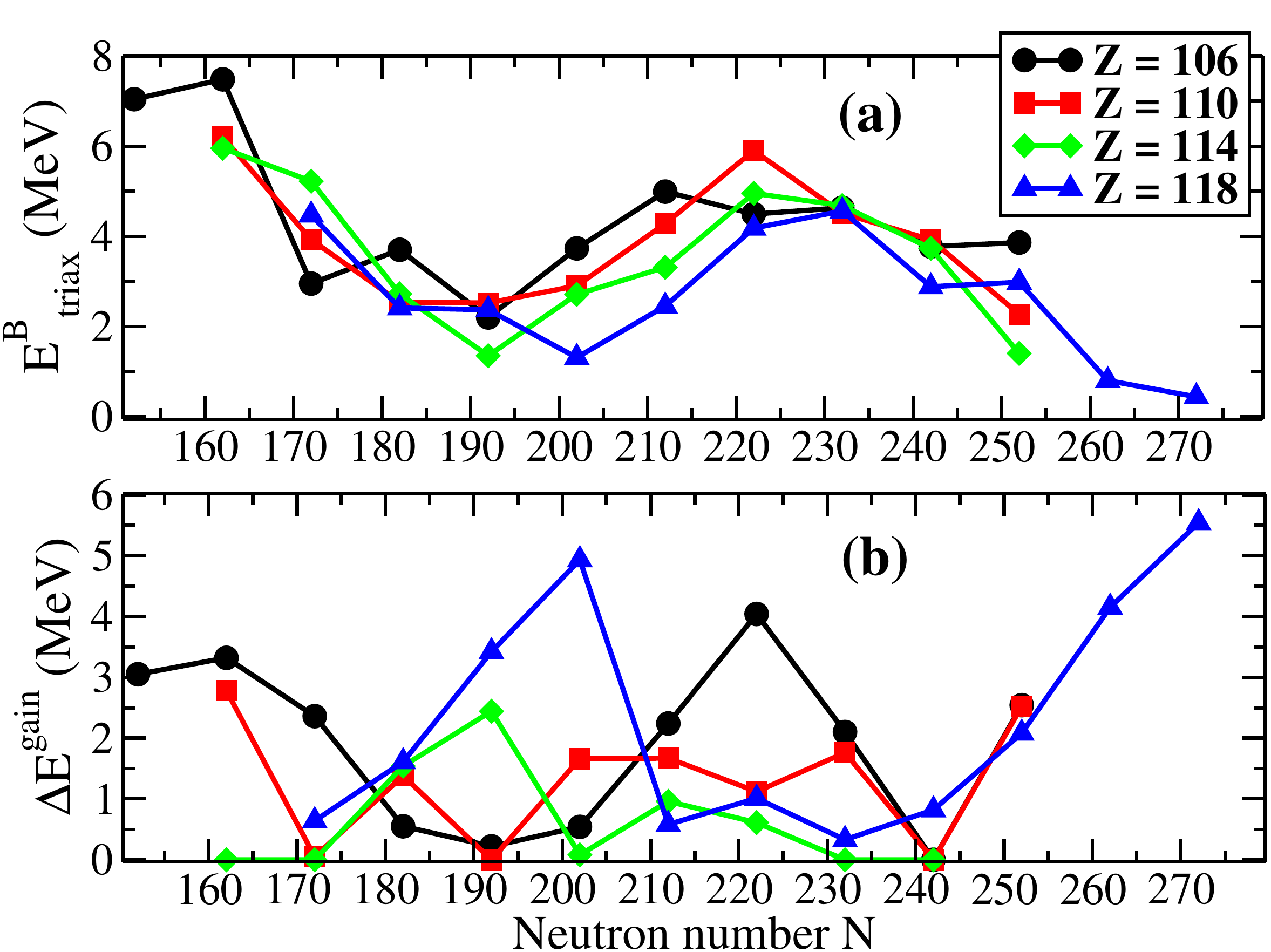}
\caption{(Color online) Inner fission barrier heights $E^B_{triax}$ obtained in the TRHB 
calculations (panel (a)) and the decrease of the fission barrier height due to
triaxiality $\Delta E ^{gain}$  (panel (b)) as a function of neutron number $N$.
\label{fission_barrier_TRHB} 
}
\end{figure}

\section{Shell closures in the islands of stability
         of spherical hyperheavy nuclei}
\label{shell_closure}

  The stability of spherical hyperheavy nuclei is  defined by underlying shell 
structure. However, in general the impact of shell gaps decreases with the 
increase of particle numbers (see discussion in Sect. III of Ref.\ \cite{AANR.15}.
Fig.\ \ref{sphere-sp} shows proton and neutron single-particle spectra 
of the $^{466}$156  nucleus at spherical shape. Proton $Z=154$ and neutron 
$N=308$ shell gaps with the sizes of approximately 2 MeV are clearly visible 
for all employed functionals in this figure. Based on these spectra it is tempting to 
call the $^{462}$154 nucleus as a doubly magic hyperheavy nucleus. However, 
the analysis of two-particle separation energies reveals more complicated situation.
The $N=308$ shell closure reveals itself via substantial drop of two-neutron 
separation energies at this particle number which exists for all proton numbers 
(see Fig.\  \ref{sphere-sep_energies}a). This drop is also visible in calculated
$\alpha$-decay half-lives (see Figs. 4 and 5 in supplemental material to
Ref.\ \cite{AAG.18}). However, the impact of the $Z=154$ proton
shell closure on two-proton separation energies is substantially
smaller (see Fig.\ \ref{sphere-sep_energies}b) and it almost does
not exist for the $N=308$ nuclei.

  Although we have not performed detailed analysis of the separation
energies for the DD-ME2, PCPK1 and NL3* functionals, selected results
for $\alpha$-decay half-lives presented in Fig. 6 of supplemental material
to Ref.\ \cite{AAG.18} allow to perform the comparison with the DD-PC1 
ones. Similar to DD-PC1 the $N=308$ shell gap is expected to be seen in
two-neutron separation energies also for the DD-ME2 functional but its 
impact is somewhat smaller as compared with DD-PC1. On the contrary, 
the NL3* and PC-PK1 results presented for the $Z=156$ and $Z=160$ isotopic 
chains in Fig. 6 of supplemental material to Ref.\ \cite{AAG.18} do not reveal 
the impact of the $N=308$ shell gap.  These observations suggest that 
in the center of the $Z\sim 156, N\sim 310$ island of stability of hyperheavy
nuclei total shell correction energies at spherical shape are more negative 
for the DD-PC1 and  DD-ME2 CEDFs as compared with the NL3* and PC-PK1
ones. This explains why fission barriers (and thus the size of the
islands of stability) [see Fig. 6 in Ref.\ \cite{AAG.18}] are larger for the DD-PC1 
and  DD-ME2 CEDFs as compared with the NL3* and PC-PK1 ones.

 Figs.\  \ref{sphere-sp-138}  and fig.\  \ref{sphere-sp-174}  show single-particle 
 spectra for the $^{366}$138 and $^{580}$174 nuclei located in the vicinity of central 
 nuclei of two regions of spherical
 hyperheavy nuclei. Although the $N=228$ and $N=406$ neutron gaps with the size
 of approximately 2 MeV are seen in these figures, there are no substantial proton
 gaps at respective particle numbers. Similar to Fig.\ \ref{sphere-sep_energies}, these 
 neutron gaps are seen in two-neutron separation energies, but two-proton separation 
 energies are quite smooth as a function of proton number and do not reveal proton
 gaps. 
 
  The features observed for proton subsystems of the nuclei under discussion together 
with clear localization of the islands of stability of spherical hyperheavy nuclei in the 
$(Z,N)$ plane strongly suggests that the shell  effects at deformed shapes leading to 
negative shell correction energies at some deformation and thus to fission barriers  play
also an important role in the stabilization of spherical hyperheavy nuclei in discussed 
regions.

   Some of discussed shell gaps appear as the gaps between the
members of the spin-orbit doublets. These are the $N=308$ shell gap 
between the $2i_{13/2}$ and $2i_{11/2}$ orbitals in the $^{466}$156 
nucleus (Fig.\ \ref{sphere-sp}b), the $N=228$ shell gap between the 
$2h_{11/2}$ and $2h_{9/2}$ orbitals in the $^{366}$138 nucleus 
(Fig.\ \ref{sphere-sp-138}b) and the $N=406$ shell gap between 
the $2j_{15/2}$ and $2j_{13/2}$ orbitals in the $^{580}$174 nucleus 
(Fig.\ \ref{sphere-sp-174}b). The energy splitting between the spin-orbit 
partner orbitals depends on the profile of the density distribution in the surface 
region (see discussion in Ref.\ \cite{BRRMG.99}). Indeed, the above mentioned 
gaps are similar in the DD-PC1/DD-ME2 and NL3*/PCPK1 pairs of the functionals 
reflecting the similarities and differences in their density distributions (see 
discussion in Sec.\ \ref{sect-density}). Note that the gaps obtained in the  
DD-PC1/DD-ME2 functionals are larger than those seen in the NL3*/PC-PK1 
ones.

\begin{table*}[htb]
\begin{center}
\caption{The heights of the fission barriers along the fission 
paths from different minima obtained in axial and triaxial RHB
calculations. The columns $3-5$ show the results of the axial
RHB calculations. Here  $\beta_{min}$, $\beta_{saddle}$ and $E_{ax}^B$ 
are the equilibrium quadrupole  deformation of the global 
minimum, the quadrupole deformation and the energy
of the saddle along respective fission path.  The results of the  
triaxial RHB calculations are provided in the columns 
$6-8$. Note that the allowance of triaxial deformation could shift
the position of the local minimum in the deformation plane
and in absolute majority of the cases shifts the positions 
of the saddle points.  Thus,  $(\beta,\gamma)_{min}$, 
$(\beta,\gamma)_{saddle}$ and $E_{triax}^{B}$ show the deformations
of the minima, the deformations of saddle points and their
energies obtained in triaxial RHB calculations. 
The neutron numbers of the nuclei in which superdeformed minimum
with $\beta_2 \sim 0.5$ is lower than normal-deformed or spherical ones
are marked by an asterisk.  With exception of these nuclei, the values 
in parentheses show either the deformation of superdeformed minimum 
or the deformation of the saddle of outer fission barrier or the height of  
outer fission barrier. Note that these values are shown only when the 
lowest height of the outer fission barrier obtained in the TRHB and RHB+OCT 
calculations is higher than 2 MeV. 
The column 3 shows the $\Delta E^{gain} = E_{ax}^{B} - E_{triax}^{B}$ quantity which is the 
decrease of the height of respective fission barrier due to triaxiality.
\label{table-fission}}
\begin{tabular}{|c|c|c|c|c|c|c|c|c|} \hline \multicolumn{2}{|c|}{  } & \multicolumn{3}{c|}{Axial RHB} & \multicolumn{4}{c|}{Triaxial RHB} \\ \hline
 $Z$   &  $N$  & $\beta_{min}$  & $\beta_{saddle}$ & $E_{ax}^{B}$ & $(\beta,\gamma)_{min}$ & $(\beta,\gamma)_{saddle}$ &  $E_{triax}^{B}$ & $\Delta E^{gain}$  \\ \hline
          1   &   2   &      3        &         4   &       5        &         6             &          7               &       8  & 9                     \\ \hline
106 (Sg) &  152  &      0.29    &      0.57   &     10.09      &      0.29, 0          &           0.62, 12.4      &   7.04 &   3.05   \\ \hline
              &  162  &      0.26    &      0.65   &     10.70      &      0.26, 0          &           0.68, 8.4      &   7.48   &   3.32  \\ \hline
              &  172  &      0.14    &      0.69   &      5.31      &      0.14, 0          &           0.71, 3.5      &   2.95    & 2.36   \\ \hline
              &  182  &     -0.05 (0.49)   & 0.27 (0.73)  & 4.25 (3.42)     &      0.05, 60.0 (0.49, 0) & 0.47, 23.6 (0.81, 8.0)     &   3.70 (2.47) & 0.55 (0.95) \\ \hline
             &  192  &      0.39    &      0.59   &      2.42      &      0.40, 0          &           0.61, 6.3      &   2.20     &   0.22  \\  \hline
             &  202  &      0.28    &      0.59   &      4.27      &      0.29, 9.5        &           0.66, 13.2      &   3.73    &  0.54    \\ \hline
             &  212  &      0.25    &      0.54   &      7.23      &      0.25, 0          &           0.69, 10.3      &   4.99   &   2.24    \\  \hline
             &  222  &      0.25    &      0.55   &      8.53      &      0.250, 0          &           0.70, 8.7      &   4.49    &   4.04  \\ \hline    
            &  232  &      0.23    &      0.65   &      6.73      &      0.23, 0          &           0.62, 10.4      &   4.63    &  2.10  \\  \hline
            &  242 &      0.13  (0.45)   & 0.25 (0.65)  & 3.77 (5.08)     &      0.13, 0  (0.45, 0)   & 0.25, 0 (0.60, 6.9)     &   3.77 (2.37) &  0.0 (2.71)\\ \hline            
           &  252  &     -0.06  (0.45)  & 0.25 (0.69)  & 6.40 (5.68)     &      0.06, 59.1 (0.45, 0)  & 0.42, 26.7 (0.69,10.3)  &   3.86 (3.20) & 2.54 (2.48)  \\  \hline
110 (Ds) &  162  &      0.24    &      0.66   &      8.98      &      0.242,0          &           0.65, 6.9      &   6.20    &   2.78   \\ \hline
              &  172  &      0.15  (0.46)    & 0.30 (0.70)  & 3.97 (5.75)     &      0.15, 0  (0.45, 0)    & 0.30, 0.0 (0.74, 5.6)        &   3.92 (3.24) &   0.05 (2.51) \\ \hline
              &  182  &   -0.14      &    0.26     &   3.92         &    0.139, 60          &  0.36, 35.6    &  2.54   &  1.38        \\ \hline             
              &  192  &      0.41    &      0.58   &      2.52      &      0.417,0          &        0.62, 5.8         &   2.52   & 0.0       \\ \hline
              &  202  &      0.38    &      0.56   &      4.56      &      0.385,0          &        0.73, 14.4         &   2.90   &   1.66     \\ \hline
              &  212  &      0.26    &      0.51   &      5.95      &      0.262,0          &        0.54, 15.7         &   4.28    &    1.67    \\   \hline
              &  222  &      0.24    &      0.54   &      7.02      &      0.243,0          &        0.36, 28.5         &   5.90   &   1.12      \\ \hline
              &  232  &      0.22    &      0.60   &      6.27      &      0.220,0          &        0.61, 7.0         &   4.51   &    1.76     \\ \hline
              &  242  &      0.14  (0.47)   & 0.27 (0.66)  & 3.92 (5.06)     &      0.17, 0   (0.46, 0)        & 0.27, 0 (0.68, 6.0)        &   3.92 (2.51)  &   0.0 (2.55)   \\ \hline
              &  252  &      0.44    &      0.70   &      4.79      &      0.444,0          &        0.72, 11.6         &   2.27  &     2.52      \\ \hline
114 (Fl)  &  162  &  0.23 (0.52)        & 0.40 (0.66)  & 5.95 (6.00)     &          0.23, 0  (0.54, 0)    &      0.40, 0 (0.57, 5.1)   &    5.95 (3.85)  & 0.0 (2.15) \\ \hline
              &  172  &  0.15  (0.50)      & 0.32 (0.73)  & 5.22 (4.76)     &          0.15, 0  (0.51, 0)     &   0.32,0 (0.73, 6.0)   &    5.22 (3.33)  & 0.0 (1.43) \\ \hline
               &  182  &  -0.14        & 0.26         & 4.26            &          0.14, 60       &    0.41, 38.5   &   2.72     &    1.54 \\ \hline
              &  192  & -0.38        & 0.15        & 3.79           &          0.40, 42     &      0.42, 33.4           &    1.35   &   2.44  \\ \hline
              &  202  &  0.38        & 0.54        & 2.79           &          0.38, 0      &      0.54, 3.7           &    2.71   &    0.08 \\ \hline
              &  212  &  0.27        & 0.49        & 4.27           &          0.28, 0      &      0.40, 23.1           &    3.31  &    0.96 \\ \hline
              &  222  &  0.24        & 0.41        & 5.56           &          0.24, 0      &      0.36, 26.3           &    4.95  &    0.61  \\ \hline
              &  232  &  0.21        & 0.35        & 4.68           &          0.21, 0      &      0.35, 0               &    4.68   &  0.0  \\  \hline             
              &  242  &  0.14  & 0.27  & 3.73     &          0.142,0    &      0.27,0   &    3.73 & 0.0 \\ \hline
              &  252*  &  0.458 (0.0)  &  0.72 (0.23)  & 3.31 (4.68)  &    0.459,0 (0.0, 0)   &   0.63, 8.0 (0.46, 23.9)  &  2.27 (1.40)   &  1.04 (3.28)      \\ \hline
118 (Og)  &  172  &  0.0        & 0.34        & 5.11           &          0.000,0      &      0.49, 28.2           &    4.47  &   0.64  \\ \hline  
               &  182*  &  0.57 (-0.27)     &  0.84 (0.26)       &  4.02 (5.32)          &          0.58,0 (0.27, 60)      &      0.66, 9.5 (0.42, 36.4)      &    2.41 (3.57)  &  1.61 (1.75)   \\ \hline
               &  192  & -0.39        & 0.15        & 5.79           &          0.40, 74     &      0.33, 43.3           &    2.37   &   3.42   \\ \hline
               &  202  & -0.43        & 0.07        & 6.24           &          0.43, 60     &      0.46, 45.0           &    1.31   &   4.93   \\ \hline
               &  212  &  0.29        & 0.44        & 3.03           &          0.30, 11     &      0.64, 21.3           &    2.45   &   0.58   \\   \hline
              &  222  &  0.24        & 0.39        & 5.19           &          0.24, 0      &      0.37, 27.2           &    4.18    &   1.01   \\ \hline
              &  232  &  0.22        & 0.35        & 4.89           &          0.22, 0      &      0.35, 0              &    4.56   &   0.33    \\ \hline
              &  242  & -0.20        & 0.31        &  3.70          &          0.21, 60.0   &      0.47, 32.4           &    2.88  &  0.82      \\ \hline
              &  252  & -0.19        & 0.19        & 5.06           &          0.20, 60     &      0.41, 38.5           &    2.98   &   2.08   \\ \hline
              &  262  & -0.23        & 0.15        & 4.95           &          0.24, 59.9   &      0.32, 41.6           &    0.80   &   4.15    \\ \hline
              &  272  & -0.49        &-0.01        & 5.98           &          0.49, 59.6   &      0.48, 51.7           &    0.44    &   5.54    \\ \hline
\end{tabular}
\end{center}
\end{table*}

\section{The stability of spherical hyperheavy nuclei with respect of
         octupole and triaxial distortions}
\label{stability_spherical}

  Fig. 6 in Ref.\ \cite{AAG.18} summarizes the heights of the fission 
barriers for the nuclei forming the islands of stability of spherical 
hyperheavy nuclei.  They represent the lowest in energy barriers 
amongst those obtained on oblate and prolate sides of spherical 
minimum in axial RHB calculations. However, one should investigate 
the stability of spherical minimum with respect of octupole and triaxial 
distortions to have a full understanding of the situation.  Such studies 
are very time-consuming and thus we present their results obtained with 
DD-PC1 CEDF only for the nuclei located in the centers of the islands 
of stability of spherical hyperheavy nuclei.

 Fig.\ \ref{oct-pes-spher-nuclei} shows the potential energy surfaces of 
 these nuclei in the $(\beta_2, \beta_3)$ plane. One can see that in the 
 $^{368}$138 and $^{584}$174 nuclei, the barriers on the oblate and prolate 
 sides have the lowest values at $\beta_3=0$. The same is true also for the 
 barrier on the prolate side of spherical minimum in the $^{466}$156 nucleus.
 However, potential energy surface is soft in octupole deformation in 
 the region of the barrier located on the oblate side of spherical 
 minimum in this nucleus. Thus, the saddle of the fission barrier
 is shifted from $\beta_2=-0.196, \beta_3=0.0$ (as obtained in axial
 RHB calculations without octupole deformation) to  
 $\beta_2=-0.198, \beta_3=0.091$  when octupole deformation is included
 in the calculations. This also leads to the decrease of the height of the
 barrier on the oblate side from 10.81 MeV down to  9.83 MeV. 
  However, this decrease has very little impact on the total stability 
 of this nucleus with respect of octupole deformation since resulting
 value of 9.83 MeV is only 120 keV lower than the height 
 $E_B=9.95$ MeV  of the barrier on the prolate side.  Note that 
 the barrier on the prolate side is the lowest one in the axial RHB
 calculations without octupole deformation. Thus, one can conclude 
 that above discussed nuclei are relatively stable with respect of 
 octupole distortions.  

  It is also necessary to mention that potential energy surfaces shown in Fig. \ 
\ref{oct-pes-spher-nuclei} do not suggest that cluster radioactivity from spherical 
$^{368}$138 and $^{466}156$ hyperheavy nuclei plays an important role. Super-asymmetric 
fission channel leading to a cluster radioactivity shows itself as a narrow fission path 
in the $(\beta_2,\beta_3)$ plane separate from main fission path (see example in Fig. 1 
of Ref.\ \cite{WZR.18}). No such path is visible in Figs.\ \ref{oct-pes-spher-nuclei}a 
and b. However, because of convergence problems one cannot define whether
cluster radioactivity is important in the $^{584}$174 nucleus.

 Fig.\ \ref{triaxial-pes} shows the potential energy surfaces obtained
in the TRHB calculations. In the $^{584}$174 nucleus, the axial saddles are located at 
$\beta_2 = 0.17, \gamma = 60^{\circ}$ (the barrier on the oblate side of spherical
minimum) and at  $\beta_2 = 0.197, \gamma = 0^{\circ}$ (the barrier on the prolate 
side of spherical minimum). Their heights are 6.389 and 7.709 MeV, respectively.  The 
potential energy surface is somewhat soft in triaxial deformation so that the saddle of 
fission barrier is shifted to $\beta_2=0.16, \gamma =36^{\circ}$ when the triaxiality is included 
in the calculations. However, the impact of triaxiality on the height of fission barrier is 
rather modest leading to its decrease (as compared with the lowest axial barrier located 
on oblate side) by only 0.26 MeV. Similar features are also seen in the $^{368}$138 and 
$^{466}$156 nuclei. The reduction in the height of fission barrier is 0.71 and 0.58 MeV 
in these nuclei.  However, as compared with the heights of 5.88 and 10.81 MeV of the 
lowest axial barrier  (located at the oblate side  of spherical minumum), these are relatively 
modest reductions which do not decrease the stability of nuclei in a substantial way. 
 
  These features could be understood in the following way. The topology 
of potential  energy surfaces of the nuclei under study are similar to those 
of volcanos.  The central area around spherical minimum is similar to caldera, the rim
of which is represented by the fission barrier.  The area beyond the rim
(fission barrier) is fast down-sloping as a function of quadrupole deformation
$\beta_2$. The saddles of axial fission barriers (on oblate and prolate sides
of spherical minimum) are located at modest quadrupole deformation of 
$\beta_2 \sim 0.2$. As a result, the distance between these two saddles
in the $(\beta_2, \gamma)$ plane is relatively small, so that large
changes in binding  energy due to triaxiality for nearly constant $\beta_2$
values could not develop.  As a
consequence, the lowest fission barrier around spherical minimum obtained in 
axial RHB calculations is a  good approximation to the barrier obtained
in the TRHB calculations. The TRHB results discussed here clearly indicate
that spherical minima of the nuclei under study are relatively stable with respect of 
triaxial distortions.

\section{Pairing interaction in hyperheavy nuclei}
\label{sect-pairing}

The magnitude of pairing interaction could be accessed via two calculated quantities: average pairing gap and pairing energy  \cite{RS.80}.  There are several definitions of average pairing gaps in literature (see discussion in Section IV  of 
Ref.\ \cite{AARR.14}). The pairing gap
\begin{equation}
\Delta_{\rm uv}=\frac{\sum_k u_kv_k\Delta_k}{\sum_k u_kv_k} ,
\end{equation}
which is related to the average of the state dependent gaps over the pairing 
tensor, is used in the present study. The analysis of Ref.\ \cite{AARR.14}
showed that the $\Delta_{\rm uv}$ gap is a better measure of pairing
correlations as compared with other definitions for average pairing gap.

 The pairing energy in the RHB calculations  is defined via
\begin{eqnarray}
E_{pairing} = -\frac{1}{2} Tr (\Delta \kappa)   
\end{eqnarray}   
where $\Delta$ and $\kappa$ are pairing field and pairing tensor, respectively 
\cite{RS.80}.  Note that $E_{pairing}$ mixes particle-particle and particle-hole
channels of the model. As a consequence, its absolute value is typically by an 
order of magnitude larger than the gain in binding due to pairing correlations
which is equal to the difference of binding energies obtained in the calculations 
with and without pairing correlations (see Refs.\ \cite{CRHB,AER.00}).

  Fig.\ \ref{Pairing-axial} shows the evolution of neutron and proton pairing 
energies $E_{pairing}$ and pairing gaps $\Delta_{uv}$ as a function of
$\beta_2$ for the lowest in energy solution in the $^{466}$156 nucleus.
The oscillating behavior of these quantities as a function of $\beta_2$
is due to the change of the density of the single-particle states in the 
vicinity of the Fermi level with deformation (see detailed discussion in 
Ref.\ \cite{KALR.10}). In regions of high/low level density it is easier/more 
difficult for the quasi-particles to spread  around the Fermi surface and 
therefore the size of the pairing correlations depends strongly on the 
level density. As a consequence,  the low/high values of the $|E_{pairing}|$ and
$\Delta_{uv}$ corresponds to low/high density of the single-particle states
in the vicinity of the Fermi level 
(see Ref.\ \cite{KALR.10}). Indeed, in the nucleus under consideration 
there is a substantial reduction of neutron $|E_{pairing}|$ and $\Delta_{uv}$ 
values near spherical shape which is attributable to the presence of large 
$N=308$ shell gap (see Fig.\ \ref{sphere-sp}). Note also that there is a collapse 
of neutron pairing correlations at toroidal shapes in the $\beta_2$ range
from $-4.2$ up to $-3.95$. This range corresponds to the minimum of the 
deformation energy curve at toroidal shapes  and its vicinity (see Fig.\ 
\ref{trunc_basis}d). The local minima in potential energy surfaces correspond 
to low density of the single-particle states in the vicinity of the Fermi level \cite{Strutinsky1967_NPA95-420,Strutinsky1968_NPA122-1}; at above quoted deformations this density is low enough to trigger the collapse of pairing 
correlations in the RHB calculations. However, this collapse of pairing in neutron subsystem is not critical since the treatment of pairing by more sophisticated 
methods, which preserve exact particle number, would only lead to moderate 
additional binding due to pairing at quoted above deformations. 
 
  Particle number dependencies of proton and neutron pairing energies at 
spherical shape of the nuclei forming the $(Z\sim 156, N\sim 310)$ island of 
stability of spherical  hyperheavy nuclei are shown in Fig.\ \ref{Pairing-spherical}.
Apart of the $N=308$ isotopes, the absolute values of neutron pairing energies 
$|E_{pairing}|$ are larger than 4 MeV. In the $N=308$ isotones, they are
smaller than 4 MeV and there is a collapse of neutron pairing due to large 
size of the $N=308$ shell gap in the nuclei with $Z=158-174$.
The treatment of pairing by the methods which include exact particle 
number projection (see, for example, Ref.\ \cite{AER.02}) will restore the pairing 
correlations in these nuclei.  However, in no way it will affect the conclusions
of the present work and of Ref.\ \cite{AAG.18}. In reality, it is expected that
such methods most likely will increase the fission barriers around spherical
minimum because of additional binding due to pairing at spherical shape. 
Proton pairing energies are displayed in Fig.\ \ref{Pairing-spherical}b; only 
in the $Z=154, N=292-298$ nuclei their absolute values are below 4 MeV.
However,  there is no proton pairing collapse in any of the nuclei shown in 
this figure. In other nuclei, proton pairing energies are quite large with 
$|E_{pairing}| > 6.0$ MeV and for the $Z\geq 168$ nuclei  the $|E_{pairing}|$ 
values exceed 20 MeV.

  Figure \ref{Pairing-spherical} shows the evolution of proton and neutron
pairing energies as a function of the $\beta_2$ and $\gamma$ deformations
in the $^{466}$156 nucleus. Similar to the calculations in axially symmetric 
case, there are substantial oscillations of the pairing energies as a function of 
deformation parameters which are due to underlying changes in the 
single-particle level density in the vicinity of the Fermi level. However, the 
topology of  these oscillations is more complex because of the presence of 
two deformation parameters. Note  that no pairing collapse is observed at 
any deformation point covered by these TRHB calculations.

\section{The impact of octupole deformation on the stability
of prolate superdeformed minima}
\label{sect-oblate}

 It is well known fact that outer fission barriers exist in superheavy 
nuclei for many CEDFs when the calculations are restricted to axial 
reflection symmetric shapes \cite{BBM.04,AAR.12}.  This is illustrated in 
Fig.\ \ref{def-energy-curves} which shows the competition in energy of 
two local minima, namely, prolate superdeformed minimum with $\beta_2 \sim 0.6$ 
and oblate one with $\beta_2 \sim -0.5$. Although the prolate superdeformed minima 
are the lowest in energy, their stability depends on the properties of 
outer fission barriers. It turns out that in absolute majority of the $Z>120$ super- and 
hyperheavy nuclei such local minima either do not exist or the heights of outer fission 
barrier are less than 2 MeV. The latter is not sufficient for the 
stabilization of prolate superdeformed  minimum (see discussion in Ref.\ \cite{AAR.12}). 
Note also that such local prolate minima do not exist in the results of 
axial reflection symmetric RHB calculations for hyperheavy nuclei with 
$Z > 140$ (see examples shown in Fig. 1 of Ref.\ \cite{AAG.18}).

 For the cases in which the heights of outer fission barriers are higher 
than 2 MeV in axial reflection symmetric RHB calculations (as those 
shown in Fig.\ \ref{def-energy-curves}), we have performed the calculations 
in axial octupole deformed code RHB-OCT developed in Ref.\ \cite{AAR.16}. 
The results of such calculations are illustrated in Fig.\ \ref{SD_minima_octup}. 
One can see that the inclusion of octupole deformation leads to the 
instability of the prolate superdeformed $\beta_2 \sim 0.6$ minima with respect of 
octupole deformation. In the $^{316,320}124$ nuclei this instability will 
lead to asymmetric fission, while the competition of symmetric and 
asymmetric fissions is possible in the $^{296,300}122$ nuclei. 
The systematic calculations for the cases in which the heights of 
outer fission barriers are higher than 2 MeV in axial reflection 
symmetric calculations clearly show that the inclusion of octupole 
deformation leads either to complete disappearance of outer fission
barrier (as seen in the cases of  $^{296}122$ and $^{316,320}124$
nuclei in Fig.\ \ref{SD_minima_octup}) or to a significant reduction 
of the heights of outer fission barriers to the values which are 
substantially lower than 2 MeV (as seen for the $^{300}122$ nucleus 
in Fig.\ \ref{SD_minima_octup}).  Thus, the prolate superdeformed minima are 
expected to be unstable in all $Z>120$ nuclei. Note that outer 
fission barriers in super- and hyperheavy nuclei could also be 
affected by triaxiality (see Ref.\ \cite{AAR.12}). However, the 
present analysis shows that prolate minima in the $Z>120$ nuclei 
are already unstable in axially symmetric calculations. This is 
a reason why nuclear landscape in the $Z=122-130$ nuclei is 
dominated by the oblate ground states (see Fig.\ 3 in Ref.\ 
\cite{AAG.18}).

\section{Systematic analysis of the results of calculations for the
$Z=138$ isotopic chain.}
\label{Z=138_chain}

  In order to illustrate the variation of the properties of 
the nuclei with neutron number, a detailed analysis of the 
results of the calculations for the $Z=138$ isotopic chain is 
presented in this section.

 Deformation energy curves of even-even $Z=138$ hyperheavy nuclei 
obtained in axial RHB calculations are shown in Fig.\ \ref{Z138-E}. 
The minimum of deformation energy curve is located at $\beta_2 \sim -4.0$ 
for proton-rich nuclei with $N=186-190$. This minimum appear at 
$\beta_2 \sim -2.5$ for the $N=194-230$ nuclei. Further increase of 
neutron number leads to the growth (in absolute sense) of the 
$\beta_2$ values: the nuclei with $N=246-262$ have minima at 
$\beta_2 \sim -3.0$. The minima of the deformation energy curves are 
located at $\beta_2 \sim -3.5$ for neutron-rich nuclei with $N=266-326$.
As discussed in details in Ref.\ \cite{AAG.18}, the nuclei have
toroidal shapes in such minima. The competing local minima with
$\beta_2 \sim -0.8$ (which corresponds to oblate ellipsoidal shape, 
see Ref.\ \cite{AAG.18}) are located at high excitation energies 
of $10-30$ MeV. This excitation energy depends on the nucleus but 
its maximum is reached at $N=286-290$ and minimum at $N\sim 206$.
Note the complex pattern of deformation energy curves at low 
deformation with a number of coexistent local minima. 
 
   Similar to few nuclei discussed in Ref.\ \cite{AAG.18}, these minima 
in deformation energy curves with toroidal shapes are potentially 
unstable with respect of the transition to prolate shape via 
$\gamma$-plane and subsequent fission since prolate 
shapes with corresponding quadrupole deformations are located 
at lower energies (compare dashed lines [which represent mirror 
reflection of the positive $\beta_2$ part of deformation energy curve 
onto  negative $\beta_2$ values] with solid ones in Figs. 1c and d of 
Ref.\ \cite{AAG.18}).  However, as discussed in Ref.\ \cite{AAG.18}
and in Sec.\ \ref{toroidal_stability} below, some of these minima 
could be stable.

  Proton and neutron chemical potentials for the solutions, 
displayed in Fig.\ \ref{Z138-E}, are shown in Fig.\ \ref{Z138-lambda}. 
They behave differently as a function of $\beta_2$. Neutron chemical 
potential on average is almost flat as a function of $\beta_{2}$. The 
magnitude of the fluctuations of the  $\lambda_n(\beta_2)$ values 
with respect of average values decreases on approaching two-neutron 
drip line. On the contrary, there is a pronounced slope in the 
$\lambda_p(\beta_2)$ values: they on average decrease with 
increasing absolute value of $\beta_2$. Note that this slope 
is especially pronounced in proton-rich nuclei. As a consequence, 
in the nuclei with $N=186-210$ there is the range of the 
$\beta_{2}$ values in which the proton chemical potential is 
positive (see top row of Fig.\ \ref{Z138-lambda}) and the shape 
of the nucleus in respective local minimum is oblate ellipsoidal. 
Even-even nuclei with $\lambda_p>0$ are typically expected 
to be unstable with respect of proton emission\footnote{
The discussion of the two-proton drip line in terms of 
proton chemical potential has its own meaning. Strictly
speaking the two-proton drip line is reached when two-proton 
separation energy $S_{2n}$ becomes negative. Alternatively (but 
less strictly) the position of the two-proton drip line is 
defined via the proton chemical potential $\lambda_p=dE/dZ$
as a point (nucleus) of the transition from negative $\lambda_p$ 
(“bound” nuclei) to positive $\lambda_p$ (“unbound” nuclei)
values. This definition depends on the employed pairing model. 
In addition, it presents a linear approximation in a Taylor 
expansion and, therefore, it ignores nonlinear effects like shape 
changes on going from the $(Z-2,N)$ to the $(Z,N)$ nucleus and 
their contribution to $S_{2p}$. However, even in the case of 
two-neutron drip line (which extremely sensitively depends on 
the fine details of the calculations) this definition leads in 
approximately two-thirds of the cases to the same two-neutron 
drip line as obtained in the definition of the two-neutron 
drip line via the separation energies \cite{AARR.15}. In the 
remaining one-third of the cases, it leads to a two-neutron
drip line which is two neutrons short of the two-neutron drip
line defined via the separation energies; the nucleus which
is unbound (as defined via the chemical potential) has in
most of the cases a low positive value of $\lambda_n \sim  0.05$ MeV.
These results were obtained in Ref.\ \cite{AARR.15} from the 
calculations of Refs.\ \cite{AARR.13,AARR.14} by analyzing 
the two-neutron drip line positions of 60 isotopic chains for 
four different CEDF’s.}.

 On the contrary, for the $N=186-210$ nuclei proton chemical 
potential is negative below $\beta_2 \sim -1.5$ and it becomes 
more negative with the increase of absolute value of $\beta_2$.
As a consequence, toroidal shapes in these nuclei are expected 
to be stable with respect of particle emission. This feature 
in the behavior of the proton chemical potential as a function
of $\beta_2$ leading to instability/stability of ellipsoidal/toroidal 
shapes in the same nucleus with respect of particle emission 
is a source of unusual shift in the position of two proton-drip 
line towards more proton rich nuclei (as compared with general 
trend seen in the $(Z,N)$ plane for the $Z < 120$ nuclei) which 
is clearly visible in Fig.\ \ref{nuclear_landscape_final} below. 
Note that such shift is absent for two-neutron drip line most 
likely because of above mentioned flatness of neutron chemical 
potential as a function of $\beta_2$.

\section{Toroidal shapes: stability and shape evolution
along the fission path.}
\label{toroidal_stability}

    The investigations of Ref.\ \cite{AAG.18} showed that some toroidal shapes 
 could be stable with respect of triaxial distortions. Fig.\ \ref{fission-path} shows 
 potential energy surfaces  of the $^{354}$134 and $^{348}$138 nuclei around 
 minima of such configurations  located at $\beta_2  \sim 2.3, \beta_4 \sim +1.5, 
 \gamma = 60^{\circ}$. The saddle points of the first fission barriers of these 
 configurations are located  at 4.4 and 8.54 MeV, respectively. However, physical 
 reasons for such stability of toroidal shapes have not been discussed in Ref.\ 
 \cite{AAG.18}.
 
   To understand these reasons the evolution of toroidal shapes along the 
fission path of the configuration in the $^{354}$134  nucleus (shown in Fig.
\ref{fission-path}) is displayed in Fig.\ \ref{shape-evol}. The toroid and its 
tube are fully symmetric at the minimum (Fig.\ \ref{shape-evol}a-c).  The deviations 
from axial symmetry lead to the distortions which are already seen at 
$\gamma=50^{\circ}$; the toroid is stretched out in the direction of the
axis of symmetry  and squeezed in perpendicular direction (Fig.\ \ref{shape-evol}f)).
However, the tubes of the toroid still remain cylindrical (Fig.\ \ref{shape-evol}d,e)).
Figs.\ \ref{shape-evol}g,h,i show the density distributions at the deformations
corresponding to the saddle point. One can see further increase of the asymmetry
of torus and the decrease of the area of toroid hole.   Thus, one can 
conclude that the barrier against fission emerges because these deviations 
from symmetrical shape of toroid cost the energy.

    Further decrease of the $\gamma$- and $\beta_2$ deformations leads to
increasing  distortion of the geometry of toroid ((Fig.\ \ref{shape-evol}l)) the
outer shape of which looks now similar to parallelogram and the shape of 
toroid  hole comes closer to square.  In addition, Figs.\ \ref{shape-evol}j,k  reveal 
visible deviations from cylindrical shape of the tube of toroid.  However, 
these changes reduce the total energy of the configuration as compared
with the one at the saddle point. 

 Above discussed changes in shapes and total energies along the fission 
path are the consequences of a delicate balance of liquid drip and shell correction 
energy contributions.

\section{The impact of triaxial deformation on the fission barriers of
              neutron-rich superheavy nuclei}
\label{fission-barriers}

    Although oblate minima of high-$Z$ ($Z>120$) superheavy and low-$Z$ 
hyperheavy nuclei are relatively stable with respect of axial reflection symmetric and 
asymmetric deformations (see Ref.\ \cite{AAG.18} and Sec.\ \ref{sect-oblate} 
in the present manuscript), that is not necessary the case with respect of triaxial 
deformation. The impact of triaxiality on the fission paths and the differences in 
the stability of super- and hyperheavy elements is illustrated in Fig.\ 
\ref{PES_3D_landscape} on the example of superheavy $^{268}$Sg and 
$^{332}$Ds nuclei and hyperheavy $^{360}$130 and $^{432}$134 nuclei.
 
  Both in super- and hyperheavy nuclei the potential energy surfaces (PES) represent
the canyon in which some local valleys and hills are located. However, there are
two principal differences between super- and hyperheavy nuclei. The canyon  
is quite narrow in superheavy nuclei which prevents the formation of local minima
at large oblate deformation and limits the role of triaxial deformation. One can see
that normal deformed minima are prolate in superheavy $^{268}$Sg and 
$^{332}$Ds nuclei and fission paths from these minima is located not far away
from the $\gamma=0^{\circ}$ axis. In addition, the bottoms of the canyons  in PES are
on average flat.

   On the contrary, in hyperheavy nuclei the walls of the canyon with very rapid 
raise of energy with deformation are located at larger separation (so only right wall 
is seen in the bottom  panels of Fig.\ \ref{PES_3D_landscape}) as compared with
superheavy nuclei and the mountain centered around $\beta_2 \sim 0$ is formed 
in this canyon.  The slope of the mountain in the direction of the $\beta_2$-deformation 
at  $\gamma=0^{\circ}$ is very high. This indicates higher instability of hyperheavy 
nuclei against fission as compared with superheavy ones.  The larger separation of 
the canyon walls leads to an increased role of triaxiality in hyperheavy nuclei: local 
minima are formed either at oblate superdeformation (see example of the $^{360}$130 
nucleus in Fig.\ \ref{PES_3D_landscape}) or at very large $\gamma$-deformation (see 
example of the $^{432}$134 nucleus in Fig.\  \ref{PES_3D_landscape}).   In addition,
the fission paths from these minima proceed at larger $\gamma$-deformations
as compared with superheavy nuclei.  Not only the fission through the $\gamma$-plane 
gets more energetically favored, but also  the fission path through $\gamma$-plane 
becomes much shorter than the  one through the $\gamma=0^{\circ}$ axis (see
also the discussion in Ref.\ \cite{AAG.18}).
    
   The general conclusion is that the barriers along the fission paths 
emerging from the oblate minima located within the $-1.0 < \beta_2 \leq 0.0$ range 
decrease with increasing proton number (see Ref.\ \cite{AAG.18}). As a result (see
discussion in  Sect.\ \ref{landscape-extension} below), the fission barriers for  oblate 
ellipsoidal shapes become consistently lower than 2 MeV  above some $(Z, N)$ line 
in nuclear landscape so the nuclei in ellipsoidal shapes cease to exist for these
particle numbers.

  However, in order to delineate this borderline additional information on the impact
of triaxiality on the fission barrier heights of the superheavy $Z=106-118$ nuclei 
located between two-proton and two-neutron drip lines is needed.  So far, such 
information is available only for  actinides and superheavy nuclei with $Z \leq 120$ 
and $N\leq 184$  \cite{AAR.10,WERP.02,SBDN.09,MSI.09,AAR.12,SR.16} .
These nuclei are either 
prolate or spherical in their ground states and thus the impact of triaxiality is limited: 
the lowering of inner fission barriers in actinides due to triaxiality is typically on the 
level of 1-3 MeV. This reduction is substantially smaller as compared with the one
typically seen in hyperheavy nuclei \cite{AAG.18}.

  Unfortunately, even nowadays fully systematic triaxial  RHB calculations are 
extremely computationally demanding.  Thus, in order to get at least rough 
outline of the impact of low fission barriers on potential topology of nuclear 
landscape, we have performed triaxial RHB calculations for selected 
nuclei with $Z=106, 110, 114$ and 118 and with $N=152 + \Delta N$, where 
$\Delta N = 0, 10, 20, ... $ and only the nuclei between two-proton and 
two-neutron drip lines are considered here. Note that in some nuclei
there are two fission barriers. If the outer fission barrier is lower than 2 MeV
in axial RHB calculations, it is ignored and the TRHB calculations are focused on 
the inner fission barrier.  If the outer fission barrier is higher than 2 MeV, then 
we first perform TRHB calculations in order to see whether triaxiality lowers outer
fission barrier below 2 MeV. If that is no a case, we carry out RHB-OCT calculations
in order to see whether octupole deformation lowers outer fission barrier below 2 
MeV. The results of such calculations, which provide information on the highest
fission barrier in the nuclei under consideration, are summarized in Table 
\ref{table-fission}. Together with the results of the TRHB calculations presented in 
Refs.\ \cite{AARR.17,AAG.18}  for superheavy $Z=122$ and hyperheavy 
$Z=126, 130$ and 134 nuclei they allow in an approximate way to delineate the 
impact of fission on the boundaries 
of nuclear landscape\footnote{The boundaries of nuclear landscape in heavy nuclei
with ellipsoidal shapes in the ground states
are defined by spontaneous fission and not by the particle emission as in lower 
$Z$ nuclei (see Ref.\ \cite{AAG.18} and the discussion in Sec.\ \ref{landscape-extension}).
This fact has been ignored in many studies of nuclear landscape in 
neutron-rich actinides and superheavy nuclei (such as Refs.\  \cite{Eet.12,AARR.14}) 
since the boundaries of nuclear landscape were defined via two-neutron 
and two-proton separation energies. The reasons for such a choice are obvious: 
such calculations require only binding energies of the ground states which are 
relatively easy to compute. On the contrary, the calculations of fission barriers in 
triaxial  DFT codes are by approximately three orders of magnitude more 
numerically time-consuming. } in Sec.\ \ref{landscape-extension} below.

   Note that these are  first ever triaxial calculations for fission barriers in
neutron-rich ($N>200$) 
superheavy nuclei and as such they can be very useful for understanding
the role of the fission in the r-process calculations. The fission of heavy systems
affects this process via fission recycling \cite{AGT.07,PMZ.07}. However, so far 
the fission barriers for such nuclei were calculated only in theoretical frameworks 
restricted to axial symmetry \cite{HFB14-BSk14,ELLMR.12,GMR.18}. 

   The results for inner fission barrier heights $E^B_{triax}$ and the decrease
 of the height of inner fission barrier due to triaxiality $\Delta E^{gain}$, presented
 in Table \ref{table-fission},  are summarized in Fig.\ \ref{fission_barrier_TRHB}.  The 
 $E^B_{triax}$ values show oscillatory behavior as a function of neutron number $N$
 with maxima seen at $N\sim 162$ and $N\sim 222$ (at $N=172$ and 232 in the 
 $Z=118$  nuclei) and minima at $N=192$ and $N\sim 252$ (at $N=202$ and 272 in 
 the $Z=118$  nuclei). More erratic behavior is  seen for the $\Delta E^{gain}$ values; 
 note that the triaxiality has no effect on the heights of inner fission barriers for  
 $\Delta E^{gain}=0$ MeV.  In some nuclei the triaxiality lowers inner fission barrier
 by more than 4 MeV. These features seen in $E^B_{triax}$  and  $\Delta E^{gain}$ 
 curves are the consequences of underlying shell structure and its evolution with proton 
 and neutron numbers (see discussion in Ref.\ \cite{AAR.10}). A general trend of the 
 lowering of the height of inner fission barrier with the increase of proton number is 
 clearly seen in Table \ref{table-fission} and Fig.\ \ref{fission_barrier_TRHB}a. 
 
   Table \ref{table-fission} clearly indicates three regions of instability based on the fission barrier heights;
in these regions the height of the highest fission barrier is below 2 MeV.  These are two islands of 
instability centered around ($Z\sim 114$, $N\sim 192$) and ($Z\sim 118$, $N\sim 202$) 
which are shown in Fig.\ \ref{nuclear_landscape_1}b. In addition, very neutron rich nuclei near and above 
$N=252$ in the $Z=114$ isotopic chain as well as near and above $N=262$ in the 
$Z=118$ isotopic chain are unstable with respect of fission. Such instability against fission 
is also seen in very neutron rich hyperheavy nuclei (see supplemental material to Ref.\ \cite{AAG.18}). 
These factors together lead to a substantial reduction of the region of potentially stable elliposoidal-like 
nuclei in the $N\geq 258$ region (compare panels (a) and (b) in Fig.\ \ref{nuclear_landscape_1} below).

\section{Extension of nuclear landscape to hyperheavy nuclei}
\label{landscape-extension}

  One of important goals of the present manuscript is the 
extension of nuclear landscape to the limits of extreme $Z$ values.
There are numerous studies of the limits of nuclear landscape at 
the neutron and proton drip lines for the $Z<120$ nuclei carried 
out in different theoretical frameworks (see Refs.\ 
\cite{MNMS.95,DGLGHPPB.10,Eet.12,TOV-min,AARR.13,AARR.14} and,
in particular,  the compilation presented in  Sec. VIII of Ref.\ 
\cite{AARR.14}). The studies of Refs.\ \cite{Eet.12,AARR.13,AARR.14}
also define systematic theoretical uncertainties in the position of 
two-proton and two-neutron drip lines. On the contrary, 
nothing was known about the nuclear landscape in hyperheavy nuclei 
and its limits before our previous publication in Ref.\  
\cite{AAG.18}. The goal of this section is to present a comprehensive 
summary on the structure and limits of nuclear landscape with 
special emphasis on the region of hyperheavy nuclei.

  The results of Ref.\ \cite{AAG.18} and the present study clearly 
show that critical distinction between the parts of nuclear chart  
are related  to the dominance of two different types of shapes: 
ellipsoidal-like and toroidal ones. Note that concave disk shapes, 
appearing at large oblate deformation, belong to ellipsoidal-like shapes. 
Fig. 1a shows the region of nuclear chart which is 
dominated by ellipsoidal-like shapes. Note that for $Z<120$ LEMAS 
obtained in reflection-symmetric RHB calculations  typically 
correspond to the ground states since only few nuclei in their 
ground states  are affected by $\gamma$-deformation (see Ref.\ 
\cite{MBCOISI.08} for the results obtained in microscopic+macroscopic method) 
and octupole deformation shows up in the ground states of the nuclei in 
few localized regions \cite{MBCOISI.08,AAR.16,AA.17-oct}. 

  The situation changes in the $Z>120$ nuclei which are typically  soft with 
respect of triaxial deformation up to the point that many ground 
states possess triaxial deformation (see Table 1 in supplemental
material of Ref.\ \cite{AAG.18} and Table \ref{table-fission} in 
the present manuscript). This softness also leads to a substantial 
reduction of the heights of the fission barriers in many nuclei. 
If the barrier height is less than 2 MeV, the nucleus is typically 
considered unstable against fission (see discussion in Ref.\ 
\cite{AAR.12}). If we take this 
fact into account, the region of nuclear chart with ellipsoidal-like 
shapes will be considerably reduced at high-$Z$ values; this is 
illustrated in Fig. 1b. Note that in some nuclei eliminated on 
transition from panel (a) to panel 
(b) the local minima (which are otherwise excited ones) with toroidal 
shapes could become the lowest in energy solutions if ellipsoidal-like 
shapes are unstable with respect of fission. Finally, two-proton 
and two-neutron drip lines for toroidal shapes are added in Fig.\ 1c.

  While there is the coexistence of ellipsoidal-like and toroidal 
shapes in the $Z=120-140$ part of nuclear chart (see Fig. 1 and 
Fig. 3 in Ref.\ \cite{AAG.18}), with increasing proton number 
beyond $Z=140$, the LEMAS always have toroidal shapes (see 
discussion in Sec. \ref{sect-truncation} and in Ref.\ 
\cite{AAG.18}). The nuclear chart extended up to $Z=180$ displays 
the two-proton and two-neutron drip lines for toroidal nuclei 
outlining the potential  limits of nuclear landscape (Fig. \ 
\ref{nuclear_landscape_final}).

The transition from ellipsoidal to toroidal shapes is driven by Coulomb
repulsion and has a lot of similarities to Coulomb frustration phenomenon
seen in nuclear pasta phase of neutron stars.  Fig.\ \ref{Coulomb} shows
the calculated Coulomb energies $E_{Coul}$ as a function of the $\beta_2$
values. One can see that in all nuclei the largest Coulomb energy is 
calculated at spherical shape which is the most compact shape for a given 
volume. The deviation from sphericity decreases the 
Coulomb energy and for a given absolute value of $\beta_2$ this effect  
is especially pronounced for negative $\beta_2$ values. Moreover, 
the magnitude of $E_{Coul}$ and its slope with deformation
for negative $\beta_2$ values drastically increases with increasing proton 
number (see Fig.\ \ref{Coulomb}).  This is also quantified in Fig.\ \ref{Coulomb} 
by the $\Delta E_{Coul}$ quantity which is the 
reduction of the Coulomb energy due to the transition from spherical
shape to typical toroidal one. The $\Delta E_{Coul}$ increases from 
346 MeV in $^{208}$Pb to 721, 874 and 1126 MeV in the $^{354}$134,
$^{466}$156 and $^{426}$176 nuclei, respectively. This clearly tells
that toroidal shapes are more energetically favored by Coulomb interaction 
than spherical (or ellipsoidal-like) ones. However, only in hyperheavy nuclei 
the Coulomb energy becomes strong enough to trigger the transition to toroidal 
shapes in the lowest in energy solutions of axial RHB calculations (see Fig. 1 
in Ref.\ \cite{AAG.18}).

    It is clear that nuclear landscape shown in Fig.\ 
\ref{nuclear_landscape_final} is not complete because 
it does not take into account the potential instabilities 
of toroidal shapes with respect of different types of 
distortions. Although it was shown in Ref.\ \cite{AAG.18} 
that some toroidal nuclei are potentially stable with 
respect of triaxial distortions, this was illustrated 
only for two nuclei. The underlying mechanism of
their stability is discussed in Sec.\ \ref{toroidal_stability}. 
The problem is that with increasing proton number the extreme 
sizes of fermionic basis (up to $N_F=30$ for nuclei at $Z=156$ 
and beyond, see discussion in Sec.\ \ref{sect-truncation}) 
are required for the description of toroidal nuclei. Neither 
triaxial nor axial reflection asymmetric calculations are 
possible nowadays for such sizes of basises. 

 The investigations of Ref.\ \cite{Wong.73} suggest that 
toroidal shapes are expected to be unstable with respect of 
so-called sausage deformations which make a torus thicker in one 
section(s) and thinner in another section(s). They are 
expected to lead to multifragmentation\footnote{There are 
also some experimental indications of the role of 
multifragmentation in toroidal nuclei, but they are restricted 
to a single  $^{86}$Kr\,+\,$^{93}$Nb reaction \cite{multifrag.97}.}.  However, these 
investigations are performed in the liquid drop model which neglects
potential stabilizing role of the shell effects. In 
addition, they do not extend beyond the region
of superheavy nuclei. To clarify the situation
the DFT studies of potential stability of toroidal shapes
with respect of triaxial distortions similar to those performed 
for $^{354}134$ and $^{348}138$ nuclei in Ref.\ \cite{AAG.18} 
(see also Sec.\ \ref{toroidal_stability} in the present 
manuscript) are needed for higher $Z$ values. Unfortunately, 
as mentioned above they are not possible nowadays because of 
extreme sizes of basises.

   If the toroidal shapes are unstable (and, from our point of 
view, the likelihood of this scenario is high in high-$Z$ nuclei),
then the spherical shapes become the ground states in the
islands of potential stability of spherical hyperheavy nuclei
(see Ref.\ \cite{AAG.18}). These islands centered around
($Z\sim 138, N\sim 230$),  ($Z\sim 156, N\sim 310$) and 
($Z\sim 174, N\sim 410$) are shown in 
Fig.\ \ref{nuclear_landscape_final}.

\begin{figure*}[htb]
\includegraphics[angle=-90,width=14.cm]{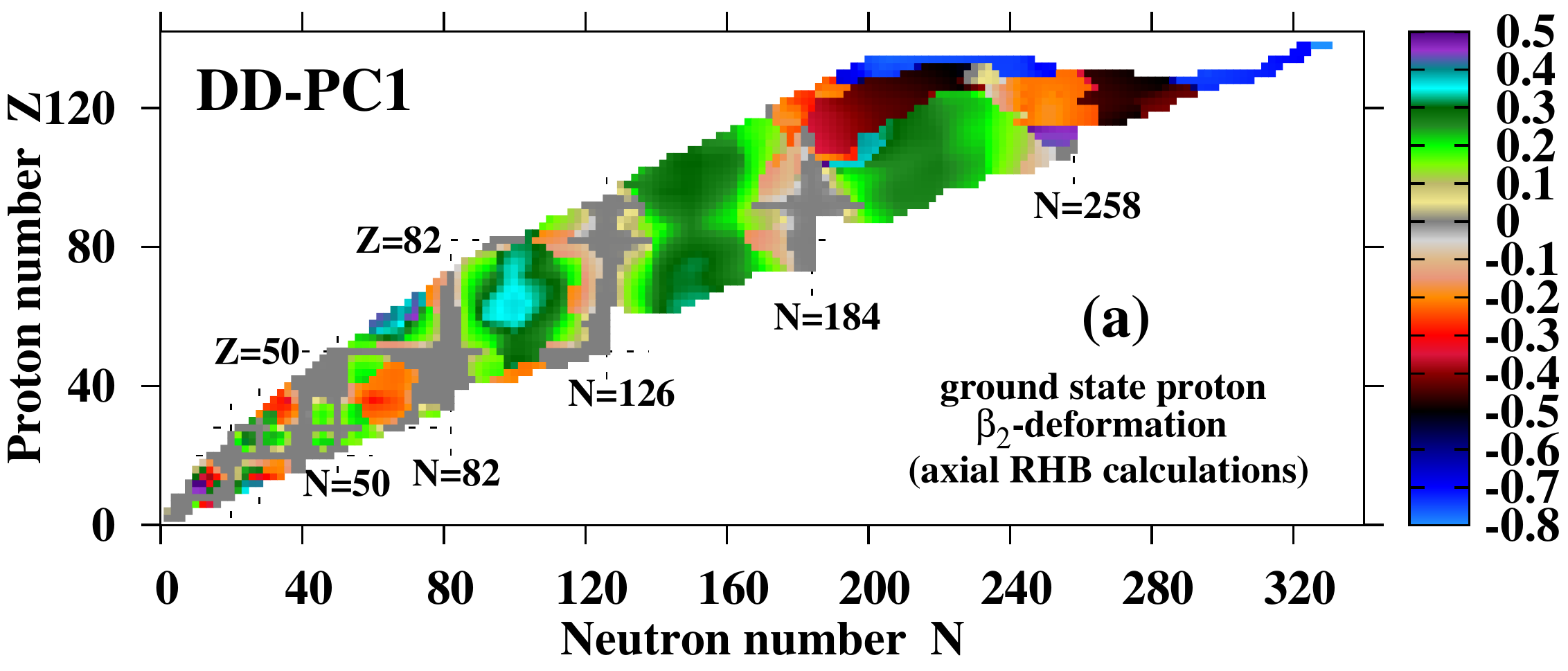}
\includegraphics[angle=-90,width=14.cm]{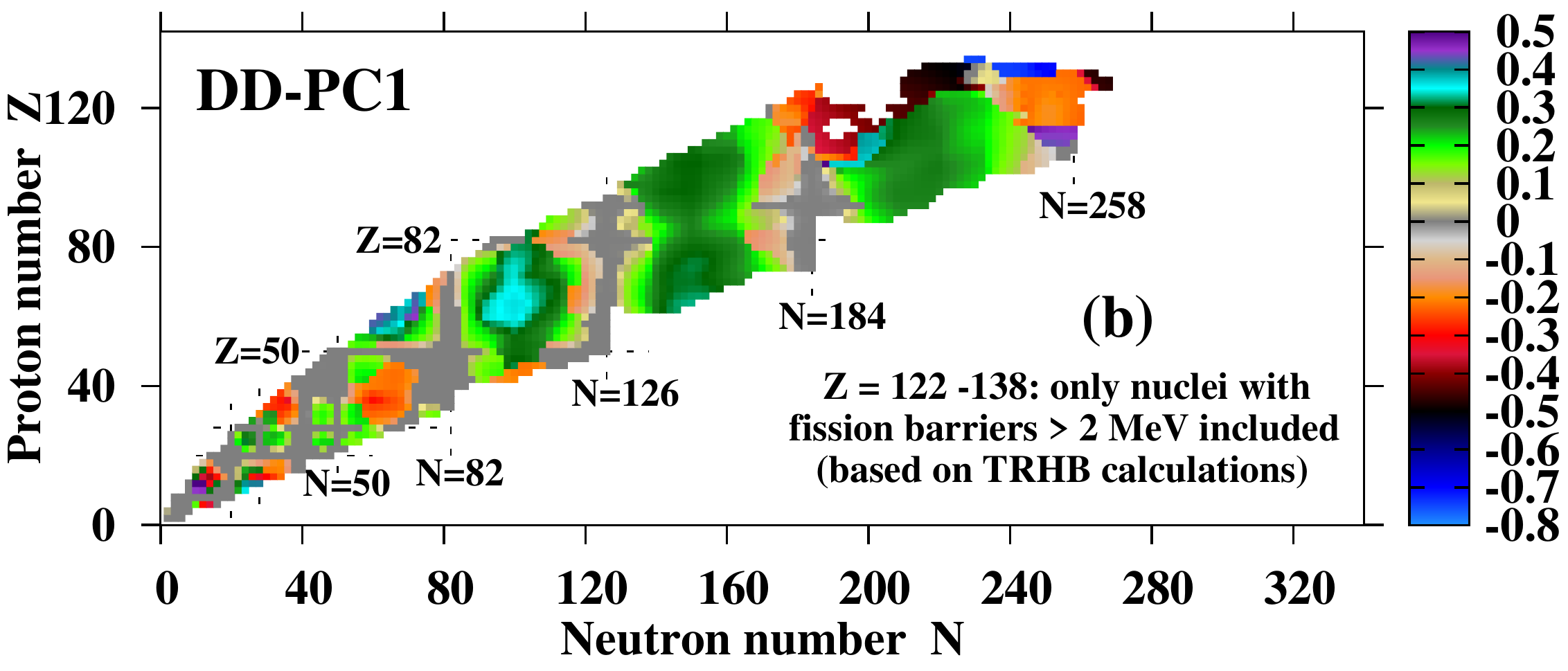}
\includegraphics[angle=-90,width=14.cm]{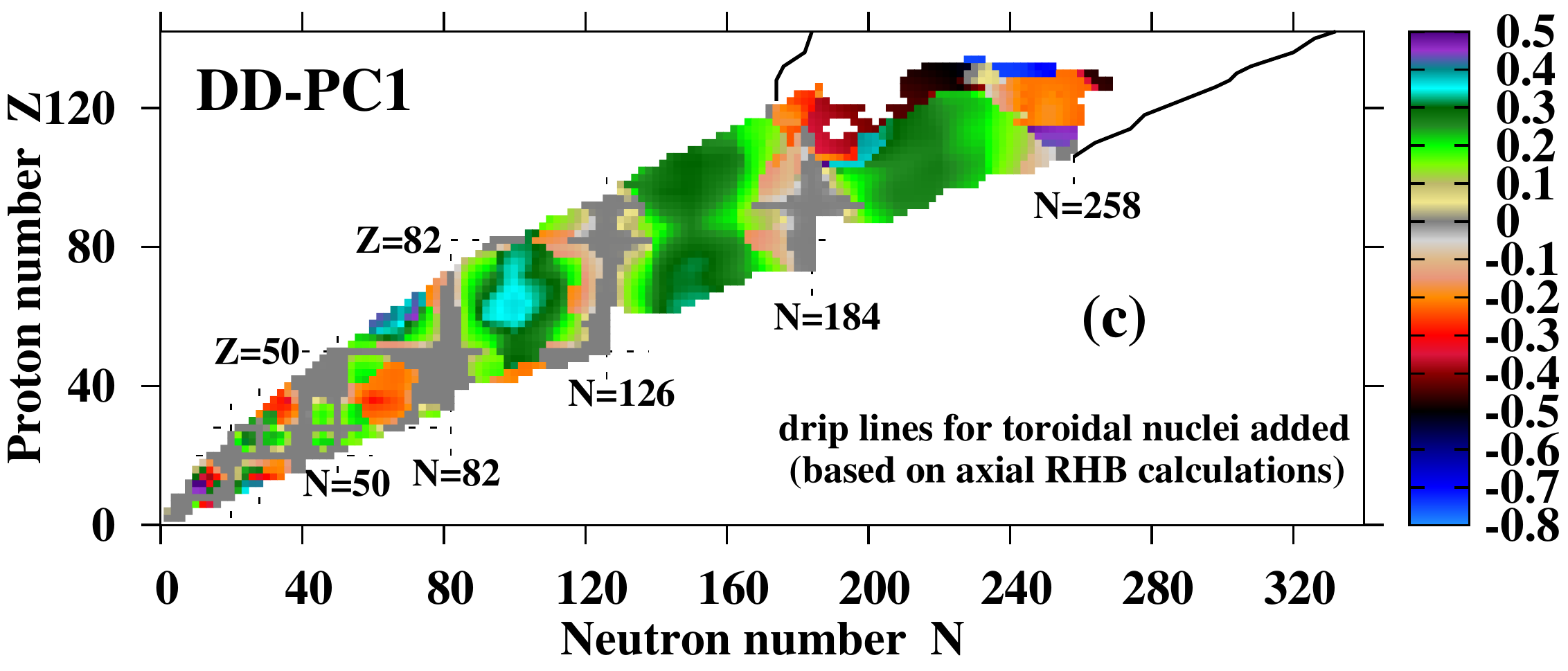}
\caption{(Color online) (a) Proton quadrupole deformations 
$\beta_2$ of the lowest in energy minima for axial symmetry (LEMAS) 
obtained in axial RHB calculations with the DD-PC1 functional.  
Based on the results presented in Fig.\ 17c of Ref.\ \cite{AARR.14} 
and Fig. 3 of Ref.\ \cite{AAG.18}. Only the nuclei whose LEMAS have 
ellipsoidal-like shapes are included here;  those who have toroidal 
shapes in LEMAS (see Fig. 3 in Ref.\ \cite{AAG.18}) are neglected.
The colormap in  the $\beta_2 = -0.4 - 0.5$ range is equivalent 
to the one of Fig.\ 17c of Ref.\ \cite{AARR.14} for consistency with 
previous results. (b) The same as panel (a) but with the nuclei, in 
which neither inner nor outer (if exist) fission barrier(s) have 
the height(s) higher than 2 MeV, excluded. Here the results of the 
calculations for fission barmvriers presented in Table 1 of supplemental 
material to Ref.\ \cite{AAG.18} and in Table \ref{table-fission} of the 
present manuscript are used for approximate deliniation of the 
boundaries of the region of nuclear chart in which fission barriers 
satisfy above mentioned condition. (c) The same as panel (b) but 
with two-proton and two-neutron drip lines (shown by solid lines),
defined from separation energies, for toroidal nuclei added. They are 
based on the results of axial RHB calculations with $N_F=26$.}
\label{nuclear_landscape_1}
\end{figure*}

\begin{figure}[htb]
\includegraphics[angle=0,width=8.5cm]{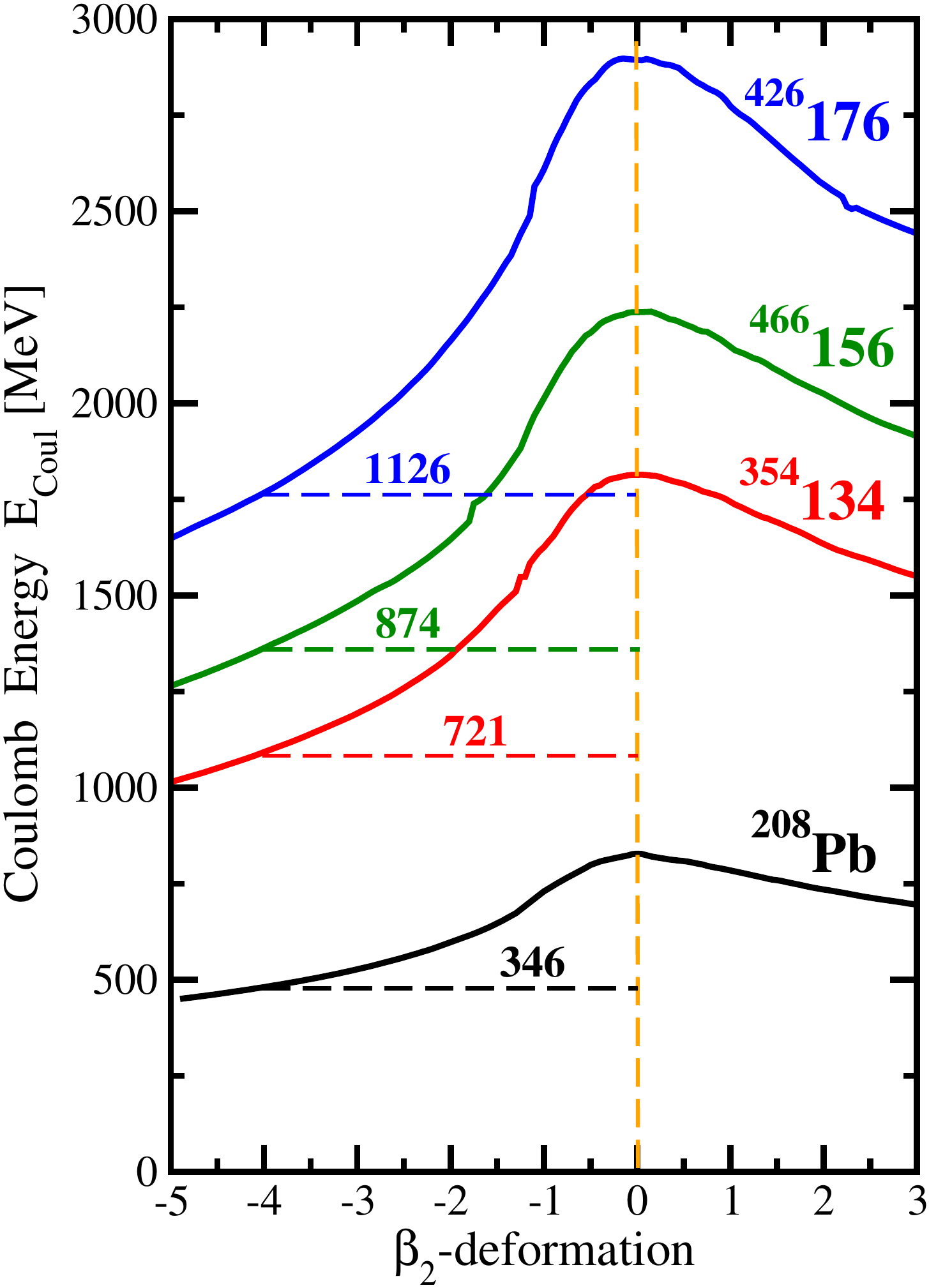}
\caption{ (Color online) Calculated Coulomb energy as a function
of the $\beta_2$ values. The results are displayed for four indicated nuclei; the total
deformation energy curves  of these nuclei are shown in Fig. \ 1 of Ref.\ \cite{AAG.18}. 
Orange vertical dashed line indicates spherical shapes. Horizontal dashed lines of
different color start at the positions of respective Coulomb energy curves at $\beta_2=-4.0$ 
and end at vertical orange dashed line. The numbers above these horizontal lines 
indicate  the difference $\Delta E_{Coul} = E_{Coul}(\beta_2 =0.0) - E_{Coul}(\beta_2 =-4.0)$
(in MeV, rounded to closest value) which is the reduction of the Coulomb energy due to 
the transition from spherical shape to toroidal shape with typical $\beta_2 = -4.0$ values 
seen at the minima of deformation energy curves of the $^{466}$156 and $^{426}$176 
nuclei (see Figs. 1c and 2 in Ref.\ 
\cite{AAG.18}.
\label{Coulomb} 
}
\end{figure}

\begin{figure*}[htb]
\includegraphics[angle=-90,width=18.cm]{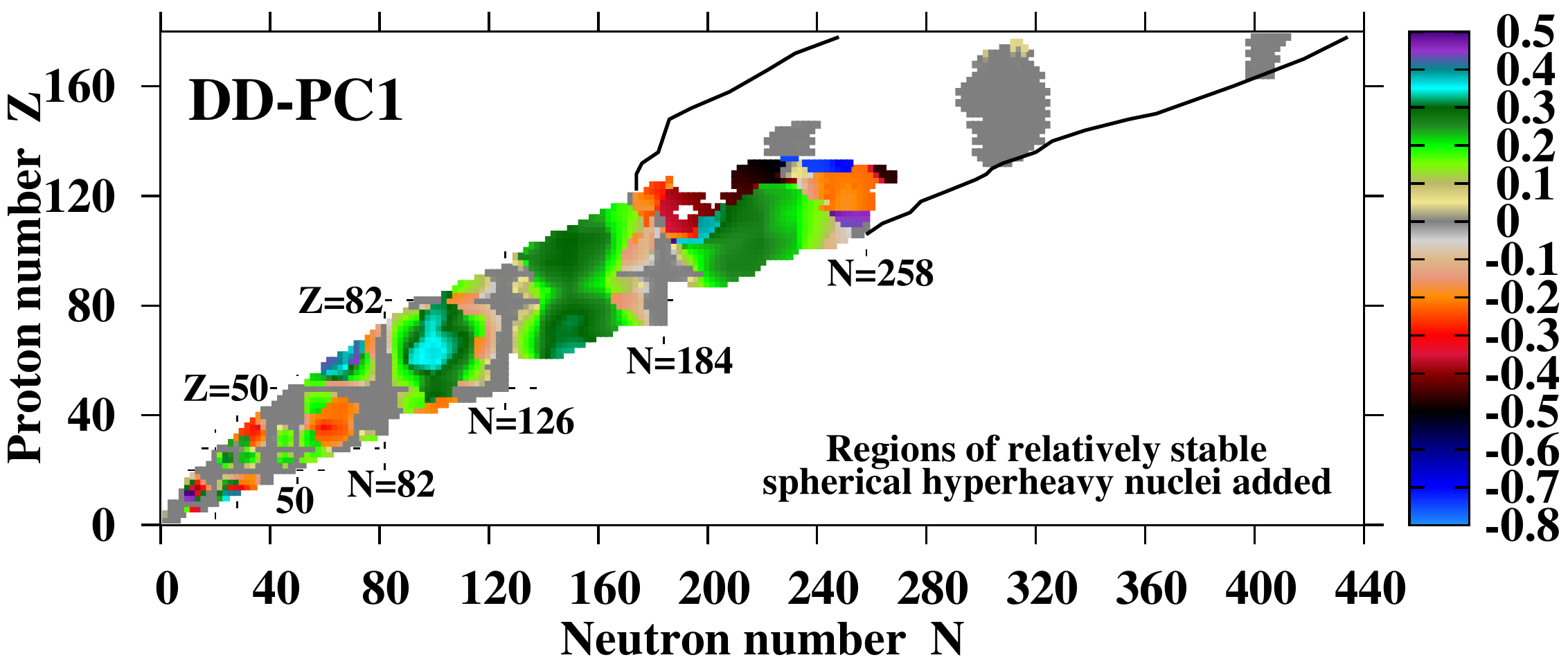}
\caption{(Color online) The same as Fig.\ \ref{nuclear_landscape_1} but 
with extended proton and neutron ranges and added regions of relatively 
stable spherical hyperheavy nuclei shown in gray.  Note that in the 
same nucleus two-neutron drip lines for spherical and toroidal shapes 
are different. This is a reason why some regions of stability of spherical 
nuclei extend beyond two-neutron drip line for toroidal shapes. 
\label{nuclear_landscape_final}
}
\end{figure*}

    The analysis of Figs.\ \ref{nuclear_landscape_1} and 
\ref{nuclear_landscape_final} clearly indicates 
that the classical structure of the nuclear landscape in 
which spherical shell closures at different particle numbers
play a defining role disappears in the $Z>120$ nuclei.
This is because the ground states are either oblate or
toroidal in axial RHB calculations.

  The extrapolation to unknown regions is definitely associated
with theoretical uncertainties \cite{DNR.14} which are especially large for
the position of two-neutron drip line 
\cite{Eet.12,AARR.13,AARR.14,AARR.15,AA.16} and fission 
barriers \cite{AARR.17,JKS.17}. In the CDFT 
framework, systematic theoretical
uncertainties due to the form of the CEDF are substantially
larger than statistical errors \cite{AAT.18}. So far, such
systematic uncertainties 
have been estimated only for fission barriers in the regions of 
potentially stable spherical hyperheavy nuclei and for the sizes
of these regions (see Fig. 6 in Ref.\ \cite{AAG.18}). Their more 
global  evaluation is extremely time consuming and at this stage,
when we try to understand the general features of hyperheavy
nuclei, is not necessary. These uncertainties will definitely
affect the stability of ellipsoidal shapes with respect of
fission and, as a consequence, the boundary of the transition from 
ellipsoidal to toroidal shapes and the two-proton and two-neutron 
drip lines for toroidal  nuclei.  However, they will not affect the 
general features.

\section{Conclusions}
\label{summary}

      In conclusion, the detailed investigation of the properties of hyperheavy 
nuclei has been performed in the framework of covariant density functional
theory. The following conclusions have been obtained:

\begin{itemize}

\item
 The stability of spherical hyperheavy nuclei located in the centers
of the  ($Z\sim 138, N\sim 230$),  ($Z\sim 156, N\sim 310$) and 
($Z\sim 174, N\sim 410$) islands of stability with respect of triaxial
and octupole distortions has been established in the RHB+OCT 
and TRHB calculations.

\item
  Proton and neutron densities, central depressions in these densities,
charge radii, and neutron skins of the nuclei  located in the centers of 
these islands of stability have been investigated  in detail. Obtained 
results clearly indicate that the accuracy of the reproduction of charge 
radii and neutron skins by the CEDFs could be an important criteria in 
favoring or disfavoring the predictions of one or another functional for 
the islands of stability of spherical hyperheavy nuclei. Among considered 
functionals, the DD-ME2 and DDPC1 functionals provide the best global 
description of charge radii and predict the highest fission barriers in these 
regions.  The results obtained in future PREX-2 experiment on neutron skin 
in $^{208}$Pb \cite{PREX-CREX} would be quite useful in helping to 
discriminate the predictions of different functionals for the islands of  
stability of spherical hyperheavy nuclei.
 
\item
  Underlying shell structure of the nuclei  located in the centers of these islands 
of  stability has been investigated  in detail. Large neutron shell gaps at 
$N=228, 308$ and 406 have a sizeable impact on two neutron-separation 
energies. On the other hand, large proton gap appear only at $Z=154$ in
the $(Z\sim 156, N\sim 310)$ island of stability of spherical hyperheavy 
nuclei.  As a result, this is the largest island of stability of spherical superheavy
nuclei found in the calculations. No significant proton gaps are seen in other 
two islands of stability. Taking into account clear localization of the islands of 
stability of spherical hyperheavy nuclei in the $(Z, N)$ plane these features 
strongly suggest that the shell effects at deformed shapes leading to negative 
shell correction energies at some deformation and thus to fission barriers play 
an  important role in the stabilization of spherical hyperheavy nuclei.

\item
The shape evolution of toroidal shapes along the fission path and the stability 
of  such shapes with respect of fission have been studied. In considered cases,
the analysis shows the transition from symmetrical toroid (at the local minimum) 
to the asymmetric one (at the saddle point). This transition cost the energy which 
is a physical reason for the formation of fission barrier and, thus, for the stability 
of such shapes.

\item
The topology of potential energy surfaces for ellipsoidal shapes of the super- and
hyperheavy nuclei has been compared. In both types of the nuclei the PES has
the form of  the 
canyon in which some local valleys and hills are located.  The canyon is quite 
narrow in superheavy nuclei which prevents the formation of local minima at 
large oblate deformation and limits the role of triaxial deformation. On the contrary,
this canyon is much wider in hyperheavy nuclei with a mountain, centered at 
$\beta_2 \sim 0$, formed in it. This leads to the formation of local minima at oblate 
superdeformation,  increased role of triaxiality and higher instability of hyperheavy 
nuclei against fission as compared with superheavy ones.

\item
  The extension of nuclear landscape  to hyperheavy nuclei  with proton numbers up 
to $Z=180$ has been performed.   With increasing proton number beyond $Z \sim 130$ 
the transition from ellipsoidal-like nuclear shapes to toroidal shapes takes place in axial
RHB calculations. The ellipsoidal ground states are affected by above-mentioned 
increased instability against fission.  Many hyperheavy nuclei with toroidal shapes (as 
the lowest in energy solutions in axial RHB calculations) are expected to be unstable 
towards multi-fragmentation.  However, it is difficult to quantify their stability or instability
since the description of toroidal shapes requires the basis which is typically significantly 
larger than the one employed for the description of ellipsoidal-like shapes. This makes
the calculations with octupole or triaxial deformation included impossible for toroidal 
shapes with extreme $\beta_2$ values. Nevertheless,  three islands of stability of spherical 
hyperheavy nuclei are predicted. The nuclei in these  islands will become the ground states in the case of instability of relevant toroidal states.

\end{itemize}

   Detailed investigation of possible mechanisms of the creation of spherical 
and toroidal hyperheavy nuclei represents an interesting topic but goes beyond the 
scope of the present manuscript. The nuclei in the ($Z\sim 138, N\sim 230$),  
($Z\sim 156, N\sim 310$) and ($Z\sim 174, N\sim 410$) islands of stability of
spherical hyperheavy nuclei have neutron to proton ratios of $N/Z \sim 1.67$, 
$N/Z \sim 1.99$ and $N/Z \sim 2.36$, respectively. Thus, they cannot be formed 
in laboratory conditions and the only possible environment in which they can 
be produced is the ejecta of the mergers of neutron stars \cite{NS-grav-wave-exp.17}.
In a similar fashion, the regions of neutron stars with nuclear pasta phases 
\cite{CSHB.15,FHS.17,KKW.17} may be a breading ground for the formation of 
toroidal nuclei in the ejecta of the merger of neutron stars.  
The two-proton drip line for toroidal nuclei is characterized by neutron to proton ratio 
of $N/Z \sim 1.25$. Thus, the stability and/or multi-fragmentation of toroidal nuclei located 
in the vicinity of two-proton drip line could possibly be studied in  nucleus-nucleus collisions 
of stable nuclei or the nuclei located close to the beta-stability line. This is similar to what 
has alrelsady been done  in the $^{86}$Kr + $^{93}$Nb reaction at incident energies ranging 
from 35 to 95 MeV/nucleon in Ref.\ \cite{multifrag.97}.

   Any extrapolation beyond known regions of nuclear chart in which the functionals have 
been fitted is associated with theoretical uncertainties \cite{Eet.12,AARR.14,AARR.15}. This 
is especially  true for the present study  with its significant extrapolations  in proton and 
neutron numbers. Despite the fact that our study is mostly based on the DD-PC1 CEDF which, according to
the results of global studies, is considered to be  the best relativistic functional, in no
way the extrapolations based on it should be considered as completely safe.  This is also
true for any relativistic or non-relativistic functional.  However, when calculated effects 
are substantially larger than the expected theoretical uncertainties one can speak about reliable theoretical 
predictions. For example, the predicted transition from ellipsoidal to toroidal shapes with 
increasing proton number is a solid prediction.  Note that according to Ref.\ \cite{Warda.07}
it appears also in  Gogny DFT as exemplified by the calculated results for two hyperheavy
nuclei. On the other  hand, the borderline in the $(Z,N)$ plane between these two types of 
the shapes is expected to depend on the details of the functional since it is defined by the 
fission properties which are subject of appreciable theoretical uncertainties \cite{AARR.17}. 
However, the lowering of the fission barrier heights for ellipsoidal shapes with increasing 
proton number $Z$, which defines this boundary,  appears both in relativistic and non-relativistic 
models (see Sect.\ 
\ref{fission-barriers} in the present paper and Refs.\ \cite{BS.13,AAG.18}).  In addition, the 
size of the regions of possible stability of spherical hyperheavy nuclei  and the stability of the
nuclei in these regions depend on the functional (see Ref.\ \cite{AAG.18}) and on the details 
of underlying shell structure (see Sect.\ \ref{shell_closure} in the present paper).  The latter
is subject of appreciable theoretical uncertainties when extrapolations are performed to
unknown regions in the $(Z,N)$ plane which are located far away from known part of 
nuclear chart in which the functionals have been fitted \cite{AARR.15}.  In addition, 
there could be hidden biases in the CDFT which could affect model predictions.  In such a 
situation, detailed investigations of hyperheavy nuclei in the framework of non-relativistic 
density functional theories based on the Skyrme and Gogny forces would be extremely 
useful for an understanding of general structure of hyperheavy nuclei, the transition from 
ellipsoidal to toroidal shapes with increasing proton number and possible existence of the 
islands of stability of spherical hyperheavy nuclei.  They will also allow to compare the 
predictions obtained in relativistic and non-relativisitc models.

\section{ACKNOWLEDGMENTS}

 This material is based upon work supported by the U.S. Department 
of Energy, Office of Science, Office of Nuclear Physics under Award 
No. DE-SC0013037 and by the Department of Energy National Nuclear 
Security Administration under Award No. DE-NA0002925.

\bibliography{references22}

\begin{thebibliography}{77}%
\makeatletter
\providecommand \@ifxundefined [1]{%
 \@ifx{#1\undefined}
}%
\providecommand \@ifnum [1]{%
 \ifnum #1\expandafter \@firstoftwo
 \else \expandafter \@secondoftwo
 \fi
}%
\providecommand \@ifx [1]{%
 \ifx #1\expandafter \@firstoftwo
 \else \expandafter \@secondoftwo
 \fi
}%
\providecommand \natexlab [1]{#1}%
\providecommand \enquote  [1]{``#1''}%
\providecommand \bibnamefont  [1]{#1}%
\providecommand \bibfnamefont [1]{#1}%
\providecommand \citenamefont [1]{#1}%
\providecommand \href@noop [0]{\@secondoftwo}%
\providecommand \href [0]{\begingroup \@sanitize@url \@href}%
\providecommand \@href[1]{\@@startlink{#1}\@@href}%
\providecommand \@@href[1]{\endgroup#1\@@endlink}%
\providecommand \@sanitize@url [0]{\catcode `\\12\catcode `\$12\catcode
  `\&12\catcode `\#12\catcode `\^12\catcode `\_12\catcode `\%12\relax}%
\providecommand \@@startlink[1]{}%
\providecommand \@@endlink[0]{}%
\providecommand \url  [0]{\begingroup\@sanitize@url \@url }%
\providecommand \@url [1]{\endgroup\@href {#1}{\urlprefix }}%
\providecommand \urlprefix  [0]{URL }%
\providecommand \Eprint [0]{\href }%
\providecommand \doibase [0]{http://dx.doi.org/}%
\providecommand \selectlanguage [0]{\@gobble}%
\providecommand \bibinfo  [0]{\@secondoftwo}%
\providecommand \bibfield  [0]{\@secondoftwo}%
\providecommand \translation [1]{[#1]}%
\providecommand \BibitemOpen [0]{}%
\providecommand \bibitemStop [0]{}%
\providecommand \bibitemNoStop [0]{.\EOS\space}%
\providecommand \EOS [0]{\spacefactor3000\relax}%
\providecommand \BibitemShut  [1]{\csname bibitem#1\endcsname}%
\let\auto@bib@innerbib\@empty
\bibitem [{\citenamefont {Afanasjev}\ \emph {et~al.}(2018)\citenamefont
  {Afanasjev}, \citenamefont {Agbemava},\ and\ \citenamefont
  {Gyawali}}]{AAG.18}%
  \BibitemOpen
  \bibfield  {author} {\bibinfo {author} {\bibfnamefont {A.~V.}\ \bibnamefont
  {Afanasjev}}, \bibinfo {author} {\bibfnamefont {S.~E.}\ \bibnamefont
  {Agbemava}}, \ and\ \bibinfo {author} {\bibfnamefont {A.}~\bibnamefont
  {Gyawali}},\ }\href@noop {} {\bibfield  {journal} {\bibinfo  {journal} {Phys.
  Lett. B}\ }\textbf {\bibinfo {volume} {782}},\ \bibinfo {pages} {533}
  (\bibinfo {year} {2018})}\BibitemShut {NoStop}%
\bibitem [{\citenamefont {Afanasjev}\ and\ \citenamefont
  {Agbemava}(2016)}]{AA.16}%
  \BibitemOpen
  \bibfield  {author} {\bibinfo {author} {\bibfnamefont {A.~V.}\ \bibnamefont
  {Afanasjev}}\ and\ \bibinfo {author} {\bibfnamefont {S.~E.}\ \bibnamefont
  {Agbemava}},\ }\href {\doibase 10.1103/PhysRevC.93.054310} {\bibfield
  {journal} {\bibinfo  {journal} {Phys. Rev. C}\ }\textbf {\bibinfo {volume}
  {93}},\ \bibinfo {pages} {054310} (\bibinfo {year} {2016})}\BibitemShut
  {NoStop}%
\bibitem [{\citenamefont {Erler}\ \emph
  {et~al.}(2012{\natexlab{a}})\citenamefont {Erler}, \citenamefont {Birge},
  \citenamefont {Kortelainen}, \citenamefont {Nazarewicz}, \citenamefont
  {Olsen}, \citenamefont {Perhac},\ and\ \citenamefont {Stoitsov}}]{Eet.12}%
  \BibitemOpen
  \bibfield  {author} {\bibinfo {author} {\bibfnamefont {J.}~\bibnamefont
  {Erler}}, \bibinfo {author} {\bibfnamefont {N.}~\bibnamefont {Birge}},
  \bibinfo {author} {\bibfnamefont {M.}~\bibnamefont {Kortelainen}}, \bibinfo
  {author} {\bibfnamefont {W.}~\bibnamefont {Nazarewicz}}, \bibinfo {author}
  {\bibfnamefont {E.}~\bibnamefont {Olsen}}, \bibinfo {author} {\bibfnamefont
  {A.~M.}\ \bibnamefont {Perhac}}, \ and\ \bibinfo {author} {\bibfnamefont
  {M.}~\bibnamefont {Stoitsov}},\ }\href@noop {} {\bibfield  {journal}
  {\bibinfo  {journal} {Nature}\ }\textbf {\bibinfo {volume} {486}},\ \bibinfo
  {pages} {509} (\bibinfo {year} {2012}{\natexlab{a}})}\BibitemShut {NoStop}%
\bibitem [{\citenamefont {Afanasjev}\ \emph {et~al.}(2013)\citenamefont
  {Afanasjev}, \citenamefont {Agbemava}, \citenamefont {Ray},\ and\
  \citenamefont {Ring}}]{AARR.13}%
  \BibitemOpen
  \bibfield  {author} {\bibinfo {author} {\bibfnamefont {A.~V.}\ \bibnamefont
  {Afanasjev}}, \bibinfo {author} {\bibfnamefont {S.~E.}\ \bibnamefont
  {Agbemava}}, \bibinfo {author} {\bibfnamefont {D.}~\bibnamefont {Ray}}, \
  and\ \bibinfo {author} {\bibfnamefont {P.}~\bibnamefont {Ring}},\ }\href@noop
  {} {\bibfield  {journal} {\bibinfo  {journal} {Phys.\ Lett. B}\ }\textbf
  {\bibinfo {volume} {726}},\ \bibinfo {pages} {680} (\bibinfo {year}
  {2013})}\BibitemShut {NoStop}%
\bibitem [{\citenamefont {Agbemava}\ \emph {et~al.}(2014)\citenamefont
  {Agbemava}, \citenamefont {Afanasjev}, \citenamefont {Ray},\ and\
  \citenamefont {Ring}}]{AARR.14}%
  \BibitemOpen
  \bibfield  {author} {\bibinfo {author} {\bibfnamefont {S.~E.}\ \bibnamefont
  {Agbemava}}, \bibinfo {author} {\bibfnamefont {A.~V.}\ \bibnamefont
  {Afanasjev}}, \bibinfo {author} {\bibfnamefont {D.}~\bibnamefont {Ray}}, \
  and\ \bibinfo {author} {\bibfnamefont {P.}~\bibnamefont {Ring}},\ }\href@noop
  {} {\bibfield  {journal} {\bibinfo  {journal} {Phys.\ Rev. C}\ }\textbf
  {\bibinfo {volume} {89}},\ \bibinfo {pages} {054320} (\bibinfo {year}
  {2014})}\BibitemShut {NoStop}%
\bibitem [{\citenamefont {Afanasjev}\ \emph {et~al.}(2015)\citenamefont
  {Afanasjev}, \citenamefont {Agbemava}, \citenamefont {Ray},\ and\
  \citenamefont {Ring}}]{AARR.15}%
  \BibitemOpen
  \bibfield  {author} {\bibinfo {author} {\bibfnamefont {A.~V.}\ \bibnamefont
  {Afanasjev}}, \bibinfo {author} {\bibfnamefont {S.~E.}\ \bibnamefont
  {Agbemava}}, \bibinfo {author} {\bibfnamefont {D.}~\bibnamefont {Ray}}, \
  and\ \bibinfo {author} {\bibfnamefont {P.}~\bibnamefont {Ring}},\ }\href@noop
  {} {\bibfield  {journal} {\bibinfo  {journal} {Phys.\ Rev.\ C}\ }\textbf
  {\bibinfo {volume} {91}},\ \bibinfo {pages} {014324} (\bibinfo {year}
  {2015})}\BibitemShut {NoStop}%
\bibitem [{\citenamefont {Gao}\ \emph {et~al.}(2013)\citenamefont {Gao},
  \citenamefont {Dobaczewski}, \citenamefont {Kortelainen}, \citenamefont
  {Toivanen},\ and\ \citenamefont {Tarpanov}}]{GDKTT.13}%
  \BibitemOpen
  \bibfield  {author} {\bibinfo {author} {\bibfnamefont {Y.}~\bibnamefont
  {Gao}}, \bibinfo {author} {\bibfnamefont {J.}~\bibnamefont {Dobaczewski}},
  \bibinfo {author} {\bibfnamefont {M.}~\bibnamefont {Kortelainen}}, \bibinfo
  {author} {\bibfnamefont {J.}~\bibnamefont {Toivanen}}, \ and\ \bibinfo
  {author} {\bibfnamefont {D.}~\bibnamefont {Tarpanov}},\ }\href@noop {}
  {\bibfield  {journal} {\bibinfo  {journal} {Phys.\ Rev. C}\ }\textbf
  {\bibinfo {volume} {87}},\ \bibinfo {pages} {034324} (\bibinfo {year}
  {2013})}\BibitemShut {NoStop}%
\bibitem [{\citenamefont {Kortelainen}\ \emph {et~al.}(2013)\citenamefont
  {Kortelainen}, \citenamefont {Erler}, \citenamefont {Nazarewicz},
  \citenamefont {Birge}, \citenamefont {Gao},\ and\ \citenamefont
  {Olsen}}]{KENBGO.13}%
  \BibitemOpen
  \bibfield  {author} {\bibinfo {author} {\bibfnamefont {M.}~\bibnamefont
  {Kortelainen}}, \bibinfo {author} {\bibfnamefont {J.}~\bibnamefont {Erler}},
  \bibinfo {author} {\bibfnamefont {W.}~\bibnamefont {Nazarewicz}}, \bibinfo
  {author} {\bibfnamefont {N.}~\bibnamefont {Birge}}, \bibinfo {author}
  {\bibfnamefont {Y.}~\bibnamefont {Gao}}, \ and\ \bibinfo {author}
  {\bibfnamefont {E.}~\bibnamefont {Olsen}},\ }\href@noop {} {\bibfield
  {journal} {\bibinfo  {journal} {Phys.\ Rev. C}\ }\textbf {\bibinfo {volume}
  {88}},\ \bibinfo {pages} {031305(R)} (\bibinfo {year} {2013})}\BibitemShut
  {NoStop}%
\bibitem [{\citenamefont {Agbemava}\ \emph {et~al.}(2019)\citenamefont
  {Agbemava}, \citenamefont {Afanasjev},\ and\ \citenamefont
  {Taninah}}]{AAT.18}%
  \BibitemOpen
  \bibfield  {author} {\bibinfo {author} {\bibfnamefont {S.~E.}\ \bibnamefont
  {Agbemava}}, \bibinfo {author} {\bibfnamefont {A.~V.}\ \bibnamefont
  {Afanasjev}}, \ and\ \bibinfo {author} {\bibfnamefont {A.}~\bibnamefont
  {Taninah}},\ }\href@noop {} {\bibfield  {journal} {\bibinfo  {journal}
  {Phys.\ Rev.\ C}\ }\textbf {\bibinfo {volume} {99}},\ \bibinfo {pages}
  {014318} (\bibinfo {year} {2019})}\BibitemShut {NoStop}%
\bibitem [{\citenamefont {Dietrich}\ and\ \citenamefont
  {Pomorski}(1998)}]{DK-prl.98}%
  \BibitemOpen
  \bibfield  {author} {\bibinfo {author} {\bibfnamefont {K.}~\bibnamefont
  {Dietrich}}\ and\ \bibinfo {author} {\bibfnamefont {K.}~\bibnamefont
  {Pomorski}},\ }\href@noop {} {\bibfield  {journal} {\bibinfo  {journal}
  {Phys. Rev. Lett.}\ }\textbf {\bibinfo {volume} {80}},\ \bibinfo {pages} {37}
  (\bibinfo {year} {1998})}\BibitemShut {NoStop}%
\bibitem [{\citenamefont {Decharg{\'e}}\ \emph {et~al.}(1999)\citenamefont
  {Decharg{\'e}}, \citenamefont {Berger}, \citenamefont {Dietrich},\ and\
  \citenamefont {Weiss}}]{DBDW.99}%
  \BibitemOpen
  \bibfield  {author} {\bibinfo {author} {\bibfnamefont {J.}~\bibnamefont
  {Decharg{\'e}}}, \bibinfo {author} {\bibfnamefont {J.-F.}\ \bibnamefont
  {Berger}}, \bibinfo {author} {\bibfnamefont {K.}~\bibnamefont {Dietrich}}, \
  and\ \bibinfo {author} {\bibfnamefont {M.}~\bibnamefont {Weiss}},\
  }\href@noop {} {\bibfield  {journal} {\bibinfo  {journal} {Phys. Lett. B}\
  }\textbf {\bibinfo {volume} {451}},\ \bibinfo {pages} {275 } (\bibinfo {year}
  {1999})}\BibitemShut {NoStop}%
\bibitem [{\citenamefont {Bender}\ \emph {et~al.}(2001)\citenamefont {Bender},
  \citenamefont {Nazarewicz},\ and\ \citenamefont {Reinhard}}]{BNR.01}%
  \BibitemOpen
  \bibfield  {author} {\bibinfo {author} {\bibfnamefont {M.}~\bibnamefont
  {Bender}}, \bibinfo {author} {\bibfnamefont {W.}~\bibnamefont {Nazarewicz}},
  \ and\ \bibinfo {author} {\bibfnamefont {P.-G.}\ \bibnamefont {Reinhard}},\
  }\href {\doibase http://dx.doi.org/10.1016/S0370-2693(01)00863-2} {\bibfield
  {journal} {\bibinfo  {journal} {Phys.\ Lett. B}\ }\textbf {\bibinfo {volume}
  {515}},\ \bibinfo {pages} {42 } (\bibinfo {year} {2001})}\BibitemShut
  {NoStop}%
\bibitem [{\citenamefont {Denisov}(2005)}]{Denisov.05}%
  \BibitemOpen
  \bibfield  {author} {\bibinfo {author} {\bibfnamefont {V.~Y.}\ \bibnamefont
  {Denisov}},\ }\href@noop {} {\bibfield  {journal} {\bibinfo  {journal} {Phys.
  At. Nuclei}\ }\textbf {\bibinfo {volume} {68}},\ \bibinfo {pages} {1133}
  (\bibinfo {year} {2005})}\BibitemShut {NoStop}%
\bibitem [{\citenamefont {Gambhir}\ \emph {et~al.}(2015)\citenamefont
  {Gambhir}, \citenamefont {Bhagwat},\ and\ \citenamefont {Gupta}}]{GBG.15}%
  \BibitemOpen
  \bibfield  {author} {\bibinfo {author} {\bibfnamefont {Y.~K.}\ \bibnamefont
  {Gambhir}}, \bibinfo {author} {\bibfnamefont {A.}~\bibnamefont {Bhagwat}}, \
  and\ \bibinfo {author} {\bibfnamefont {M.}~\bibnamefont {Gupta}},\
  }\href@noop {} {\bibfield  {journal} {\bibinfo  {journal} {J.\ Phys. G}\
  }\textbf {\bibinfo {volume} {42}},\ \bibinfo {pages} {125105} (\bibinfo
  {year} {2015})}\BibitemShut {NoStop}%
\bibitem [{\citenamefont {Ismail}\ \emph {et~al.}(2016)\citenamefont {Ismail},
  \citenamefont {Ellithi}, \citenamefont {Adel},\ and\ \citenamefont
  {Anwer}}]{IEAA.16}%
  \BibitemOpen
  \bibfield  {author} {\bibinfo {author} {\bibfnamefont {M.}~\bibnamefont
  {Ismail}}, \bibinfo {author} {\bibfnamefont {A.~Y.}\ \bibnamefont {Ellithi}},
  \bibinfo {author} {\bibfnamefont {A.}~\bibnamefont {Adel}}, \ and\ \bibinfo
  {author} {\bibfnamefont {H.}~\bibnamefont {Anwer}},\ }\href@noop {}
  {\bibfield  {journal} {\bibinfo  {journal} {J.\ Phys. G}\ }\textbf {\bibinfo
  {volume} {43}},\ \bibinfo {pages} {015101} (\bibinfo {year}
  {2016})}\BibitemShut {NoStop}%
\bibitem [{\citenamefont {Berger}\ \emph {et~al.}(2001)\citenamefont {Berger},
  \citenamefont {Bitaud}, \citenamefont {Decharg{'e}}, \citenamefont {Girod},\
  and\ \citenamefont {Dietrich}}]{BBDGK.01}%
  \BibitemOpen
  \bibfield  {author} {\bibinfo {author} {\bibfnamefont {J.~F.}\ \bibnamefont
  {Berger}}, \bibinfo {author} {\bibfnamefont {L.}~\bibnamefont {Bitaud}},
  \bibinfo {author} {\bibfnamefont {J.}~\bibnamefont {Decharg{'e}}}, \bibinfo
  {author} {\bibfnamefont {M.}~\bibnamefont {Girod}}, \ and\ \bibinfo {author}
  {\bibfnamefont {K.}~\bibnamefont {Dietrich}},\ }\href@noop {} {\bibfield
  {journal} {\bibinfo  {journal} {Nucl. Phys. A}\ }\textbf {\bibinfo {volume}
  {685}},\ \bibinfo {pages} {1c} (\bibinfo {year} {2001})}\BibitemShut
  {NoStop}%
\bibitem [{\citenamefont {Warda}(2007)}]{Warda.07}%
  \BibitemOpen
  \bibfield  {author} {\bibinfo {author} {\bibfnamefont {M.}~\bibnamefont
  {Warda}},\ }\href@noop {} {\bibfield  {journal} {\bibinfo  {journal} {Int. J.
  Mod. Phys. E}\ }\textbf {\bibinfo {volume} {16}},\ \bibinfo {pages} {452}
  (\bibinfo {year} {2007})}\BibitemShut {NoStop}%
\bibitem [{\citenamefont {Staszczak}\ and\ \citenamefont
  {Wong}(2009)}]{StaW.09}%
  \BibitemOpen
  \bibfield  {author} {\bibinfo {author} {\bibfnamefont {A.}~\bibnamefont
  {Staszczak}}\ and\ \bibinfo {author} {\bibfnamefont {C.~Y.}\ \bibnamefont
  {Wong}},\ }\href@noop {} {\bibfield  {journal} {\bibinfo  {journal} {Acta
  Phys. Pol.}\ }\textbf {\bibinfo {volume} {40}},\ \bibinfo {pages} {753}
  (\bibinfo {year} {2009})}\BibitemShut {NoStop}%
\bibitem [{\citenamefont {Brodzi\ifmmode~\acute{n}\else \'{n}\fi{}ski}\ and\
  \citenamefont {Skalski}(2013)}]{BS.13}%
  \BibitemOpen
  \bibfield  {author} {\bibinfo {author} {\bibfnamefont {W.}~\bibnamefont
  {Brodzi\ifmmode~\acute{n}\else \'{n}\fi{}ski}}\ and\ \bibinfo {author}
  {\bibfnamefont {J.}~\bibnamefont {Skalski}},\ }\href {\doibase
  10.1103/PhysRevC.88.044307} {\bibfield  {journal} {\bibinfo  {journal} {Phys.
  Rev. C}\ }\textbf {\bibinfo {volume} {88}},\ \bibinfo {pages} {044307}
  (\bibinfo {year} {2013})}\BibitemShut {NoStop}%
\bibitem [{\citenamefont {Wong}(1973)}]{Wong.73}%
  \BibitemOpen
  \bibfield  {author} {\bibinfo {author} {\bibfnamefont {C.~Y.}\ \bibnamefont
  {Wong}},\ }\href@noop {} {\bibfield  {journal} {\bibinfo  {journal} {Annals
  of Physics}\ }\textbf {\bibinfo {volume} {77}},\ \bibinfo {pages} {279}
  (\bibinfo {year} {1973})}\BibitemShut {NoStop}%
\bibitem [{\citenamefont {Staszczak}\ and\ \citenamefont {Wong}(2014)}]{SW.14}%
  \BibitemOpen
  \bibfield  {author} {\bibinfo {author} {\bibfnamefont {A.}~\bibnamefont
  {Staszczak}}\ and\ \bibinfo {author} {\bibfnamefont {C.-Y.}\ \bibnamefont
  {Wong}},\ }\href {\doibase https://doi.org/10.1016/j.physletb.2014.10.013}
  {\bibfield  {journal} {\bibinfo  {journal} {Phys. Lett. B}\ }\textbf
  {\bibinfo {volume} {738}},\ \bibinfo {pages} {401 } (\bibinfo {year}
  {2014})}\BibitemShut {NoStop}%
\bibitem [{\citenamefont {Ichikawa}\ \emph {et~al.}(2014)\citenamefont
  {Ichikawa}, \citenamefont {Matsuyanagi}, \citenamefont {Maruhn},\ and\
  \citenamefont {Itagaki}}]{IMMI.14}%
  \BibitemOpen
  \bibfield  {author} {\bibinfo {author} {\bibfnamefont {T.}~\bibnamefont
  {Ichikawa}}, \bibinfo {author} {\bibfnamefont {K.}~\bibnamefont
  {Matsuyanagi}}, \bibinfo {author} {\bibfnamefont {J.~A.}\ \bibnamefont
  {Maruhn}}, \ and\ \bibinfo {author} {\bibfnamefont {N.}~\bibnamefont
  {Itagaki}},\ }\href {\doibase 10.1103/PhysRevC.90.034314} {\bibfield
  {journal} {\bibinfo  {journal} {Phys. Rev. C}\ }\textbf {\bibinfo {volume}
  {90}},\ \bibinfo {pages} {034314} (\bibinfo {year} {2014})}\BibitemShut
  {NoStop}%
\bibitem [{\citenamefont {Kosior}\ \emph {et~al.}(2017)\citenamefont {Kosior},
  \citenamefont {Staszczak},\ and\ \citenamefont {Wong}}]{KSW.17}%
  \BibitemOpen
  \bibfield  {author} {\bibinfo {author} {\bibfnamefont {A.}~\bibnamefont
  {Kosior}}, \bibinfo {author} {\bibfnamefont {A.}~\bibnamefont {Staszczak}}, \
  and\ \bibinfo {author} {\bibfnamefont {C.-Y.}\ \bibnamefont {Wong}},\
  }\href@noop {} {\bibfield  {journal} {\bibinfo  {journal} {Acta Phys. Pol.}\
  }\textbf {\bibinfo {volume} {10}},\ \bibinfo {pages} {249} (\bibinfo {year}
  {2017})}\BibitemShut {NoStop}%
\bibitem [{\citenamefont {Nazarewicz}\ \emph {et~al.}(2002)\citenamefont
  {Nazarewicz}, \citenamefont {Bender}, \citenamefont {Cwiok}, \citenamefont
  {Heenen}, \citenamefont {Kruppa}, \citenamefont {Reinhard},\ and\
  \citenamefont {Vertse}}]{NBCHKRV.02}%
  \BibitemOpen
  \bibfield  {author} {\bibinfo {author} {\bibfnamefont {W.}~\bibnamefont
  {Nazarewicz}}, \bibinfo {author} {\bibfnamefont {M.}~\bibnamefont {Bender}},
  \bibinfo {author} {\bibfnamefont {S.}~\bibnamefont {Cwiok}}, \bibinfo
  {author} {\bibfnamefont {P.~H.}\ \bibnamefont {Heenen}}, \bibinfo {author}
  {\bibfnamefont {A.~T.}\ \bibnamefont {Kruppa}}, \bibinfo {author}
  {\bibfnamefont {P.-G.}\ \bibnamefont {Reinhard}}, \ and\ \bibinfo {author}
  {\bibfnamefont {T.}~\bibnamefont {Vertse}},\ }\href@noop {} {\bibfield
  {journal} {\bibinfo  {journal} {Nucl. Phys. A}\ }\textbf {\bibinfo {volume}
  {701}},\ \bibinfo {pages} {165c} (\bibinfo {year} {2002})}\BibitemShut
  {NoStop}%
\bibitem [{\citenamefont {Staszczak}\ \emph {et~al.}(2017)\citenamefont
  {Staszczak}, \citenamefont {Wong},\ and\ \citenamefont {Kosior}}]{SWK.17}%
  \BibitemOpen
  \bibfield  {author} {\bibinfo {author} {\bibfnamefont {A.}~\bibnamefont
  {Staszczak}}, \bibinfo {author} {\bibfnamefont {C.-Y.}\ \bibnamefont {Wong}},
  \ and\ \bibinfo {author} {\bibfnamefont {A.}~\bibnamefont {Kosior}},\ }\href
  {\doibase 10.1103/PhysRevC.95.054315} {\bibfield  {journal} {\bibinfo
  {journal} {Phys. Rev. C}\ }\textbf {\bibinfo {volume} {95}},\ \bibinfo
  {pages} {054315} (\bibinfo {year} {2017})}\BibitemShut {NoStop}%
\bibitem [{\citenamefont {Afanasjev}\ \emph {et~al.}(1999)\citenamefont
  {Afanasjev}, \citenamefont {Fossan}, \citenamefont {Lane},\ and\
  \citenamefont {Ragnarsson}}]{PhysRep-SBT}%
  \BibitemOpen
  \bibfield  {author} {\bibinfo {author} {\bibfnamefont {A.~V.}\ \bibnamefont
  {Afanasjev}}, \bibinfo {author} {\bibfnamefont {D.~B.}\ \bibnamefont
  {Fossan}}, \bibinfo {author} {\bibfnamefont {G.~J.}\ \bibnamefont {Lane}}, \
  and\ \bibinfo {author} {\bibfnamefont {I.}~\bibnamefont {Ragnarsson}},\
  }\href@noop {} {\bibfield  {journal} {\bibinfo  {journal} {Phys.\ Rep.}\
  }\textbf {\bibinfo {volume} {322}},\ \bibinfo {pages} {1} (\bibinfo {year}
  {1999})}\BibitemShut {NoStop}%
\bibitem [{\citenamefont {Agbemava}\ \emph {et~al.}(2016)\citenamefont
  {Agbemava}, \citenamefont {Afanasjev},\ and\ \citenamefont {Ring}}]{AAR.16}%
  \BibitemOpen
  \bibfield  {author} {\bibinfo {author} {\bibfnamefont {S.~E.}\ \bibnamefont
  {Agbemava}}, \bibinfo {author} {\bibfnamefont {A.~V.}\ \bibnamefont
  {Afanasjev}}, \ and\ \bibinfo {author} {\bibfnamefont {P.}~\bibnamefont
  {Ring}},\ }\href@noop {} {\bibfield  {journal} {\bibinfo  {journal} {Phys.
  Rev. C}\ }\textbf {\bibinfo {volume} {93}},\ \bibinfo {pages} {044304}
  (\bibinfo {year} {2016})}\BibitemShut {NoStop}%
\bibitem [{\citenamefont {Agbemava}\ \emph {et~al.}(2017)\citenamefont
  {Agbemava}, \citenamefont {Afanasjev}, \citenamefont {Ray},\ and\
  \citenamefont {Ring}}]{AARR.17}%
  \BibitemOpen
  \bibfield  {author} {\bibinfo {author} {\bibfnamefont {S.~E.}\ \bibnamefont
  {Agbemava}}, \bibinfo {author} {\bibfnamefont {A.~V.}\ \bibnamefont
  {Afanasjev}}, \bibinfo {author} {\bibfnamefont {D.}~\bibnamefont {Ray}}, \
  and\ \bibinfo {author} {\bibfnamefont {P.}~\bibnamefont {Ring}},\ }\href
  {\doibase 10.1103/PhysRevC.95.054324} {\bibfield  {journal} {\bibinfo
  {journal} {Phys. Rev. C}\ }\textbf {\bibinfo {volume} {95}},\ \bibinfo
  {pages} {054324} (\bibinfo {year} {2017})}\BibitemShut {NoStop}%
\bibitem [{\citenamefont {Abusara}\ \emph {et~al.}(2010)\citenamefont
  {Abusara}, \citenamefont {Afanasjev},\ and\ \citenamefont {Ring}}]{AAR.10}%
  \BibitemOpen
  \bibfield  {author} {\bibinfo {author} {\bibfnamefont {H.}~\bibnamefont
  {Abusara}}, \bibinfo {author} {\bibfnamefont {A.~V.}\ \bibnamefont
  {Afanasjev}}, \ and\ \bibinfo {author} {\bibfnamefont {P.}~\bibnamefont
  {Ring}},\ }\href@noop {} {\bibfield  {journal} {\bibinfo  {journal} {Phys.\
  Rev. C}\ }\textbf {\bibinfo {volume} {82}},\ \bibinfo {pages} {044303}
  (\bibinfo {year} {2010})}\BibitemShut {NoStop}%
\bibitem [{\citenamefont {Nik\v{s}i\'{c}}\ \emph {et~al.}(2008)\citenamefont
  {Nik\v{s}i\'{c}}, \citenamefont {Vretenar},\ and\ \citenamefont
  {Ring}}]{DD-PC1}%
  \BibitemOpen
  \bibfield  {author} {\bibinfo {author} {\bibfnamefont {T.}~\bibnamefont
  {Nik\v{s}i\'{c}}}, \bibinfo {author} {\bibfnamefont {D.}~\bibnamefont
  {Vretenar}}, \ and\ \bibinfo {author} {\bibfnamefont {P.}~\bibnamefont
  {Ring}},\ }\href@noop {} {\bibfield  {journal} {\bibinfo  {journal} {Phys.\
  Rev. C}\ }\textbf {\bibinfo {volume} {78}},\ \bibinfo {pages} {034318}
  (\bibinfo {year} {2008})}\BibitemShut {NoStop}%
\bibitem [{\citenamefont {Agbemava}\ \emph {et~al.}(2015)\citenamefont
  {Agbemava}, \citenamefont {Afanasjev}, \citenamefont {Nakatsukasa},\ and\
  \citenamefont {Ring}}]{AANR.15}%
  \BibitemOpen
  \bibfield  {author} {\bibinfo {author} {\bibfnamefont {S.~E.}\ \bibnamefont
  {Agbemava}}, \bibinfo {author} {\bibfnamefont {A.~V.}\ \bibnamefont
  {Afanasjev}}, \bibinfo {author} {\bibfnamefont {T.}~\bibnamefont
  {Nakatsukasa}}, \ and\ \bibinfo {author} {\bibfnamefont {P.}~\bibnamefont
  {Ring}},\ }\href {\doibase 10.1103/PhysRevC.92.054310} {\bibfield  {journal}
  {\bibinfo  {journal} {Phys. Rev. C}\ }\textbf {\bibinfo {volume} {92}},\
  \bibinfo {pages} {054310} (\bibinfo {year} {2015})}\BibitemShut {NoStop}%
\bibitem [{\citenamefont {Prassa}\ \emph {et~al.}(2012)\citenamefont {Prassa},
  \citenamefont {Nik\ifmmode \check{s}\else \v{s}\fi{}i\ifmmode~\acute{c}\else
  \'{c}\fi{}}, \citenamefont {Lalazissis},\ and\ \citenamefont
  {Vretenar}}]{PNLV.12}%
  \BibitemOpen
  \bibfield  {author} {\bibinfo {author} {\bibfnamefont {V.}~\bibnamefont
  {Prassa}}, \bibinfo {author} {\bibfnamefont {T.}~\bibnamefont {Nik\ifmmode
  \check{s}\else \v{s}\fi{}i\ifmmode~\acute{c}\else \'{c}\fi{}}}, \bibinfo
  {author} {\bibfnamefont {G.~A.}\ \bibnamefont {Lalazissis}}, \ and\ \bibinfo
  {author} {\bibfnamefont {D.}~\bibnamefont {Vretenar}},\ }\href@noop {}
  {\bibfield  {journal} {\bibinfo  {journal} {Phys. Rev. C}\ }\textbf {\bibinfo
  {volume} {86}},\ \bibinfo {pages} {024317} (\bibinfo {year}
  {2012})}\BibitemShut {NoStop}%
\bibitem [{\citenamefont {Lu}\ \emph {et~al.}(2012)\citenamefont {Lu},
  \citenamefont {Zhao},\ and\ \citenamefont {Zhou}}]{LZZ.12}%
  \BibitemOpen
  \bibfield  {author} {\bibinfo {author} {\bibfnamefont {B.-N.}\ \bibnamefont
  {Lu}}, \bibinfo {author} {\bibfnamefont {E.-G.}\ \bibnamefont {Zhao}}, \ and\
  \bibinfo {author} {\bibfnamefont {S.-G.}\ \bibnamefont {Zhou}},\ }\href@noop
  {} {\bibfield  {journal} {\bibinfo  {journal} {Phys.\ Rev. C}\ }\textbf
  {\bibinfo {volume} {85}},\ \bibinfo {pages} {011301} (\bibinfo {year}
  {2012})}\BibitemShut {NoStop}%
\bibitem [{\citenamefont {Lalazissis}\ \emph {et~al.}(2005)\citenamefont
  {Lalazissis}, \citenamefont {Nik{\v{s}}i{\'{c}}}, \citenamefont {Vretenar},\
  and\ \citenamefont {Ring}}]{DD-ME2}%
  \BibitemOpen
  \bibfield  {author} {\bibinfo {author} {\bibfnamefont {G.~A.}\ \bibnamefont
  {Lalazissis}}, \bibinfo {author} {\bibfnamefont {T.}~\bibnamefont
  {Nik{\v{s}}i{\'{c}}}}, \bibinfo {author} {\bibfnamefont {D.}~\bibnamefont
  {Vretenar}}, \ and\ \bibinfo {author} {\bibfnamefont {P.}~\bibnamefont
  {Ring}},\ }\href@noop {} {\bibfield  {journal} {\bibinfo  {journal} {Phys.\
  Rev. C}\ }\textbf {\bibinfo {volume} {71}},\ \bibinfo {pages} {024312}
  (\bibinfo {year} {2005})}\BibitemShut {NoStop}%
\bibitem [{\citenamefont {Zhao}\ \emph {et~al.}(2010)\citenamefont {Zhao},
  \citenamefont {Li}, \citenamefont {Yao},\ and\ \citenamefont
  {Meng}}]{PC-PK1}%
  \BibitemOpen
  \bibfield  {author} {\bibinfo {author} {\bibfnamefont {P.~W.}\ \bibnamefont
  {Zhao}}, \bibinfo {author} {\bibfnamefont {Z.~P.}\ \bibnamefont {Li}},
  \bibinfo {author} {\bibfnamefont {J.~M.}\ \bibnamefont {Yao}}, \ and\
  \bibinfo {author} {\bibfnamefont {J.}~\bibnamefont {Meng}},\ }\href@noop {}
  {\bibfield  {journal} {\bibinfo  {journal} {Phys.\ Rev. C}\ }\textbf
  {\bibinfo {volume} {82}},\ \bibinfo {pages} {054319} (\bibinfo {year}
  {2010})}\BibitemShut {NoStop}%
\bibitem [{\citenamefont {Lalazissis}\ \emph {et~al.}(2009)\citenamefont
  {Lalazissis}, \citenamefont {Karatzikos}, \citenamefont {Fossion},
  \citenamefont {Arteaga}, \citenamefont {Afanasjev},\ and\ \citenamefont
  {Ring}}]{NL3*}%
  \BibitemOpen
  \bibfield  {author} {\bibinfo {author} {\bibfnamefont {G.~A.}\ \bibnamefont
  {Lalazissis}}, \bibinfo {author} {\bibfnamefont {S.}~\bibnamefont
  {Karatzikos}}, \bibinfo {author} {\bibfnamefont {R.}~\bibnamefont {Fossion}},
  \bibinfo {author} {\bibfnamefont {D.~P.}\ \bibnamefont {Arteaga}}, \bibinfo
  {author} {\bibfnamefont {A.~V.}\ \bibnamefont {Afanasjev}}, \ and\ \bibinfo
  {author} {\bibfnamefont {P.}~\bibnamefont {Ring}},\ }\href@noop {} {\bibfield
   {journal} {\bibinfo  {journal} {Phys.\ Lett.}\ }\textbf {\bibinfo {volume}
  {B671}},\ \bibinfo {pages} {36} (\bibinfo {year} {2009})}\BibitemShut
  {NoStop}%
\bibitem [{\citenamefont {Ring}\ and\ \citenamefont {Schuck}(1980)}]{RS.80}%
  \BibitemOpen
  \bibfield  {author} {\bibinfo {author} {\bibfnamefont {P.}~\bibnamefont
  {Ring}}\ and\ \bibinfo {author} {\bibfnamefont {P.}~\bibnamefont {Schuck}},\
  }\href@noop {} {\bibfield  {journal} {\bibinfo  {journal} {{\em The Nuclear
  Many-Body Problem} (Springer-Verlag, Berlin)}\ } (\bibinfo {year}
  {1980})}\BibitemShut {NoStop}%
\bibitem [{\citenamefont {Bonche}\ \emph {et~al.}(2005)\citenamefont {Bonche},
  \citenamefont {Flocard},\ and\ \citenamefont {Heenen}}]{BFH.05}%
  \BibitemOpen
  \bibfield  {author} {\bibinfo {author} {\bibfnamefont {P.}~\bibnamefont
  {Bonche}}, \bibinfo {author} {\bibfnamefont {H.}~\bibnamefont {Flocard}}, \
  and\ \bibinfo {author} {\bibfnamefont {P.~H.}\ \bibnamefont {Heenen}},\
  }\href@noop {} {\bibfield  {journal} {\bibinfo  {journal} {Comp.\ Phys.
  Comm.}\ }\textbf {\bibinfo {volume} {171}},\ \bibinfo {pages} {49} (\bibinfo
  {year} {2005})}\BibitemShut {NoStop}%
\bibitem [{\citenamefont {Afanasjev}\ and\ \citenamefont
  {Frauendorf}(2005)}]{AF.05-dep}%
  \BibitemOpen
  \bibfield  {author} {\bibinfo {author} {\bibfnamefont {A.~V.}\ \bibnamefont
  {Afanasjev}}\ and\ \bibinfo {author} {\bibfnamefont {S.}~\bibnamefont
  {Frauendorf}},\ }\href {\doibase 10.1103/PhysRevC.71.024308} {\bibfield
  {journal} {\bibinfo  {journal} {Phys. Rev. C}\ }\textbf {\bibinfo {volume}
  {71}},\ \bibinfo {pages} {024308} (\bibinfo {year} {2005})}\BibitemShut
  {NoStop}%
\bibitem [{\citenamefont {Karatzikos}\ \emph {et~al.}(2010)\citenamefont
  {Karatzikos}, \citenamefont {Afanasjev}, \citenamefont {Lalazissis},\ and\
  \citenamefont {Ring}}]{KALR.10}%
  \BibitemOpen
  \bibfield  {author} {\bibinfo {author} {\bibfnamefont {S.}~\bibnamefont
  {Karatzikos}}, \bibinfo {author} {\bibfnamefont {A.~V.}\ \bibnamefont
  {Afanasjev}}, \bibinfo {author} {\bibfnamefont {G.~A.}\ \bibnamefont
  {Lalazissis}}, \ and\ \bibinfo {author} {\bibfnamefont {P.}~\bibnamefont
  {Ring}},\ }\href@noop {} {\bibfield  {journal} {\bibinfo  {journal} {Phys.\
  Lett. B}\ }\textbf {\bibinfo {volume} {689}},\ \bibinfo {pages} {72}
  (\bibinfo {year} {2010})}\BibitemShut {NoStop}%
\bibitem [{\citenamefont {Tian}\ \emph {et~al.}(2009)\citenamefont {Tian},
  \citenamefont {Ma},\ and\ \citenamefont {Ring}}]{TMR.09}%
  \BibitemOpen
  \bibfield  {author} {\bibinfo {author} {\bibfnamefont {Y.}~\bibnamefont
  {Tian}}, \bibinfo {author} {\bibfnamefont {Z.~Y.}\ \bibnamefont {Ma}}, \ and\
  \bibinfo {author} {\bibfnamefont {P.}~\bibnamefont {Ring}},\ }\href@noop {}
  {\bibfield  {journal} {\bibinfo  {journal} {Phys.\ Lett. B}\ }\textbf
  {\bibinfo {volume} {676}},\ \bibinfo {pages} {44} (\bibinfo {year}
  {2009})}\BibitemShut {NoStop}%
\bibitem [{\citenamefont {Afanasjev}\ and\ \citenamefont
  {Abdurazakov}(2013)}]{AO.13}%
  \BibitemOpen
  \bibfield  {author} {\bibinfo {author} {\bibfnamefont {A.~V.}\ \bibnamefont
  {Afanasjev}}\ and\ \bibinfo {author} {\bibfnamefont {O.}~\bibnamefont
  {Abdurazakov}},\ }\href@noop {} {\bibfield  {journal} {\bibinfo  {journal}
  {Phys.\ Rev. C}\ }\textbf {\bibinfo {volume} {88}},\ \bibinfo {pages}
  {014320} (\bibinfo {year} {2013})}\BibitemShut {NoStop}%
\bibitem [{\citenamefont {Dobaczewski}\ \emph {et~al.}(2015)\citenamefont
  {Dobaczewski}, \citenamefont {Afanasjev}, \citenamefont {Bender},
  \citenamefont {Robledo},\ and\ \citenamefont {Shi}}]{DABRS.15}%
  \BibitemOpen
  \bibfield  {author} {\bibinfo {author} {\bibfnamefont {J.}~\bibnamefont
  {Dobaczewski}}, \bibinfo {author} {\bibfnamefont {A.~V.}\ \bibnamefont
  {Afanasjev}}, \bibinfo {author} {\bibfnamefont {M.}~\bibnamefont {Bender}},
  \bibinfo {author} {\bibfnamefont {L.~M.}\ \bibnamefont {Robledo}}, \ and\
  \bibinfo {author} {\bibfnamefont {Y.}~\bibnamefont {Shi}},\ }\href {\doibase
  http://dx.doi.org/10.1016/j.nuclphysa.2015.07.015} {\bibfield  {journal}
  {\bibinfo  {journal} {Nucl. Phys. A}\ }\textbf {\bibinfo {volume} {944}},\
  \bibinfo {pages} {388 } (\bibinfo {year} {2015})}\BibitemShut {NoStop}%
\bibitem [{\citenamefont {Abrahamyan}\ \emph {et~al.}(2012)\citenamefont
  {Abrahamyan} \emph {et~al.}}]{PREX.12}%
  \BibitemOpen
  \bibfield  {author} {\bibinfo {author} {\bibfnamefont {S.}~\bibnamefont
  {Abrahamyan}} \emph {et~al.},\ }\href@noop {} {\bibfield  {journal} {\bibinfo
   {journal} {Phys.\ Rev.\ Lett.}\ }\textbf {\bibinfo {volume} {108}},\
  \bibinfo {pages} {112502} (\bibinfo {year} {2012})}\BibitemShut {NoStop}%
\bibitem [{PRE()}]{PREX-CREX}%
  \BibitemOpen
  \href@noop {} {\bibinfo  {journal} {The PREX-II proposal, unpublished,
  available at hallaweb.jlab.org/parity/prex}\ }\BibitemShut {NoStop}%
\bibitem [{\citenamefont {Schuetrumpf}\ \emph {et~al.}(2017)\citenamefont
  {Schuetrumpf}, \citenamefont {Nazarewicz},\ and\ \citenamefont
  {Reinhard}}]{SNR.18}%
  \BibitemOpen
\bibfield  {journal} {  }\bibfield  {author} {\bibinfo {author} {\bibfnamefont
  {B.}~\bibnamefont {Schuetrumpf}}, \bibinfo {author} {\bibfnamefont
  {W.}~\bibnamefont {Nazarewicz}}, \ and\ \bibinfo {author} {\bibfnamefont
  {P.-G.}\ \bibnamefont {Reinhard}},\ }\href {\doibase
  10.1103/PhysRevC.96.024306} {\bibfield  {journal} {\bibinfo  {journal} {Phys.
  Rev. C}\ }\textbf {\bibinfo {volume} {96}},\ \bibinfo {pages} {024306}
  (\bibinfo {year} {2017})}\BibitemShut {NoStop}%
\bibitem [{\citenamefont {Bender}\ \emph {et~al.}(1999)\citenamefont {Bender},
  \citenamefont {Rutz}, \citenamefont {Reinhard}, \citenamefont {Maruhn},\ and\
  \citenamefont {Greiner}}]{BRRMG.99}%
  \BibitemOpen
  \bibfield  {author} {\bibinfo {author} {\bibfnamefont {M.}~\bibnamefont
  {Bender}}, \bibinfo {author} {\bibfnamefont {K.}~\bibnamefont {Rutz}},
  \bibinfo {author} {\bibfnamefont {P.-G.}\ \bibnamefont {Reinhard}}, \bibinfo
  {author} {\bibfnamefont {J.~A.}\ \bibnamefont {Maruhn}}, \ and\ \bibinfo
  {author} {\bibfnamefont {W.}~\bibnamefont {Greiner}},\ }\href@noop {}
  {\bibfield  {journal} {\bibinfo  {journal} {Phys.\ Rev. C}\ }\textbf
  {\bibinfo {volume} {60}},\ \bibinfo {pages} {034304} (\bibinfo {year}
  {1999})}\BibitemShut {NoStop}%
\bibitem [{\citenamefont {Warda}\ \emph {et~al.}(2018)\citenamefont {Warda},
  \citenamefont {Zdeb},\ and\ \citenamefont {Robledo}}]{WZR.18}%
  \BibitemOpen
  \bibfield  {author} {\bibinfo {author} {\bibfnamefont {M.}~\bibnamefont
  {Warda}}, \bibinfo {author} {\bibfnamefont {A.}~\bibnamefont {Zdeb}}, \ and\
  \bibinfo {author} {\bibfnamefont {L.~M.}\ \bibnamefont {Robledo}},\ }\href
  {\doibase 10.1103/PhysRevC.98.041602} {\bibfield  {journal} {\bibinfo
  {journal} {Phys. Rev. C}\ }\textbf {\bibinfo {volume} {98}},\ \bibinfo
  {pages} {041602} (\bibinfo {year} {2018})}\BibitemShut {NoStop}%
\bibitem [{\citenamefont {Afanasjev}\ \emph {et~al.}(2000)\citenamefont
  {Afanasjev}, \citenamefont {Ring},\ and\ \citenamefont {K{\"o}nig}}]{CRHB}%
  \BibitemOpen
  \bibfield  {author} {\bibinfo {author} {\bibfnamefont {A.~V.}\ \bibnamefont
  {Afanasjev}}, \bibinfo {author} {\bibfnamefont {P.}~\bibnamefont {Ring}}, \
  and\ \bibinfo {author} {\bibfnamefont {J.}~\bibnamefont {K{\"o}nig}},\
  }\href@noop {} {\bibfield  {journal} {\bibinfo  {journal} {Nucl.\ Phys.}\
  }\textbf {\bibinfo {volume} {A676}},\ \bibinfo {pages} {196} (\bibinfo {year}
  {2000})}\BibitemShut {NoStop}%
\bibitem [{\citenamefont {Valor}\ \emph {et~al.}(2000)\citenamefont {Valor},
  \citenamefont {Egido},\ and\ \citenamefont {Robledo}}]{AER.00}%
  \BibitemOpen
  \bibfield  {author} {\bibinfo {author} {\bibfnamefont {A.}~\bibnamefont
  {Valor}}, \bibinfo {author} {\bibfnamefont {J.~L.}\ \bibnamefont {Egido}}, \
  and\ \bibinfo {author} {\bibfnamefont {L.~M.}\ \bibnamefont {Robledo}},\
  }\href@noop {} {\bibfield  {journal} {\bibinfo  {journal} {Nucl.\ Phys. A}\
  }\textbf {\bibinfo {volume} {665}},\ \bibinfo {pages} {46} (\bibinfo {year}
  {2000})}\BibitemShut {NoStop}%
\bibitem [{\citenamefont {Strutinsky}(1967)}]{Strutinsky1967_NPA95-420}%
  \BibitemOpen
  \bibfield  {author} {\bibinfo {author} {\bibfnamefont {V.~M.}\ \bibnamefont
  {Strutinsky}},\ }\href@noop {} {\bibfield  {journal} {\bibinfo  {journal}
  {Nucl. Phys. A}\ }\textbf {\bibinfo {volume} {95}},\ \bibinfo {pages} {420 }
  (\bibinfo {year} {1967})}\BibitemShut {NoStop}%
\bibitem [{\citenamefont {Strutinsky}(1968)}]{Strutinsky1968_NPA122-1}%
  \BibitemOpen
  \bibfield  {author} {\bibinfo {author} {\bibfnamefont {V.~M.}\ \bibnamefont
  {Strutinsky}},\ }\href@noop {} {\bibfield  {journal} {\bibinfo  {journal}
  {Nucl. Phys. A}\ }\textbf {\bibinfo {volume} {122}},\ \bibinfo {pages} {1 }
  (\bibinfo {year} {1968})}\BibitemShut {NoStop}%
\bibitem [{\citenamefont {Anguiano}\ \emph {et~al.}(2002)\citenamefont
  {Anguiano}, \citenamefont {Egido},\ and\ \citenamefont {Robledo}}]{AER.02}%
  \BibitemOpen
  \bibfield  {author} {\bibinfo {author} {\bibfnamefont {M.}~\bibnamefont
  {Anguiano}}, \bibinfo {author} {\bibfnamefont {J.}~\bibnamefont {Egido}}, \
  and\ \bibinfo {author} {\bibfnamefont {L.}~\bibnamefont {Robledo}},\ }\href
  {\doibase https://doi.org/10.1016/S0370-2693(02)02557-1} {\bibfield
  {journal} {\bibinfo  {journal} {Phys. Lett. B}\ }\textbf {\bibinfo {volume}
  {545}},\ \bibinfo {pages} {62 } (\bibinfo {year} {2002})}\BibitemShut
  {NoStop}%
\bibitem [{\citenamefont {B{\"u}rvenich}\ \emph {et~al.}(2004)\citenamefont
  {B{\"u}rvenich}, \citenamefont {Bender}, \citenamefont {Maruhn},\ and\
  \citenamefont {Reinhard}}]{BBM.04}%
  \BibitemOpen
  \bibfield  {author} {\bibinfo {author} {\bibfnamefont {T.}~\bibnamefont
  {B{\"u}rvenich}}, \bibinfo {author} {\bibfnamefont {M.}~\bibnamefont
  {Bender}}, \bibinfo {author} {\bibfnamefont {J.~A.}\ \bibnamefont {Maruhn}},
  \ and\ \bibinfo {author} {\bibfnamefont {P.-G.}\ \bibnamefont {Reinhard}},\
  }\href@noop {} {\bibfield  {journal} {\bibinfo  {journal} {Phys.\ Rev. C}\
  }\textbf {\bibinfo {volume} {69}},\ \bibinfo {pages} {014307} (\bibinfo
  {year} {2004})}\BibitemShut {NoStop}%
\bibitem [{\citenamefont {Abusara}\ \emph {et~al.}(2012)\citenamefont
  {Abusara}, \citenamefont {Afanasjev},\ and\ \citenamefont {Ring}}]{AAR.12}%
  \BibitemOpen
  \bibfield  {author} {\bibinfo {author} {\bibfnamefont {H.}~\bibnamefont
  {Abusara}}, \bibinfo {author} {\bibfnamefont {A.~V.}\ \bibnamefont
  {Afanasjev}}, \ and\ \bibinfo {author} {\bibfnamefont {P.}~\bibnamefont
  {Ring}},\ }\href@noop {} {\bibfield  {journal} {\bibinfo  {journal} {Phys.\
  Rev. C}\ }\textbf {\bibinfo {volume} {85}},\ \bibinfo {pages} {024314}
  (\bibinfo {year} {2012})}\BibitemShut {NoStop}%
\bibitem [{\citenamefont {Warda}\ \emph {et~al.}(2002)\citenamefont {Warda},
  \citenamefont {Egido}, \citenamefont {Robledo},\ and\ \citenamefont
  {Pomorski}}]{WERP.02}%
  \BibitemOpen
  \bibfield  {author} {\bibinfo {author} {\bibfnamefont {M.}~\bibnamefont
  {Warda}}, \bibinfo {author} {\bibfnamefont {J.~L.}\ \bibnamefont {Egido}},
  \bibinfo {author} {\bibfnamefont {L.~M.}\ \bibnamefont {Robledo}}, \ and\
  \bibinfo {author} {\bibfnamefont {K.}~\bibnamefont {Pomorski}},\ }\href@noop
  {} {\bibfield  {journal} {\bibinfo  {journal} {Phys. Rev. C}\ }\textbf
  {\bibinfo {volume} {66}},\ \bibinfo {pages} {014310} (\bibinfo {year}
  {2002})}\BibitemShut {NoStop}%
\bibitem [{\citenamefont {Staszczak}\ \emph {et~al.}(2009)\citenamefont
  {Staszczak}, \citenamefont {Baran}, \citenamefont {Dobaczewski},\ and\
  \citenamefont {Nazarewicz}}]{SBDN.09}%
  \BibitemOpen
  \bibfield  {author} {\bibinfo {author} {\bibfnamefont {A.}~\bibnamefont
  {Staszczak}}, \bibinfo {author} {\bibfnamefont {A.}~\bibnamefont {Baran}},
  \bibinfo {author} {\bibfnamefont {J.}~\bibnamefont {Dobaczewski}}, \ and\
  \bibinfo {author} {\bibfnamefont {W.}~\bibnamefont {Nazarewicz}},\ }\href
  {\doibase 10.1103/PhysRevC.80.014309} {\bibfield  {journal} {\bibinfo
  {journal} {Phys. Rev. C}\ }\textbf {\bibinfo {volume} {80}},\ \bibinfo
  {pages} {014309} (\bibinfo {year} {2009})}\BibitemShut {NoStop}%
\bibitem [{\citenamefont {M\"oller}\ \emph {et~al.}(2009)\citenamefont
  {M\"oller}, \citenamefont {Sierk}, \citenamefont {Ichikawa}, \citenamefont
  {Iwamoto}, \citenamefont {Bengtsson}, \citenamefont {Uhrenholt},\ and\
  \citenamefont {\AA{}berg}}]{MSI.09}%
  \BibitemOpen
  \bibfield  {author} {\bibinfo {author} {\bibfnamefont {P.}~\bibnamefont
  {M\"oller}}, \bibinfo {author} {\bibfnamefont {A.~J.}\ \bibnamefont {Sierk}},
  \bibinfo {author} {\bibfnamefont {T.}~\bibnamefont {Ichikawa}}, \bibinfo
  {author} {\bibfnamefont {A.}~\bibnamefont {Iwamoto}}, \bibinfo {author}
  {\bibfnamefont {R.}~\bibnamefont {Bengtsson}}, \bibinfo {author}
  {\bibfnamefont {H.}~\bibnamefont {Uhrenholt}}, \ and\ \bibinfo {author}
  {\bibfnamefont {S.}~\bibnamefont {\AA{}berg}},\ }\href {\doibase
  10.1103/PhysRevC.79.064304} {\bibfield  {journal} {\bibinfo  {journal} {Phys.
  Rev. C}\ }\textbf {\bibinfo {volume} {79}},\ \bibinfo {pages} {064304}
  (\bibinfo {year} {2009})}\BibitemShut {NoStop}%
\bibitem [{\citenamefont {Schunck}\ and\ \citenamefont
  {Robledo}(2016)}]{SR.16}%
  \BibitemOpen
  \bibfield  {author} {\bibinfo {author} {\bibfnamefont {N.}~\bibnamefont
  {Schunck}}\ and\ \bibinfo {author} {\bibfnamefont {L.~M.}\ \bibnamefont
  {Robledo}},\ }\href@noop {} {\bibfield  {journal} {\bibinfo  {journal} {Rep.
  Prog. Phys.}\ }\textbf {\bibinfo {volume} {79}},\ \bibinfo {pages} {116301}
  (\bibinfo {year} {2016})}\BibitemShut {NoStop}%
\bibitem [{\citenamefont {Arnould}\ \emph {et~al.}(2007)\citenamefont
  {Arnould}, \citenamefont {Goriely},\ and\ \citenamefont
  {Takahashi}}]{AGT.07}%
  \BibitemOpen
  \bibfield  {author} {\bibinfo {author} {\bibfnamefont {M.}~\bibnamefont
  {Arnould}}, \bibinfo {author} {\bibfnamefont {S.}~\bibnamefont {Goriely}}, \
  and\ \bibinfo {author} {\bibfnamefont {K.}~\bibnamefont {Takahashi}},\
  }\href@noop {} {\bibfield  {journal} {\bibinfo  {journal} {Phys.\ Rep.}\
  }\textbf {\bibinfo {volume} {450}},\ \bibinfo {pages} {97} (\bibinfo {year}
  {2007})}\BibitemShut {NoStop}%
\bibitem [{\citenamefont {Mart\'{i}nez-Pinedo}\ \emph
  {et~al.}(2007)\citenamefont {Mart\'{i}nez-Pinedo}, \citenamefont {Mocelj},
  \citenamefont {Zinner}, \citenamefont {Keli\'{c}}, \citenamefont {Langanke},
  \citenamefont {Panov}, \citenamefont {Pfeiffer}, \citenamefont {Rauscher},
  \citenamefont {Schmidt},\ and\ \citenamefont {Thielemann}}]{PMZ.07}%
  \BibitemOpen
  \bibfield  {author} {\bibinfo {author} {\bibfnamefont {G.}~\bibnamefont
  {Mart\'{i}nez-Pinedo}}, \bibinfo {author} {\bibfnamefont {D.}~\bibnamefont
  {Mocelj}}, \bibinfo {author} {\bibfnamefont {N.~T.}\ \bibnamefont {Zinner}},
  \bibinfo {author} {\bibfnamefont {A.}~\bibnamefont {Keli\'{c}}}, \bibinfo
  {author} {\bibfnamefont {K.}~\bibnamefont {Langanke}}, \bibinfo {author}
  {\bibfnamefont {I.}~\bibnamefont {Panov}}, \bibinfo {author} {\bibfnamefont
  {B.}~\bibnamefont {Pfeiffer}}, \bibinfo {author} {\bibfnamefont
  {T.}~\bibnamefont {Rauscher}}, \bibinfo {author} {\bibfnamefont {K.-H.}\
  \bibnamefont {Schmidt}}, \ and\ \bibinfo {author} {\bibfnamefont {F.-K.}\
  \bibnamefont {Thielemann}},\ }\href@noop {} {\bibfield  {journal} {\bibinfo
  {journal} {Prog.~ Part.~ Nucl.~ Phys.}\ }\textbf {\bibinfo {volume} {59}},\
  \bibinfo {pages} {199} (\bibinfo {year} {2007})}\BibitemShut {NoStop}%
\bibitem [{\citenamefont {Goriely}\ \emph {et~al.}(2007)\citenamefont
  {Goriely}, \citenamefont {Samyn},\ and\ \citenamefont
  {Pearson}}]{HFB14-BSk14}%
  \BibitemOpen
  \bibfield  {author} {\bibinfo {author} {\bibfnamefont {S.}~\bibnamefont
  {Goriely}}, \bibinfo {author} {\bibfnamefont {M.}~\bibnamefont {Samyn}}, \
  and\ \bibinfo {author} {\bibfnamefont {J.~M.}\ \bibnamefont {Pearson}},\
  }\href {\doibase 10.1103/PhysRevC.75.064312} {\bibfield  {journal} {\bibinfo
  {journal} {Phys. Rev. C}\ }\textbf {\bibinfo {volume} {75}},\ \bibinfo
  {pages} {064312} (\bibinfo {year} {2007})}\BibitemShut {NoStop}%
\bibitem [{\citenamefont {Erler}\ \emph
  {et~al.}(2012{\natexlab{b}})\citenamefont {Erler}, \citenamefont {Langanke},
  \citenamefont {Loens}, \citenamefont {Martinez-Pinedo},\ and\ \citenamefont
  {Reinhard}}]{ELLMR.12}%
  \BibitemOpen
  \bibfield  {author} {\bibinfo {author} {\bibfnamefont {J.}~\bibnamefont
  {Erler}}, \bibinfo {author} {\bibfnamefont {K.}~\bibnamefont {Langanke}},
  \bibinfo {author} {\bibfnamefont {H.~P.}\ \bibnamefont {Loens}}, \bibinfo
  {author} {\bibfnamefont {G.}~\bibnamefont {Martinez-Pinedo}}, \ and\ \bibinfo
  {author} {\bibfnamefont {P.-G.}\ \bibnamefont {Reinhard}},\ }\href@noop {}
  {\bibfield  {journal} {\bibinfo  {journal} {Phys.\ Rev. C}\ }\textbf
  {\bibinfo {volume} {85}},\ \bibinfo {pages} {025802} (\bibinfo {year}
  {2012}{\natexlab{b}})}\BibitemShut {NoStop}%
\bibitem [{\citenamefont {Giuliani}\ \emph {et~al.}(2018)\citenamefont
  {Giuliani}, \citenamefont {Mart\'{\i}nez-Pinedo},\ and\ \citenamefont
  {Robledo}}]{GMR.18}%
  \BibitemOpen
  \bibfield  {author} {\bibinfo {author} {\bibfnamefont {S.~A.}\ \bibnamefont
  {Giuliani}}, \bibinfo {author} {\bibfnamefont {G.}~\bibnamefont
  {Mart\'{\i}nez-Pinedo}}, \ and\ \bibinfo {author} {\bibfnamefont {L.~M.}\
  \bibnamefont {Robledo}},\ }\href {\doibase 10.1103/PhysRevC.97.034323}
  {\bibfield  {journal} {\bibinfo  {journal} {Phys. Rev. C}\ }\textbf {\bibinfo
  {volume} {97}},\ \bibinfo {pages} {034323} (\bibinfo {year}
  {2018})}\BibitemShut {NoStop}%
\bibitem [{\citenamefont {M{\"o}ller}\ \emph {et~al.}(1995)\citenamefont
  {M{\"o}ller}, \citenamefont {Nix}, \citenamefont {Myers},\ and\ \citenamefont
  {Swiatecki}}]{MNMS.95}%
  \BibitemOpen
  \bibfield  {author} {\bibinfo {author} {\bibfnamefont {P.}~\bibnamefont
  {M{\"o}ller}}, \bibinfo {author} {\bibfnamefont {J.~R.}\ \bibnamefont {Nix}},
  \bibinfo {author} {\bibfnamefont {W.~D.}\ \bibnamefont {Myers}}, \ and\
  \bibinfo {author} {\bibfnamefont {W.~J.}\ \bibnamefont {Swiatecki}},\
  }\href@noop {} {\bibfield  {journal} {\bibinfo  {journal} {At.\ Data Nucl.\
  Data Tables}\ }\textbf {\bibinfo {volume} {59}},\ \bibinfo {pages} {185}
  (\bibinfo {year} {1995})}\BibitemShut {NoStop}%
\bibitem [{\citenamefont {Delaroche}\ \emph {et~al.}(2010)\citenamefont
  {Delaroche}, \citenamefont {Girod}, \citenamefont {Libert}, \citenamefont
  {Goutte}, \citenamefont {Hilaire}, \citenamefont {Peru}, \citenamefont
  {Pillet},\ and\ \citenamefont {Bertsch}}]{DGLGHPPB.10}%
  \BibitemOpen
  \bibfield  {author} {\bibinfo {author} {\bibfnamefont {J.-P.}\ \bibnamefont
  {Delaroche}}, \bibinfo {author} {\bibfnamefont {M.}~\bibnamefont {Girod}},
  \bibinfo {author} {\bibfnamefont {J.}~\bibnamefont {Libert}}, \bibinfo
  {author} {\bibfnamefont {H.}~\bibnamefont {Goutte}}, \bibinfo {author}
  {\bibfnamefont {S.}~\bibnamefont {Hilaire}}, \bibinfo {author} {\bibfnamefont
  {S.}~\bibnamefont {Peru}}, \bibinfo {author} {\bibfnamefont {N.}~\bibnamefont
  {Pillet}}, \ and\ \bibinfo {author} {\bibfnamefont {G.~F.}\ \bibnamefont
  {Bertsch}},\ }\href@noop {} {\bibfield  {journal} {\bibinfo  {journal} {Phys.
  Rev. C}\ }\textbf {\bibinfo {volume} {81}},\ \bibinfo {pages} {014303}
  (\bibinfo {year} {2010})}\BibitemShut {NoStop}%
\bibitem [{\citenamefont {Erler}\ \emph {et~al.}(2013)\citenamefont {Erler},
  \citenamefont {Horowitz}, \citenamefont {Nazarewicz}, \citenamefont
  {Rafalski},\ and\ \citenamefont {Reinhard}}]{TOV-min}%
  \BibitemOpen
  \bibfield  {author} {\bibinfo {author} {\bibfnamefont {J.}~\bibnamefont
  {Erler}}, \bibinfo {author} {\bibfnamefont {C.~J.}\ \bibnamefont {Horowitz}},
  \bibinfo {author} {\bibfnamefont {W.}~\bibnamefont {Nazarewicz}}, \bibinfo
  {author} {\bibfnamefont {M.}~\bibnamefont {Rafalski}}, \ and\ \bibinfo
  {author} {\bibfnamefont {P.-G.}\ \bibnamefont {Reinhard}},\ }\href@noop {}
  {\bibfield  {journal} {\bibinfo  {journal} {Phys.\ Rev. C}\ }\textbf
  {\bibinfo {volume} {87}},\ \bibinfo {pages} {044320} (\bibinfo {year}
  {2013})}\BibitemShut {NoStop}%
\bibitem [{\citenamefont {M{\"o}ller}\ \emph {et~al.}(2008)\citenamefont
  {M{\"o}ller}, \citenamefont {Bengtsson}, \citenamefont {Carlsson},
  \citenamefont {Olivius}, \citenamefont {Ichikawa}, \citenamefont {Sagawa},\
  and\ \citenamefont {Iwamoto}}]{MBCOISI.08}%
  \BibitemOpen
  \bibfield  {author} {\bibinfo {author} {\bibfnamefont {P.}~\bibnamefont
  {M{\"o}ller}}, \bibinfo {author} {\bibfnamefont {R.}~\bibnamefont
  {Bengtsson}}, \bibinfo {author} {\bibfnamefont {B.}~\bibnamefont {Carlsson}},
  \bibinfo {author} {\bibfnamefont {P.}~\bibnamefont {Olivius}}, \bibinfo
  {author} {\bibfnamefont {T.}~\bibnamefont {Ichikawa}}, \bibinfo {author}
  {\bibfnamefont {H.}~\bibnamefont {Sagawa}}, \ and\ \bibinfo {author}
  {\bibfnamefont {A.}~\bibnamefont {Iwamoto}},\ }\href@noop {} {\bibfield
  {journal} {\bibinfo  {journal} {At. Data and Nucl. Data Tables}\ }\textbf
  {\bibinfo {volume} {94}},\ \bibinfo {pages} {758} (\bibinfo {year}
  {2008})}\BibitemShut {NoStop}%
\bibitem [{\citenamefont {Agbemava}\ and\ \citenamefont
  {Afanasjev}(2017)}]{AA.17-oct}%
  \BibitemOpen
  \bibfield  {author} {\bibinfo {author} {\bibfnamefont {S.~E.}\ \bibnamefont
  {Agbemava}}\ and\ \bibinfo {author} {\bibfnamefont {A.~V.}\ \bibnamefont
  {Afanasjev}},\ }\href {\doibase 10.1103/PhysRevC.96.024301} {\bibfield
  {journal} {\bibinfo  {journal} {Phys. Rev. C}\ }\textbf {\bibinfo {volume}
  {96}},\ \bibinfo {pages} {024301} (\bibinfo {year} {2017})}\BibitemShut
  {NoStop}%
\bibitem [{\citenamefont {Stone}\ \emph {et~al.}(1997)\citenamefont {Stone},
  \citenamefont {Bjarki}, \citenamefont {Gualtieri}, \citenamefont
  {Hannuschke}, \citenamefont {Lacey}, \citenamefont {Lauret}, \citenamefont
  {Llope}, \citenamefont {Magestro}, \citenamefont {Pak}, \citenamefont
  {Molen}, \citenamefont {Westfall},\ and\ \citenamefont {Yee}}]{multifrag.97}%
  \BibitemOpen
  \bibfield  {author} {\bibinfo {author} {\bibfnamefont {N.~T.~B.}\
  \bibnamefont {Stone}}, \bibinfo {author} {\bibfnamefont {O.}~\bibnamefont
  {Bjarki}}, \bibinfo {author} {\bibfnamefont {E.~E.}\ \bibnamefont
  {Gualtieri}}, \bibinfo {author} {\bibfnamefont {S.~A.}\ \bibnamefont
  {Hannuschke}}, \bibinfo {author} {\bibfnamefont {R.}~\bibnamefont {Lacey}},
  \bibinfo {author} {\bibfnamefont {J.}~\bibnamefont {Lauret}}, \bibinfo
  {author} {\bibfnamefont {W.~J.}\ \bibnamefont {Llope}}, \bibinfo {author}
  {\bibfnamefont {D.~J.}\ \bibnamefont {Magestro}}, \bibinfo {author}
  {\bibfnamefont {R.}~\bibnamefont {Pak}}, \bibinfo {author} {\bibfnamefont
  {A.~M.~V.}\ \bibnamefont {Molen}}, \bibinfo {author} {\bibfnamefont {G.~D.}\
  \bibnamefont {Westfall}}, \ and\ \bibinfo {author} {\bibfnamefont
  {J.}~\bibnamefont {Yee}},\ }\href@noop {} {\bibfield  {journal} {\bibinfo
  {journal} {Phys. Rev. Lett.}\ }\textbf {\bibinfo {volume} {78}},\ \bibinfo
  {pages} {2084} (\bibinfo {year} {1997})}\BibitemShut {NoStop}%
\bibitem [{\citenamefont {Dobaczewski}\ \emph {et~al.}(2014)\citenamefont
  {Dobaczewski}, \citenamefont {Nazarewicz},\ and\ \citenamefont
  {Reinhard}}]{DNR.14}%
  \BibitemOpen
  \bibfield  {author} {\bibinfo {author} {\bibfnamefont {J.}~\bibnamefont
  {Dobaczewski}}, \bibinfo {author} {\bibfnamefont {W.}~\bibnamefont
  {Nazarewicz}}, \ and\ \bibinfo {author} {\bibfnamefont {P.-G.}\ \bibnamefont
  {Reinhard}},\ }\href@noop {} {\bibfield  {journal} {\bibinfo  {journal} {J.~
  Phys.~ G}\ }\textbf {\bibinfo {volume} {41}},\ \bibinfo {pages} {074001}
  (\bibinfo {year} {2014})}\BibitemShut {NoStop}%
\bibitem [{\citenamefont {Jachimowicz}\ \emph {et~al.}(2017)\citenamefont
  {Jachimowicz}, \citenamefont {Kowal},\ and\ \citenamefont
  {Skalski}}]{JKS.17}%
  \BibitemOpen
  \bibfield  {author} {\bibinfo {author} {\bibfnamefont {P.}~\bibnamefont
  {Jachimowicz}}, \bibinfo {author} {\bibfnamefont {M.}~\bibnamefont {Kowal}},
  \ and\ \bibinfo {author} {\bibfnamefont {J.}~\bibnamefont {Skalski}},\ }\href
  {\doibase 10.1103/PhysRevC.95.014303} {\bibfield  {journal} {\bibinfo
  {journal} {Phys. Rev. C}\ }\textbf {\bibinfo {volume} {95}},\ \bibinfo
  {pages} {014303} (\bibinfo {year} {2017})}\BibitemShut {NoStop}%
\bibitem [{\citenamefont {Abbott}\ \emph {et~al.}(2017)\citenamefont {Abbott}
  \emph {et~al.}}]{NS-grav-wave-exp.17}%
  \BibitemOpen
  \bibfield  {author} {\bibinfo {author} {\bibfnamefont {B.~P.}\ \bibnamefont
  {Abbott}} \emph {et~al.},\ }\href@noop {} {\bibfield  {journal} {\bibinfo
  {journal} {Phys. Rev. Lett.}\ }\textbf {\bibinfo {volume} {119}},\ \bibinfo
  {pages} {161101} (\bibinfo {year} {2017})}\BibitemShut {NoStop}%
\bibitem [{\citenamefont {Caplan}\ \emph {et~al.}(2015)\citenamefont {Caplan},
  \citenamefont {Schneider}, \citenamefont {Horowitz},\ and\ \citenamefont
  {Berry}}]{CSHB.15}%
  \BibitemOpen
  \bibfield  {author} {\bibinfo {author} {\bibfnamefont {M.~E.}\ \bibnamefont
  {Caplan}}, \bibinfo {author} {\bibfnamefont {A.~S.}\ \bibnamefont
  {Schneider}}, \bibinfo {author} {\bibfnamefont {C.~J.}\ \bibnamefont
  {Horowitz}}, \ and\ \bibinfo {author} {\bibfnamefont {D.~K.}\ \bibnamefont
  {Berry}},\ }\href {\doibase 10.1103/PhysRevC.91.065802} {\bibfield  {journal}
  {\bibinfo  {journal} {Phys. Rev. C}\ }\textbf {\bibinfo {volume} {91}},\
  \bibinfo {pages} {065802} (\bibinfo {year} {2015})}\BibitemShut {NoStop}%
\bibitem [{\citenamefont {Fattoyev}\ \emph {et~al.}(2017)\citenamefont
  {Fattoyev}, \citenamefont {Horowitz},\ and\ \citenamefont
  {Schuetrumpf}}]{FHS.17}%
  \BibitemOpen
  \bibfield  {author} {\bibinfo {author} {\bibfnamefont {F.~J.}\ \bibnamefont
  {Fattoyev}}, \bibinfo {author} {\bibfnamefont {C.~J.}\ \bibnamefont
  {Horowitz}}, \ and\ \bibinfo {author} {\bibfnamefont {B.}~\bibnamefont
  {Schuetrumpf}},\ }\href {\doibase 10.1103/PhysRevC.95.055804} {\bibfield
  {journal} {\bibinfo  {journal} {Phys. Rev. C}\ }\textbf {\bibinfo {volume}
  {95}},\ \bibinfo {pages} {055804} (\bibinfo {year} {2017})}\BibitemShut
  {NoStop}%
\bibitem [{\citenamefont {Kycia}\ \emph {et~al.}(2017)\citenamefont {Kycia},
  \citenamefont {Kubis},\ and\ \citenamefont {W\'ojcik}}]{KKW.17}%
  \BibitemOpen
  \bibfield  {author} {\bibinfo {author} {\bibfnamefont {R.~A.}\ \bibnamefont
  {Kycia}}, \bibinfo {author} {\bibfnamefont {S.}~\bibnamefont {Kubis}}, \ and\
  \bibinfo {author} {\bibfnamefont {W.}~\bibnamefont {W\'ojcik}},\ }\href
  {\doibase 10.1103/PhysRevC.96.025803} {\bibfield  {journal} {\bibinfo
  {journal} {Phys. Rev. C}\ }\textbf {\bibinfo {volume} {96}},\ \bibinfo
  {pages} {025803} (\bibinfo {year} {2017})}\BibitemShut {NoStop}%
\bibitem [{\citenamefont {Petermann}\ \emph {et~al.}(2012)\citenamefont
  {Petermann}, \citenamefont {Langanke}, \citenamefont {Martinez-Pinedo},
  \citenamefont {Panov}, \citenamefont {Reinhard},\ and\ \citenamefont
  {Thielemann}}]{}%
  \BibitemOpen
  \bibfield  {author} {\bibinfo {author} {\bibfnamefont {I.}~\bibnamefont
  {Petermann}}, \bibinfo {author} {\bibfnamefont {K.}~\bibnamefont {Langanke}},
  \bibinfo {author} {\bibfnamefont {G.}~\bibnamefont {Martinez-Pinedo}},
  \bibinfo {author} {\bibfnamefont {I.~V.}\ \bibnamefont {Panov}}, \bibinfo
  {author} {\bibfnamefont {P.-G.}\ \bibnamefont {Reinhard}}, \ and\ \bibinfo
  {author} {\bibfnamefont {F.-K.}\ \bibnamefont {Thielemann}},\ }\href@noop {}
  {\bibfield  {journal} {\bibinfo  {journal} {Eur.\ Phys. J. A}\ }\textbf
  {\bibinfo {volume} {48}},\ \bibinfo {pages} {122} (\bibinfo {year}
  {2012})}\BibitemShut {NoStop}%
\end{thebibliography}%
\end{document}